\documentclass[fleqn,usenatbib]{mnras} 

\usepackage{newtxtext,newtxmath}

\usepackage[T1]{fontenc}

\DeclareRobustCommand{\VAN}[3]{#2}
\let\VANthebibliography\thebibliography
\def\thebibliography{\DeclareRobustCommand{\VAN}[3]{##3}\VANthebibliography}

\usepackage{graphicx}	
\usepackage[dvipsnames]{xcolor}
\usepackage{units}
\usepackage{bm}

\definecolor{mypurple}{rgb}{0.7,0.3,0.8}

\newcommand\EST{E_{\rm th}}

\newcommand{\vect}[1]{{{\mbox{\boldmath $#1$}}}}

\newcommand\dx{ {\delta x}}

\newcommand\Gyr{~ {\rm Gyr}}
\newcommand\erg{~ {\rm erg}}

\def\rmp{{Rev.~Mod.~Phys.}}             

\title[Supernova induced processing of ISM dust]{Supernova induced processing of interstellar dust: impact of ISM gas density and gas turbulence}

\author[Kirchschlager et al.]{
Florian Kirchschlager,$^{1}$\thanks{E-mail: f.kirchschlager@ucl.ac.uk}
Lars Mattsson,$^{2}$
 and
Frederick A. Gent$^{3,4}$
\\
$^{1}$ Department of Physics and Astronomy, University College London, Gower Street, London WC1E 6BT, UK\\
$^{2}$Nordita, KTH Royal Institute of Technology and Stockholm University, Hannes Alfv\'ens v\"ag 
12, SE-106 91 Stockholm, Sweden\\
$^{3}$ Astroinformatics, Department of Computer Science, Aalto University, PO Box 15400, FI-00076 Espoo, Finland\\
$^{4}$ School of Mathematics, Statistics and Physics,
      Newcastle University, NE1 7RU, UK 
}

\date{Accepted 2021 October 15. Received 2021 October 6; in original form 2021 August 31.}

\pubyear{2021}

\begin{document}
\label{firstpage}
\pagerange{\pageref{firstpage}--\pageref{lastpage}}
\maketitle

\begin{abstract}
Quantifying the efficiency of dust destruction in the interstellar medium (ISM)
due to supernovae (SNe) is crucial %
  for the understanding of 
galactic dust evolution.  We present 3D hydrodynamic
simulations of an SN blast wave propagating through the ISM.
 The interaction between the forward shock of the remnant and the surrounding
ISM leads to destruction of ISM dust by the shock heated gas.
 We consider the dust processing due to ion sputtering, accretion of
 atoms/molecules and grain-grain collisions. 
 Using 2D slices from the simulation timeseries, we apply post-processing
 calculations using the \textsc{Paperboats} code. We find that
efficiency of dust destruction depends strongly on the rate of grain shattering
due to grain-grain collisions. 
The effective dust destruction is similar to previous theoretical estimates 
when grain-grain collisions are omitted, but
with grain shattering included, the net destruction efficiency is roughly one
order of magnitude higher.
 This result
  indicates
that the dust destruction
rate in the ISM may have been severely underestimated in previous work, which
only 
  exacerbates
the dust-budget crises seen in galaxies at high redshifts.
\end{abstract}




 
\section{Introduction}
Dust grains 
  comprising
silicates, iron oxides and carbonaceous material are found throughout the
interstellar medium (ISM). It is well established that evolved stars and
supernovae (SNe) produce dust (\citealt{Zhukovska08, Matsuura09,  Gomez12a,
Wesson2015, Niculescu2021}), but dust grains are also consumed in star
formation (e.g.~\citealt{Testi2014}) or can be destroyed by shock waves from
SNe (\citealt{Slavin2015, Slavin2020, Martinez2018, Martinez2019, Hu2019}). In
the latter case destruction is believed to be the result of SN shocks hitting
the gas and dust of the ambient ISM and thus leading to an increased rate of
ion sputtering and grain-grain collisions.
These reduce the dust mass, change
the shape of the grain-size distribution, and increase the gas-to-dust mass
ratio \citep{McKee89,Draine90,Jones94,Slavin04,Bocchio2014, Lakicevic2015, Mattsson16}.
Consequently, high star formation rates cause high rates of dust destruction,
which seems at odds with the existence of massive starburst galaxies with very
large dust masses at high redshifts
\citep{Bertoldi03,Gall11a,Gall11b,Mattsson11b,Michalowski10a,Michalowski10b,Watson15}.

Dust destruction by SN induced processes is hard to quantify \citep[see,
e.g.,][]{Jones11}. The most commonly adopted calibration \citep[based on the
prescription by][]{McKee89} for the solar neighbourhood suggests that 
the masses 
$M_{\rm cl.\,gas} = 800 M_\odot$ (corresponding to a present-day time-scale $\tau_{d,0}
= 0.8 \Gyr$) for carbonaceous dust and $M_{\rm cl.\,gas} = 1200 M_\odot$
($\tau_{d,0} = 0.6 \Gyr$) for silicates are the gas masses that are cleared of
dust by a single SN event (\citealt{Jones94,Tielens94}). Similar gas masses are
found by \cite{Slavin2015} and \cite{Hu2019} while the destruction time-scales
can vary between $0.44$ and $\unit[3.2]{Gyr}$ for carbonaceous dust and between
$0.35$ and $\unit[2]{Gyr}$ for silicates.

Dust grains can also grow by accretion of molecules, coagulation or ion trapping, thus providing a replenishment mechanism or even a dominant channel of dust formation \citep{Draine90,Mattsson11b,Kuo12, Kirchschlager2020}. Radial distributions of dust in late-type galaxies suggest efficient grain growth by condensation in the ISM \citep{Mattsson12a,Mattsson12b,Mattsson14b} and the overall levels of dust depletion in damped $Ly-\alpha$ systems and quasar host galaxies point in the same direction \citep{DeCia13, DeCia16,Kuo12,Mattsson14a}. Overall, the interactions between the different types of grain processing mentioned above can be decisive for the average efficiency of dust condensation in the ISM and, obviously, also have a profound effect on the grain-size distribution.

In this paper, we study the survival rate of ISM dust when a single SN blast
wave expands into an initially homogeneous ISM. 
We consider uniform ambient gas number density of 1 and 0.1 cm$^{-3}$, at rest
and modestly perturbed, and compare effects on the dust
transport, dust destruction and grain growth. Hydrodynamic
simulations with the \textsc{Pencil} code (\citealt{Pencil-JOSS}) are used to
mimic the expansion of the SN in the gas phase of the ISM, and post-processing
simulations with the \textsc{Paperboats} code (\citealt{Kirchschlager2019}) are
used to model the dust evolution.


 
\section{Hydrodynamic simulations}\label{sect:HD}

The simulations of the dust are applied to a background ISM gas modelling the
hydrodynamics of an explosion of an SN. Each model is within a 3D periodic
domain large enough to evolve the SN remnant beyond $\unit[1]{Myr}$.
The models have
an equidistant grid resolution of 0.5\,pc along each edge, sufficiently well
resolved to obtain well converged solutions \citep{GMKSH20,GMKS21} and also to
exhibit instabilities resembling Vishniac-Ostriker-Bertschinger (VOB)
overstability \citep{Vishniac83,VOB85}, Rayleigh-Taylor (RT) or potentially
Richtmyer-Meshkov (RM) instability \citep{Bro02}. 

The SN remnant is initialised by the injection of thermal energy $\EST = 10^{51}\erg$ within a sphere of initial nominal radius of {$R_0$}, following a radial profile as 
  \begin{equation}  
     E(r) = E_0\exp\left(-\left[
     r/{R_0}
  \right]^{{2}}\right),
  \end{equation}
  where $r$ is the radial distance to the explosion origin, $R_0 = 2.5$\,pc, and $E_0$ is the normalising coefficient set such that the volume integral of $E(r)$ is equal to $\EST$.
  This smooth profile has been found to avoid numerical artifacts, which
more easily emerge  when applying stepped or much steeper profiles in the
injected energy distributions, while still recovering an expansion closely
resembling the Sedov-Taylor solution well in advance of the snowplough phase of relevance to this study.
  The remnant origin is located on a grid point in the center of the domain.
  The initial condition creates a blast wave, which rapidly evacuates the ISM
  from the origin of the sphere and continues to expand into the ambient ISM.
  Within a short time, still within the adiabatic expansion phase of the 
  SN lifespan, the properties of the remnant converge to a solution fitting
  the Sedov-Taylor analytic solution \citep{Taylor50,Sedov59},
  \begin{equation}
    \label{eq:sed}
    R= \left(\kappa\frac{\EST}{\rho_{0}}\right)^{{1}/{5}}t^{{2}/{5}},
  \end{equation}
  where $R$ is the remnant radius, $\rho_{0}$ is the
  ambient gas density, and ${\kappa \approx 2.026}$ is the dimensionless parameter for
  $\gamma=5/3$ \citep{Ostriker88}.
  We neglect any additional mass from the SN ejecta.

 Non-adiabatic heating $\Gamma$ and cooling $\Lambda (T)$ are included
 \citep{Gent:2013b}.
 Cooling by radiative losses follow \citet{Wolfire:1995} and
 \citet{Sarazin:1987}, using a piecewise power law dependence of the cooling
 coefficient on temperature.
 The contribution from FUV heating follows
 \citet[][see \citealt{Gent:2013a}]{Wolfire:1995}, which vanishes for
 temperatures somewhat $\gtrsim10^{4}$\,K.
 When radiative cooling processes are included the SN evolution changes.
 As the remnant expands and the shock front accumulates more gas from the
 ambient ISM, cooling becomes more efficient in the increasingly dense shell.
 Loss of energy more rapidly reduces the shell speed.  
 As demonstrated by \citet{GMKSH20} at resolution better than 1\,pc our 
 solution converges to the semi-analytical non-adiabatic solutions obtained by 
 \citet{Cioffi88}, slightly modified to account for cooling in our model
 continuing to apply for $T<10^4$\,K.

 To integrate these solutions we use the \textsc{Pencil} code\footnote{https://github.com/pencil-code} \citep{Pencil-JOSS} for the system of non-ideal, compressible, non-isothermal HD equations
  \begin{eqnarray}
  \label{eq:mass}
    \frac{D\rho}{Dt} &=& 
    -\rho \vect{\nabla} \cdot \vect{u}
    +\vect\nabla \cdot\zeta_D\vect\nabla\rho,
  \end{eqnarray}
  \begin{eqnarray}
  \label{eq:mom}
    \rho\frac{D\vect{u}}{Dt} &=& 
    -\rho c_{\rm s}^2\vect\nabla\left({s}/{c_{\rm p}}+\ln\rho\right)
    \nonumber\\
    &+&\vect\nabla\cdot \left(2\rho\nu{\mathbfss W}\right)
    +\rho\,\vect\nabla\left(\zeta_{\nu}\vect\nabla \cdot \vect{u} \right)
    \nonumber\\
    &+&\vect\nabla\cdot \left(2\rho\nu_3{\mathbfss W}^{(3)}\right)
  {-\vect u\vect{\nabla}\cdot\left(\zeta_D\vect{\nabla}\rho\right)},
  \end{eqnarray}
  \begin{eqnarray}
  \label{eq:ent}
    \rho T\frac{D s}{Dt} &=&
     \EST\dot\sigma +\rho\Gamma
    -\rho^2\Lambda 
    \nonumber\\
    &+&2 \rho \nu\left|{\mathbfss W}\right|^{2}
    +\rho\,\zeta_{\nu}\left(\vect\nabla \cdot \vect{u} \right)^2
    \nonumber\\
    &+&\vect\nabla\cdot\left(\zeta_\chi\rho T\vect\nabla s\right)
    +\rho T\chi_3\vect\nabla^6 s
    \nonumber\\
    &-& {c_{\rm{v}}\,T \left(
    \zeta_D\nabla^2\rho + \vect\nabla\zeta_D\cdot\vect\nabla\rho\right)},
  \end{eqnarray}
 with the ideal gas equation of state closing the system.
 Most variables take their usual meanings, with ${\mathbfss W}$ being the 
 first order traceless rate of strain tensor and $\left|{\mathbfss W}\right|^{2}\equiv W_{ij}W_{ij}$.
 Terms containing $\zeta_D{=2.5},\,\zeta_\nu{=6.25}$ and $\zeta{_\chi=4.0}$
 are applied to resolve shock discontinuities with artificial diffusion of
 mass, momentum, and energy proportional to shock strength
 \citep[see][for details]{GMKSH20}.
 Equations~(\ref{eq:mom}) and (\ref{eq:ent}) include terms with $\zeta_D$
 to provide momentum and energy conserving corrections for the
 artificial mass diffusion applying in Equation~(\ref{eq:mass}).
 Terms containing $\nu_3,\,\chi_3$ and $\eta_3$ apply sixth-order hyperdiffusion,
 in which ${\mathbfss W}^{(3)}$ is the fifth order rate of strain tensor, 
 to resolve grid-scale instabilities \citep[see, e.g.,][]{ABGS02,HB04,GMKS21},
 with mesh Reynolds number set to be $\simeq1$ for each $\dx$.
 The incidence of SN denoted by $\dot\sigma$ occurs once at $t=0$.


  \begin{table}
 \centering
 \caption{Overview of the hydrodynamic simulation setup. The index ({A-D}) refers to a certain simulation setup. $n_{\rm gas,0}$ is the gas number density of the unperturbed ambient medium at $t=0$, $n_\textrm{\rm grid}$ is the number of grid cells in one dimension, and $l_\textrm{\rm box}$ is the size of the domain in one dimension
}
 \begin{tabular}{c c c c c}
 \hline\hline
 Index&$n_{\rm gas,0}\,[\textrm{cm}^{-3}]$&turbulence&$n_\textrm{\rm grid}$& $l_\textrm{\rm box}\,[pc]$\\\hline 
 A&$0.1$ & no &512&256\\ 
 B&$1.0$ & no &320&160\\  
 C&$0.1$ & yes&512&256\\  
 D&$1.0$& yes&320&160\\\hline   
 \end{tabular}
 \label{List_hydrosimulations}
 \end{table} 

 In order to study the effects of gas density and turbulence on the evolution
of the ISM gas impacted by a blast wave, we model the hydrodynamics of four
different scenarios (Table~\ref{List_hydrosimulations}). Initially, the gas
temperature of the pre-shock gas amounts to $T_\textrm{gas}=\unit[10^4]{K}$,
the mean molecular weight is $\mu_\textrm{gas}=0.531$ representing an ionised
hydrogen gas, and the gas density is homogeneous with $n_{\rm
gas,0}=\unit[0.1]{cm^{-3}}$ or $\unit[1]{cm^{-3}}$, respectively. Both shock
velocity and gas temperature decrease with time as a result of expansion. For
simulations without turbulence the pre-shock gas is in rest, while we adopt
velocity fluctuations of $\unit[0.2]{km/s}$ in the turbulent scenario which
cause density inhomogeneities with ongoing time. The weak subsonic
perturbation is intended to induce the turbulent density, structure primarily
through thermal instabilities from differential cooling.  The effects of
ambient turbulence in the subsonic, transonic and supersonic regimes shall be
more extensively considered in a future analysis.
 
 The spatial resolution of the simulations is $\unit[0.5]{pc}$ and the temporal resolution of the output frames is $\unit[250]{yr}$.  The total simulation time is for all scenarios $\unit[1]{Myr}$ which is similar to the expected time scale for the existence of a SN blast wave before it is obliterated and cancelled out  in the ISM due to density inhomogeneities and further (SN) shock waves in the near vicinity. The given box sizes of $l_\textrm{box}= \unit[256]{pc}$ and $\unit[160]{pc}$, respectively, ensure that the blast wave does not reach the domain edges 
  within
$\unit[1]{Myr}$.

 
\section{Dust processing}\label{sect:DP}

 We use our post-processing code \textsc{Paperboats} (\citealt{Kirchschlager2019}) to study the dust evolution in the turbulent or non-turbulent ISM gas when impacted by an SN shock wave. Based on the temporally and spatially resolved gas density, gas velocity, and gas temperature output of \textsc{Pencil}, we investigate the dust transport and derive the dust destruction rate. We give here a short overview of the processes considered and refer to \cite{Kirchschlager2019} for a detailed description.

\subsection{Dust processes} 
\textsc{Paperboats} is a dust processing code which simulates the dust transport as well as grain destruction and growth processes in a gaseous medium. The dust is accelerated by the streaming gas taking into account both collisional and plasma drag (\citealt{Baines65, Draine79}). The calculation of the grain charges follows the analytical description derived by \cite{Fry2020} with respect to impinging plasma particles, secondary electrons, transmitted plasma particles, and field emission. The gas is assumed to be fully ionised. Destruction processes include thermal and non-thermal (kinematic) sputtering (e.g., \citealt{Barlow78, Shull1978, Tielens94}) as well as fragmentation and vaporisation in grain-grain collisions (e.g.,
\citealt{Jones96, Hirashita09}). Growth processes comprise the coagulation (sticking) in grain-grain collisions and the accretion of gas onto the surface of
the grains which are present either at low relative velocities or when the impact energy of a gas particle is below a certain threshold. The effect of ion trapping where ions that are responsible for sputtering events penetrate into the grain and get trapped (\citealt{Kirchschlager2020}) is not included in this study. Coulomb interactions between a gas ion and a charged dust grain in a sputtering event as well as between two charged grains in a grain-grain collision are taken into account as is the size-dependence of the sputtering yield (\citealt{Bocchio2012}). Although \textsc{Paperboats} is able to treat 3D simulations, we consider here only the central slice through the middle of the box of the \textsc{Pencil} output (a single cell in z-direction) due to the large computational effort for highly resolved 3D post-processing simulations.

\subsection{Dust model}
The grains in our simulations are made of compact silicate material
and assumed to be spherical with radius $a$. The initial grain size distribution follows a power-law $a^{-\gamma}$ with index $\gamma = 3.5$ and minimum and maximum grain size $a_{\rm min}=\unit[5]{nm}$ and $a_{\rm max}=\unit[250]{nm}$, respectively (MRN distribution; \citealt{Mathis77}). The grain sizes are binned in 20 log-spaced size bins which range from $\unit[0.6]{nm}$ to $\unit[350]{nm}$, allowing the dust to get partly destroyed or to grow due to destruction and growth processes. Dust material with sizes below  $\unit[0.6]{nm}$ is treated as completely destroyed and is assigned to the gas phase, known as "dusty gas" (see \citealt{Kirchschlager2019}). The material parameters required for the dust post-processing (for sputtering, grain-grain collisions, and charging) are given in Table 2 of \cite{Kirchschlager2019}. Initially, the dust is at rest and homogeneously distributed in the domain with a gas-to-dust mass ratio of $\Delta_{\rm gd} = 100$ (for comparison, in Appendix \ref{apx:delta10} we show the results for $\Delta_{\rm gd} = 10$).

 
 \section{Results and discussion}\label{sec:results}

   \begin{figure*}
        \resizebox{\hsize}{!}{
      \includegraphics[trim=1.0cm 1.1cm 3.9cm 1.34cm, clip=true]{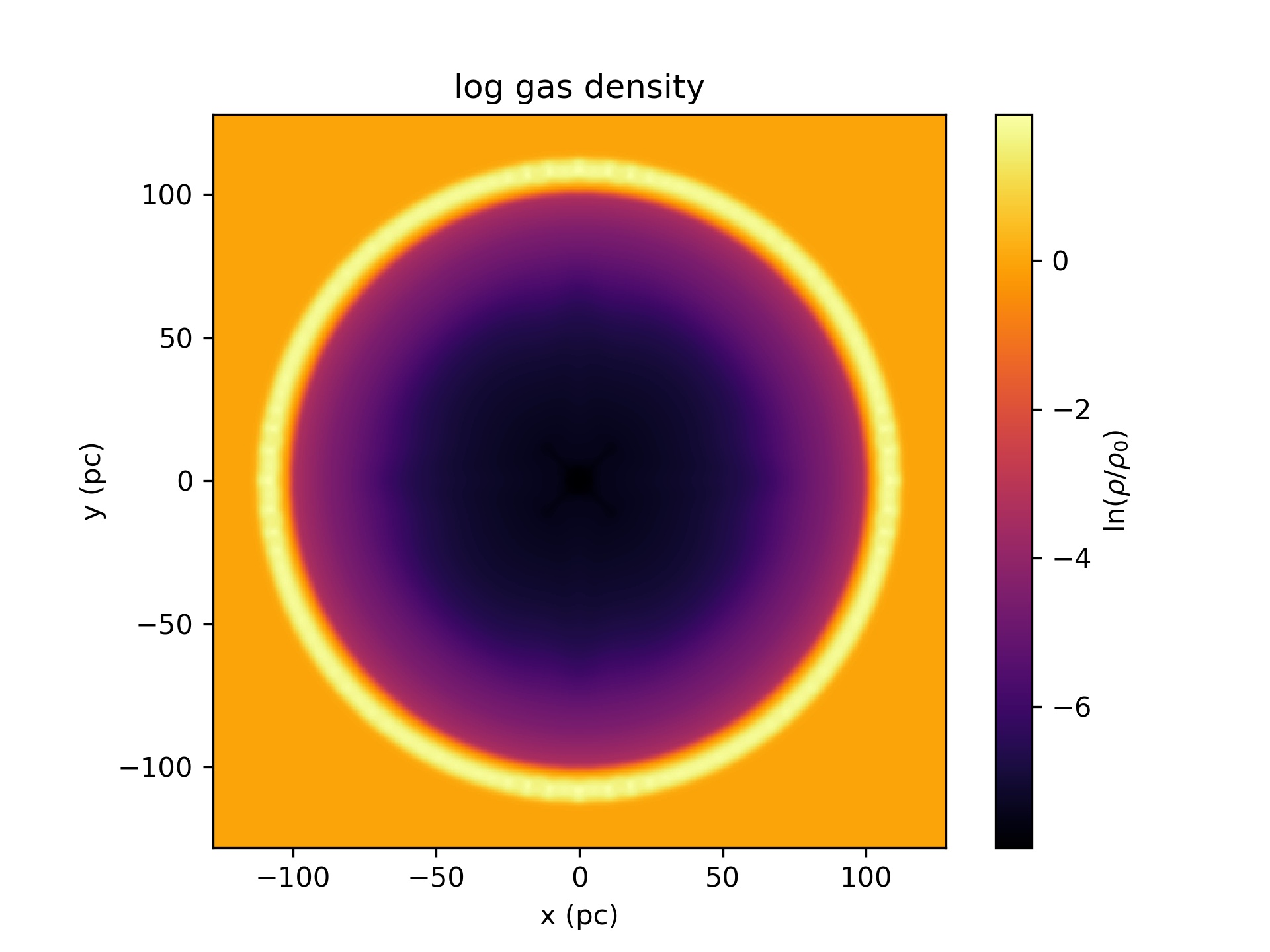}
      \includegraphics[trim=2.6cm 1.1cm 1.5cm 1.34cm, clip=true]{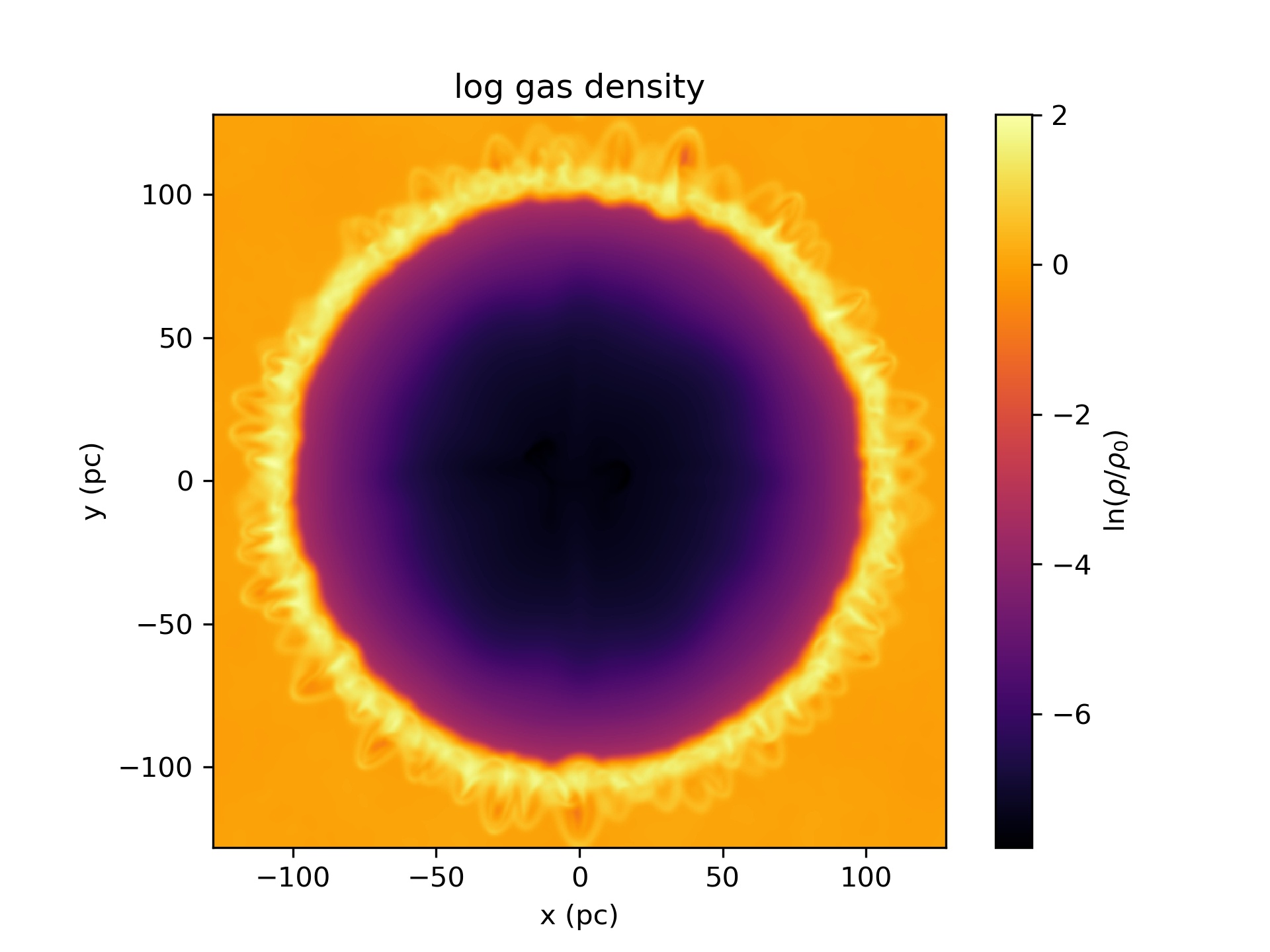}
       }
      \resizebox{\hsize}{!}{
      \includegraphics[trim=1.0cm 0.0cm 3.9cm 1.34cm, clip=true]{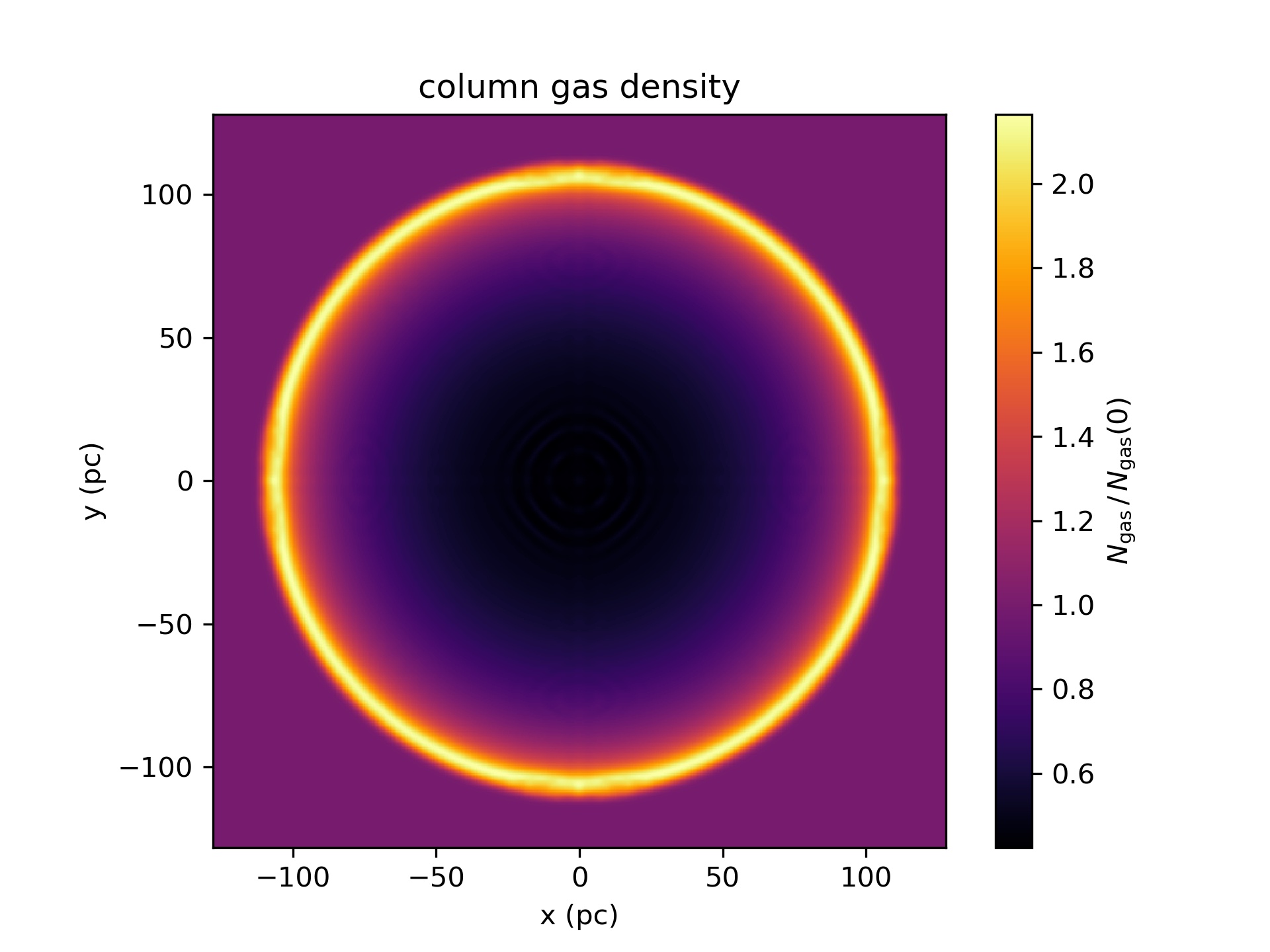}
      \includegraphics[trim=2.6cm 0.0cm 1.5cm 1.34cm, clip=true]{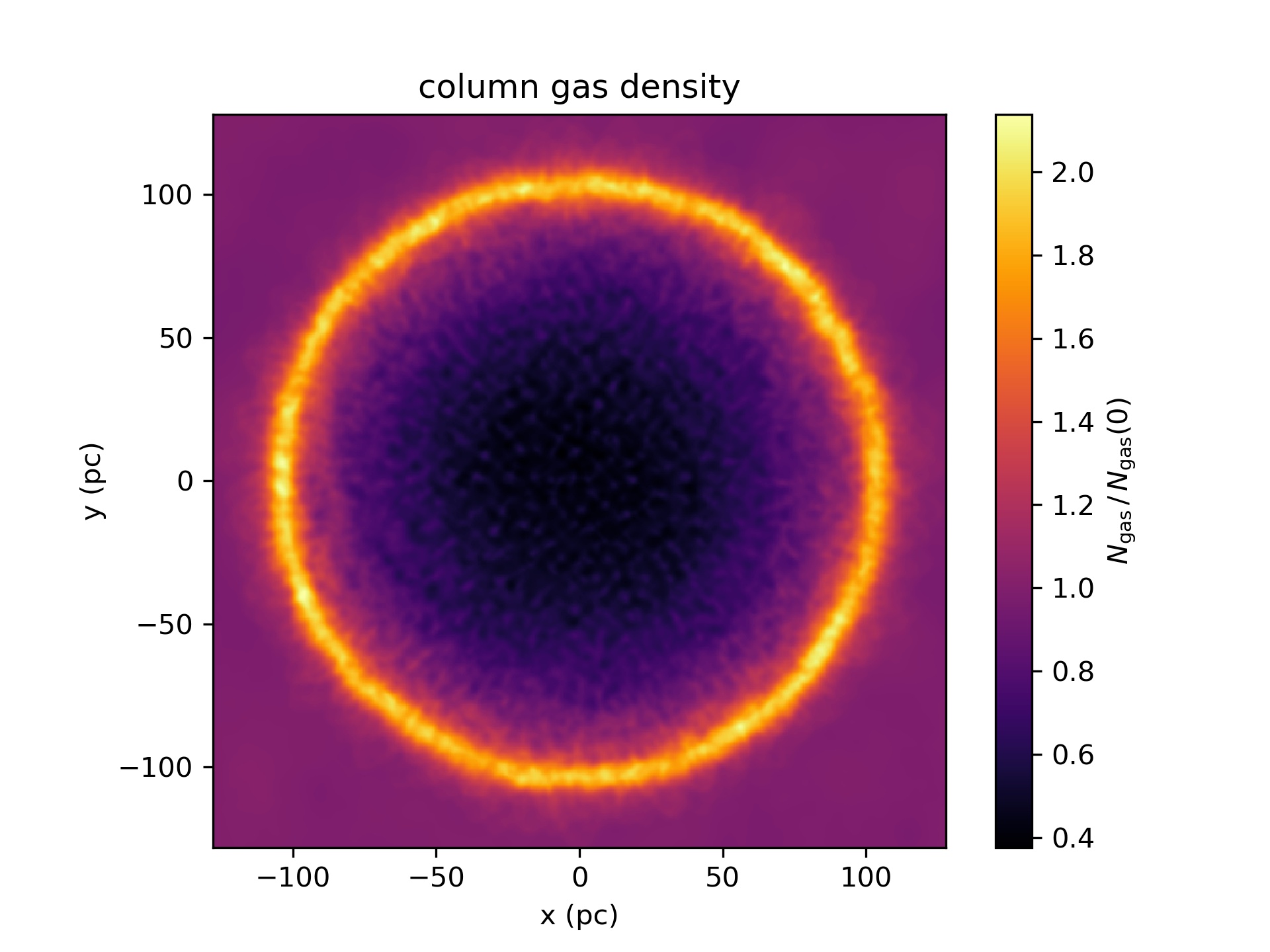}}
  \caption{Gas shell structure at $t = \unit[1]{Myr}$ for a blast wave propagating through an ISM gas with a uniform density
$n_{\rm gas,0}=\unit[0.1]{cm^{-3}}$ at $t = 0$. Upper row: Gas density for a non-turbulent (left) or a turbulent medium (right). Lower row: Column density integrated along the line of sight for a non-turbulent (left) or a turbulent medium (right).\label{3Dsedov_n01}}
  \end{figure*}

   \begin{figure*}
        \resizebox{\hsize}{!}{
      \includegraphics[trim=1.0cm 1.1cm 3.9cm 1.4cm, clip=true]{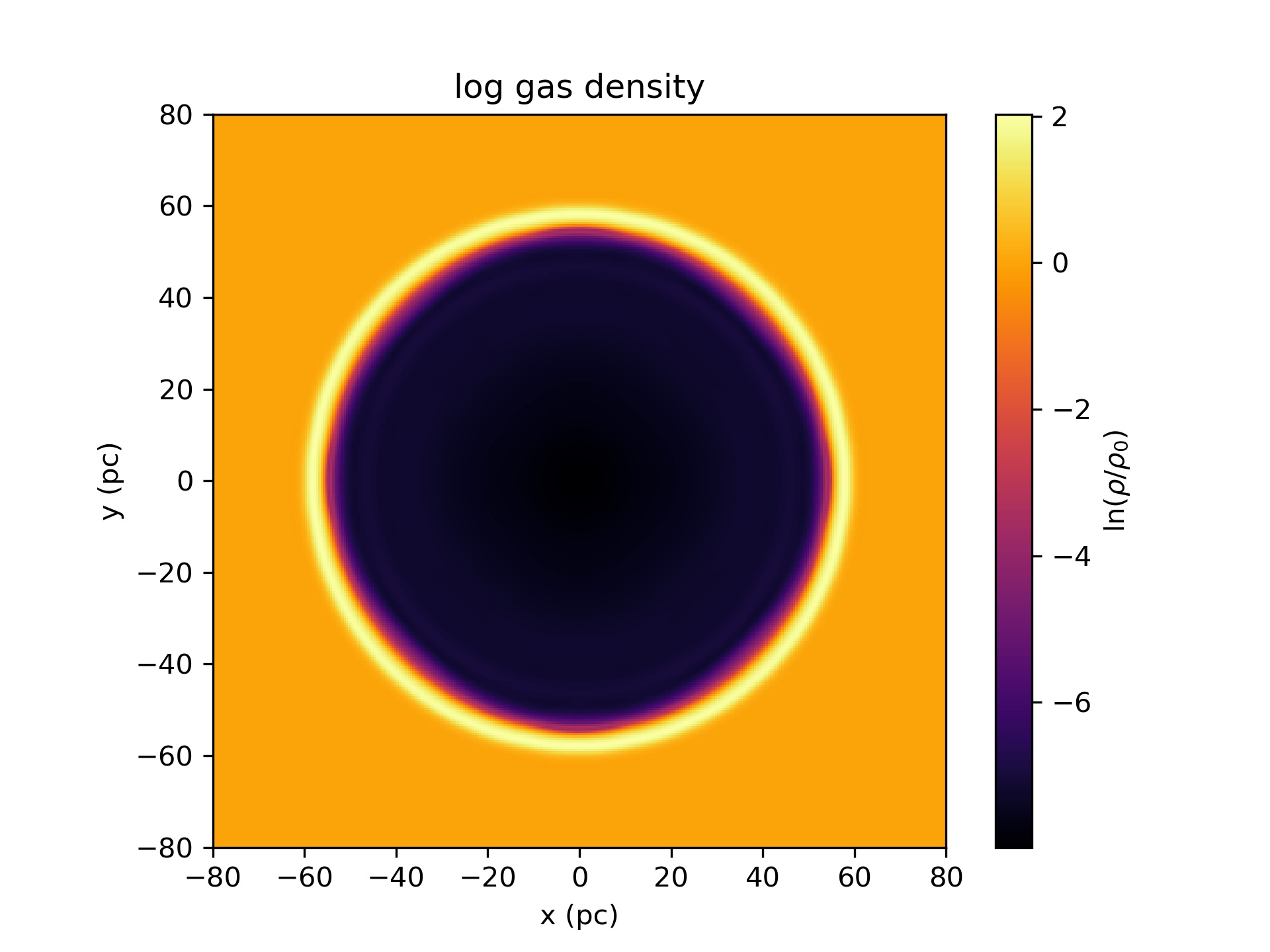}
       \includegraphics[trim=2.6cm 1.1cm 1.5cm 1.4cm, clip=true]{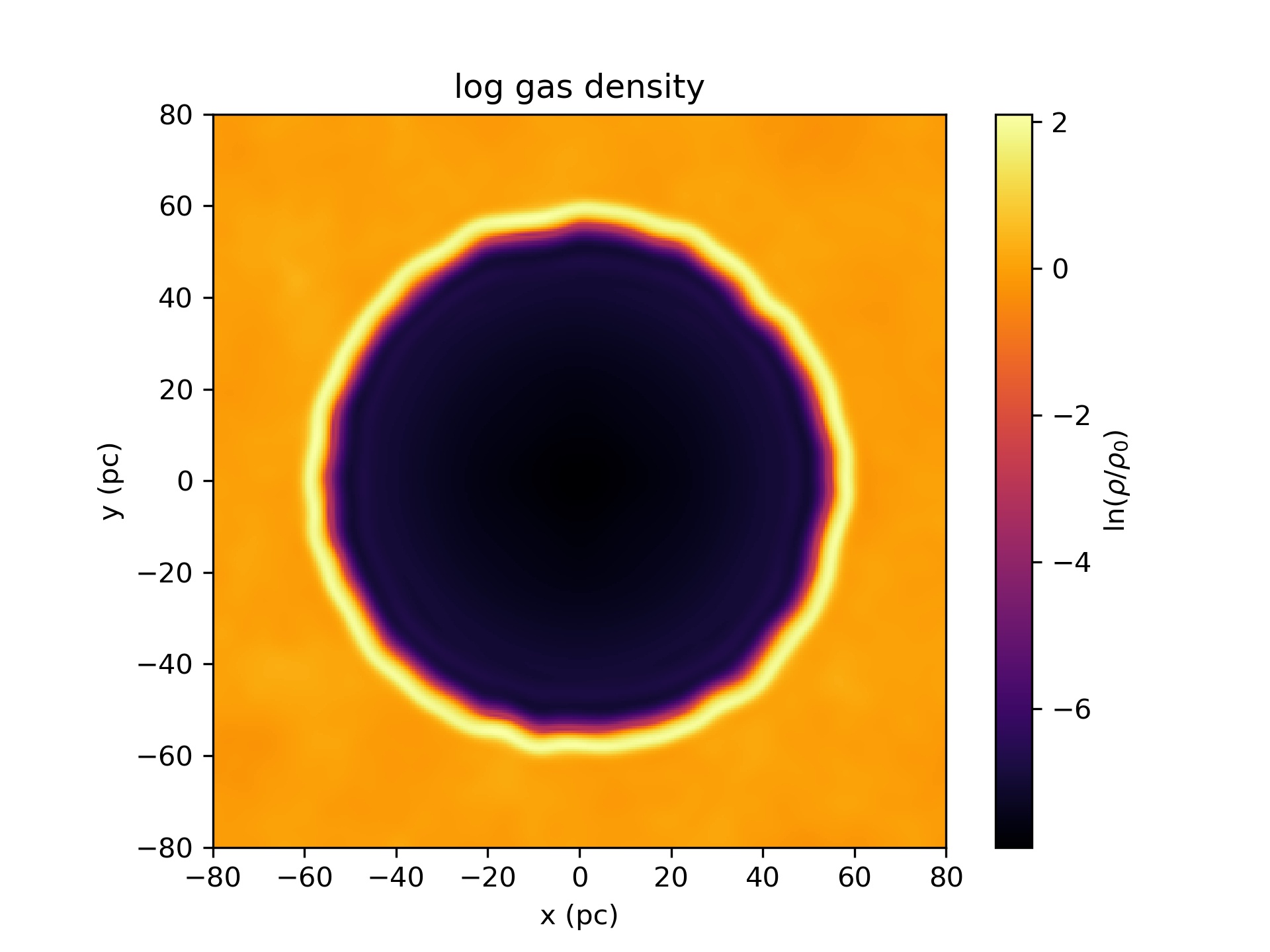}
      }
      \resizebox{\hsize}{!}{
      \includegraphics[trim=1.0cm 0.0cm 3.9cm 1.34cm, clip=true]{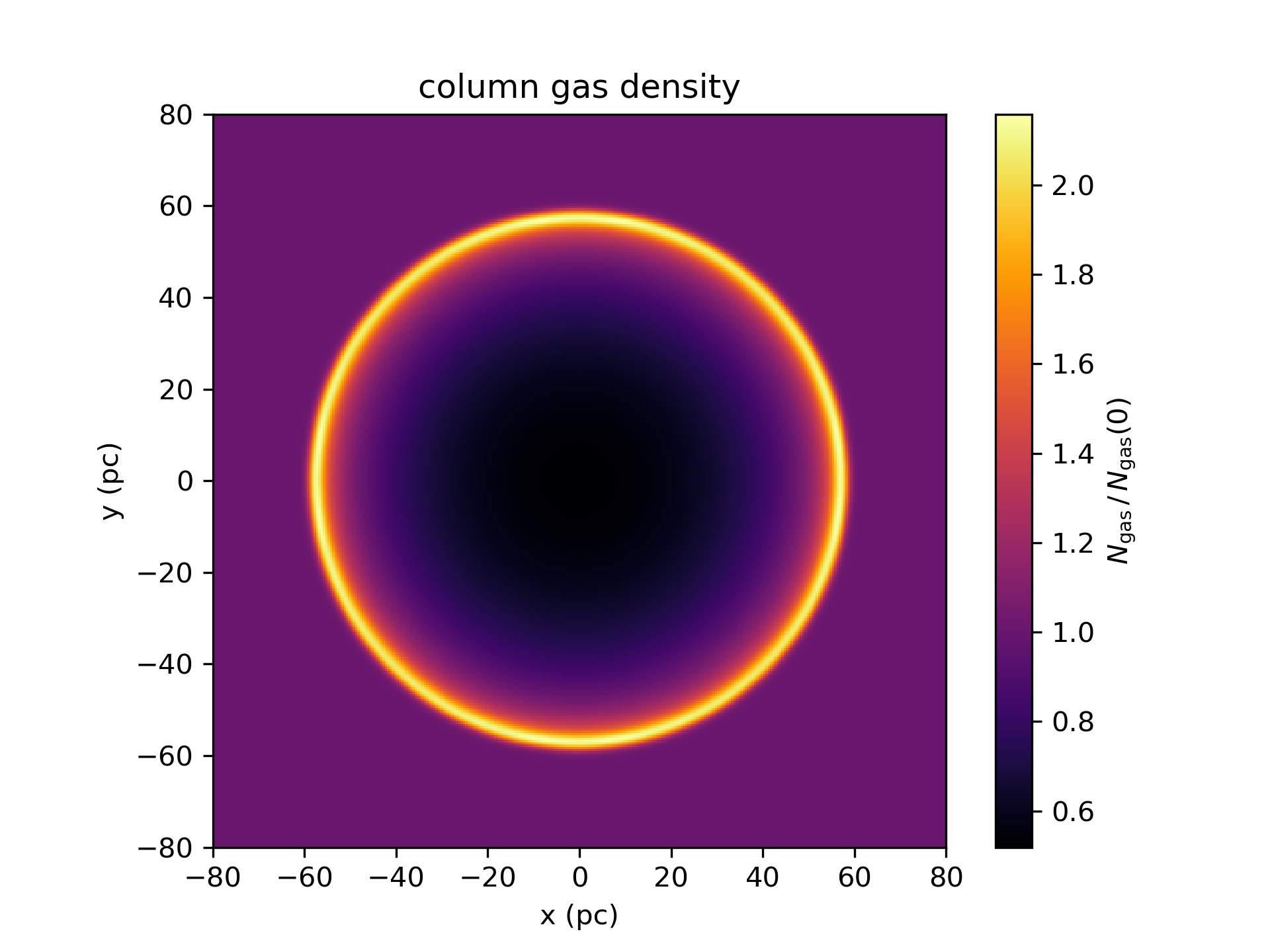}
      \includegraphics[trim=2.6cm 0.0cm 1.5cm 1.34cm, clip=true]{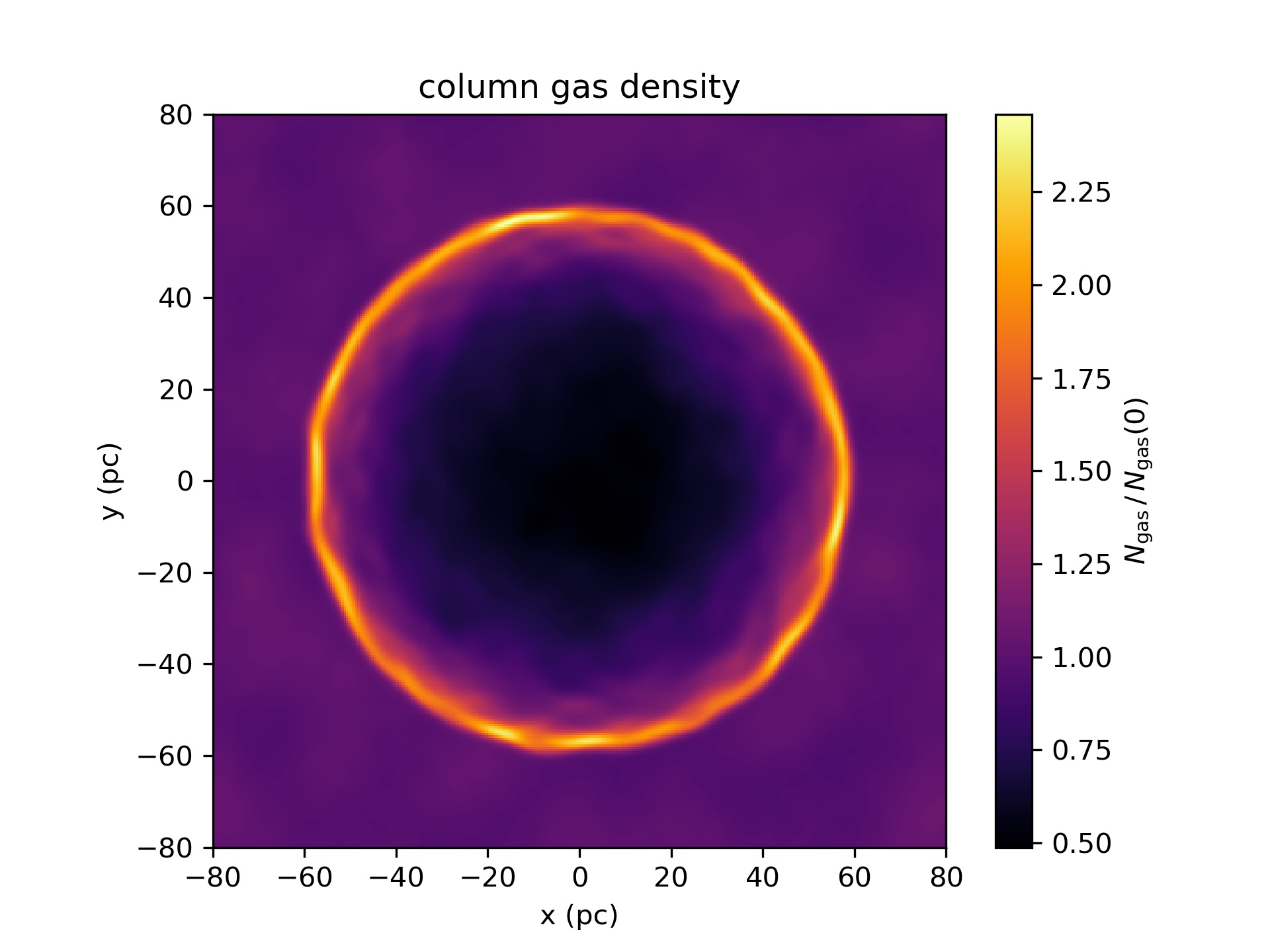}
      }
  \caption{Same as Fig.~\ref{3Dsedov_n01} but for the uniform density $n_{\rm gas,0}=\unit[1]{cm^{-3}}$ at $t = 0$.\label{3Dsedov}}
  \end{figure*}
  
     \begin{figure*}
        \includegraphics[clip=true,page=1,height = 3.4cm]{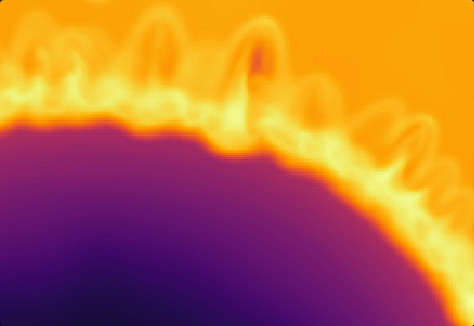}\hspace*{0.1cm}
%
        \includegraphics[clip=true,page=1,height = 3.4cm]{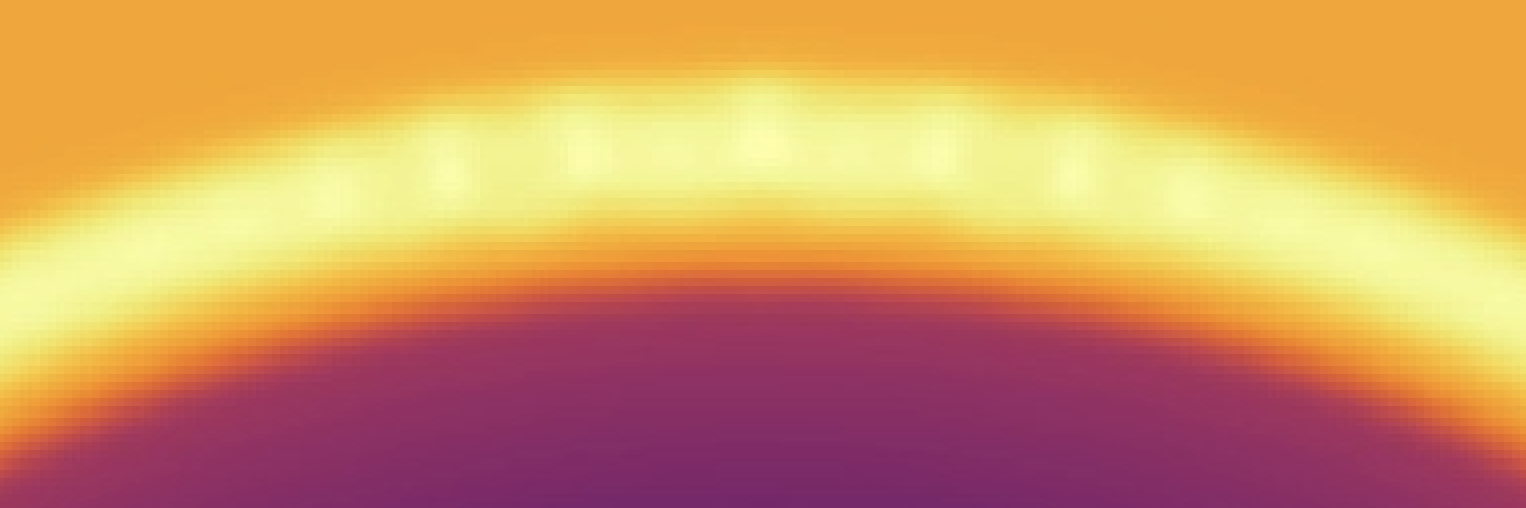}
  \caption{Cutouts from upper right panels of Figs.~\ref{3Dsedov_n01} (left) and \ref{3Dsedov}  (right)  showing the instabilities forming in more detail. \label{cutout}}
  \end{figure*}
  

\subsection{3D hydrodynamic simulations of SN blast waves}
We have performed simulations of an SN blast wave expanding in a 3D domain using the \textsc{Pencil Code} as described in Section \ref{sect:HD}. We present four different runs (see Table \ref{List_hydrosimulations}) where we consider both a low and a high-density ambient ISM ($n_{\rm gas,0} = 0.1$~g~cm$^{-3}$  and $n_{\rm gas,0} = 1.0$~g~cm$^{-3}$, respectively) and explore possible effects due to turbulence induced instabilities (and overstabilities). 

\subsubsection{Low-density ISM}
In Fig. \ref{3Dsedov_n01} we show the resultant shell structure at $t = 1$~Myr for a blast wave propagating through an ISM gas with a uniform density $n_{\rm gas,0} = 0.1$~g~cm$^{-3}$ at $t=0$. Upper panels show a logarithmic density slice through the middle of the simulation box, while the lower panels show projected linear density, similar to what would be detected by observations. The left panels show the case of a homogeneous ambient medium (simulation A, see index in Table~\ref{List_hydrosimulations}) and the right panels show the results including the turbulence-like initial velocity field (simulation C). The physical size of the shell at $t = 1$~Myr is $\sim 220$~pc across, which is somewhat larger than most observed SN remnants and about twice the size of the shell for the high-density case (see below). Moreover, we note that the low-density runs are prone to develop instabilities. The case without ``turbulence'' (simulation A) shows only small perturbations in the shell, which we interpret as an example of artificial VOB overstability induced by limitations of the Cartesian grid. With ``turbulence'' included (simulation C, right panels in Fig. \ref{3Dsedov_n01}) we see a clear case of VOB overstability after $\sim 500$ kyr (also reflected in the dust, see Fig. \ref{fig_C}). The VOB ``wiggles'' appear to induce instabilities, which we interpret as being of RM type  \citep{Bro02}, although it should be emphasised that it is difficult to tell from the simulation result, since the scale of the VOB ``wiggles'' are of the order a few pc and thus the whole phenomenon is not sufficiently resolved (grid resolution: $\unit[0.5]{pc}$). 

\subsubsection{High-density ISM}
The typical average gas density in the local ISM of the Galaxy is $n_{\rm
gas,0} \sim 1.0$~g~cm$^{-3}$. Hence, we also present simulations of a blast
wave propagating through an ISM gas with a uniform density $n_{\rm gas,0} =
1.0$~g~cm$^{-3}$ at $t=0$. Higher density means that the blast wave slows down
faster, i.e., the shell structure is smaller ($\sim\unit[140]{pc}$ across) and
less evolved at $t = 1$~Myrs.
We chose not to continue the simulations (B and D) beyond this point, because an
SNR will eventually merge with the ISM and is unlikely to survive much longer
than 1~Myr (the oldest known SNR is the Lambda Orionis Ring with an estimated age of $\unit[1]{Myr}$; \citealt{Dolan2002}).
 In
Fig. \ref{3Dsedov} (same as Fig. \ref{3Dsedov_n01}, but for $n_{\rm gas,0} =
1.0$~g~cm$^{-3}$) we note that the characteristic ``wiggles'' of the VOB
overstability are present, but no secondary instabilities (``plumes'' of RT or RM type) appear to have formed at 1~Myr. The lower right panel of Fig. \ref{3Dsedov} also shows regions of enhanced density in the projected shell, which is consistent with what is seen in observations.

\subsubsection{Weakly turbulent ISM}
In the context of dust processing, the most important difference between the
$n_{\rm gas,0} = 0.1$~g~cm$^{-3}$ and $n_{\rm gas,0} = 1.0$~g~cm$^{-3}$
simulations is of course the density contrast. The shells of the shocked
material in simulations B and D contain much more matter than the shells
forming in simulations A and C. Compared to the high density simulation with turbulence (D) the low density
simulation (C) shows more small-scale structure at and near the remnant
shell, which may potentially amplify the dust processing (more about this in Section~\ref{sec:dustproc} below). In order to emphasize the instabilities induced by turbulence, we present in Fig.~\ref{cutout} cutouts of the shell structure at
$\unit[1]{Myr}$.

To get a handle on how much variance is caused by the turbulence in the
background ISM, we have performed simulations corresponding to simulation C
and D but without the SN blast wave (see Appendix \ref{apx:faketurb}). In
Fig.~\ref{fig:fake} we show the resultant gas-density structure (slices through
the middle of the box) after 1~Myr of decay from the initial state with uniform
density and a \citet{Kolmogorov41} velocity spectrum. We see that the velocity
variations rapidly produce density fluctuations in the background and that the
simulation corresponding to case D ($n_{\rm gas,0} = 1.0$~g~cm$^{-3}$)
shows about twice as much relative variance in $\rho$ compared to case C
($n_{\rm gas,0} = 0.1$~g~cm$^{-3}$). For the processing of dust grains the
kinetic-energy variance of the background ISM is more important and we note
that it is ${\sim}25\,$\% larger
for  $n_{\rm gas,0} = 0.1$~g~cm$^{-3}$ compared to $n_{\rm gas,0} = 1.0$~g~cm$^{-3}$.

 \begin{figure*}
 \includegraphics[trim=3.9cm 3.4cm 7.3cm 2.0cm, clip=true,page=1,height = 3.15cm]{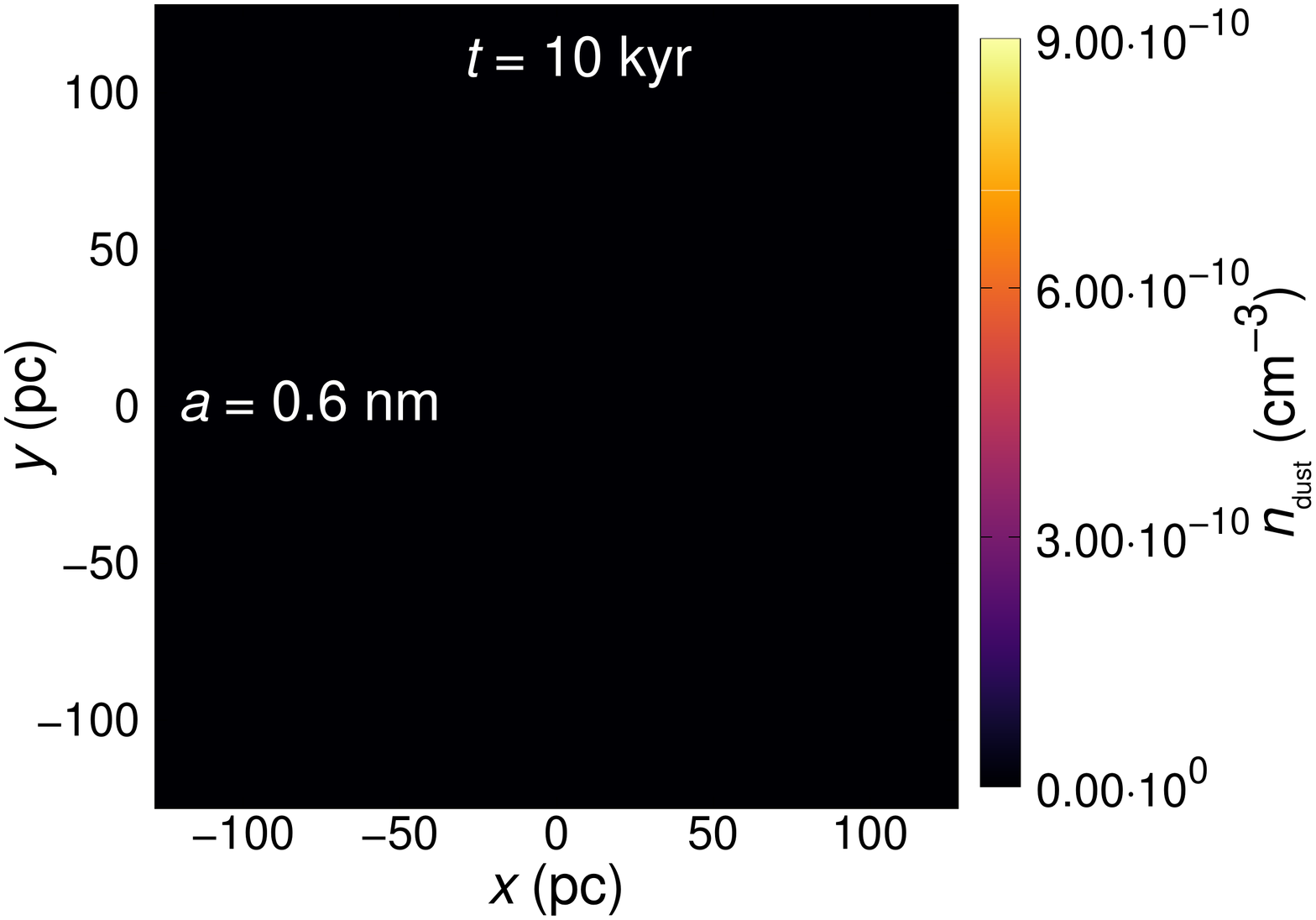}\hspace*{-0.05cm}
 \includegraphics[trim=6.7cm 3.4cm 7.3cm 2.0cm, clip=true,page=1,height = 3.15cm]{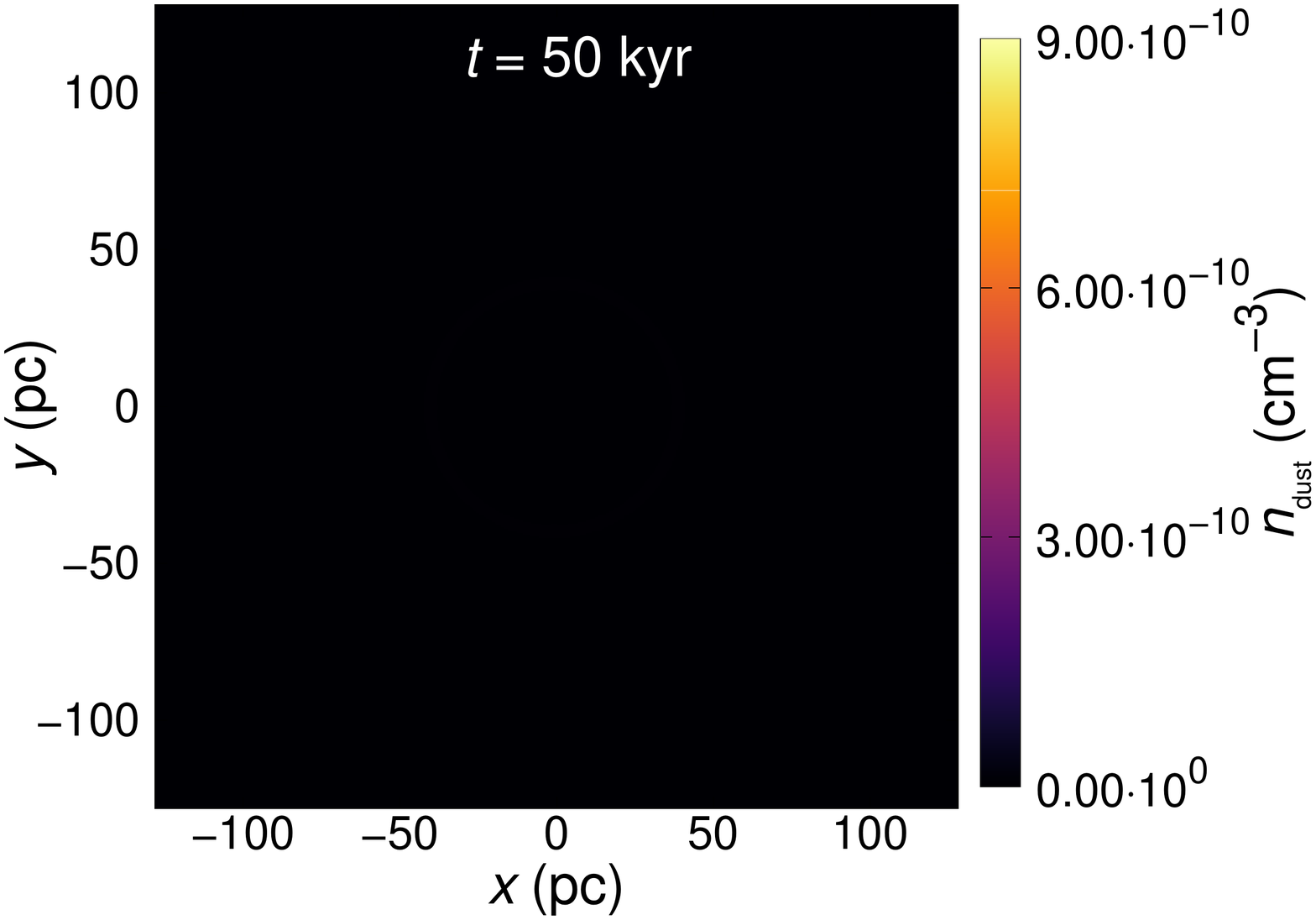}\hspace*{-0.05cm}
 \includegraphics[trim=6.7cm 3.4cm 7.3cm 2.0cm, clip=true,page=1,height = 3.15cm]{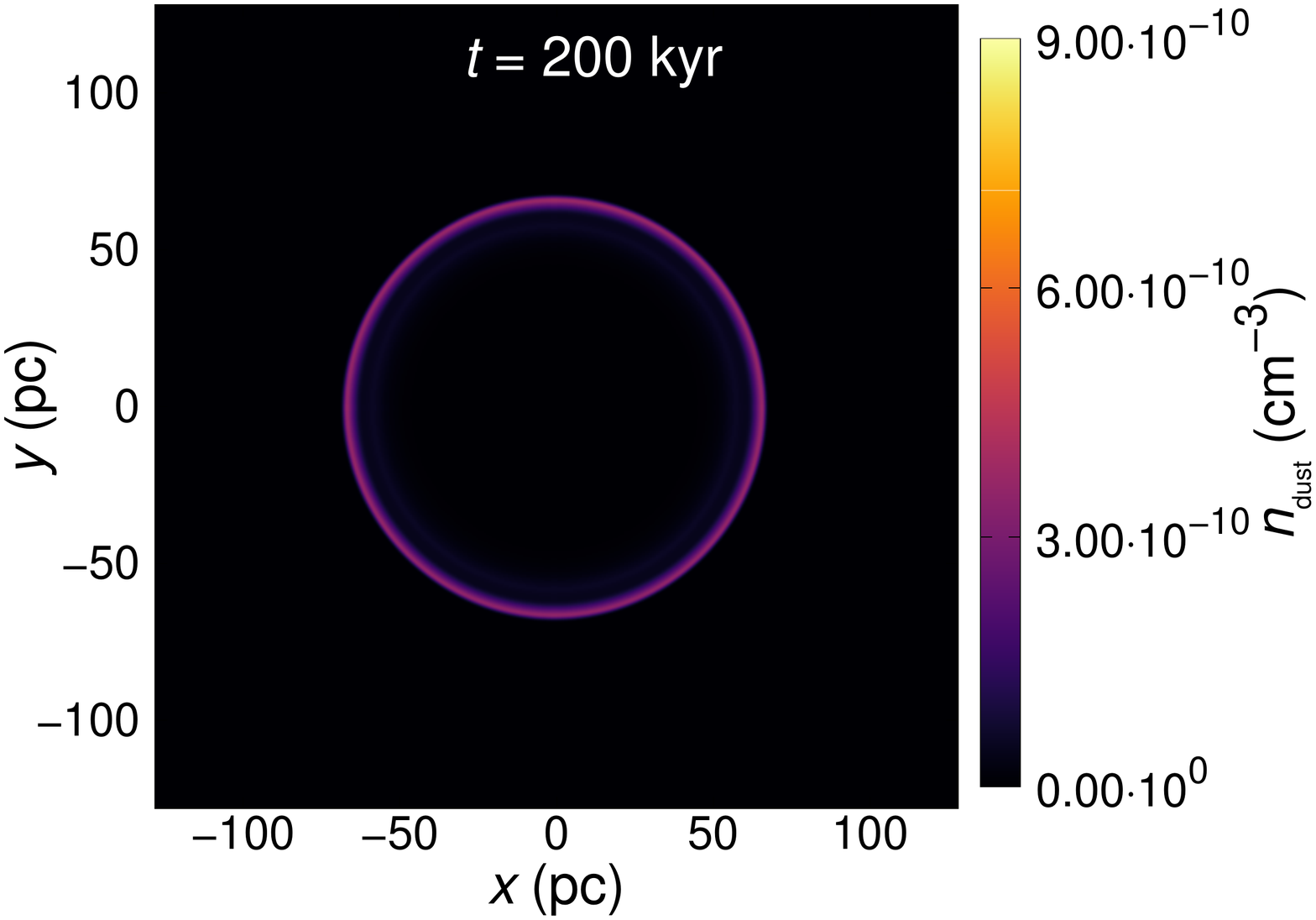}\hspace*{-0.05cm}
 \includegraphics[trim=6.7cm 3.4cm 7.3cm 2.0cm, clip=true,page=1,height = 3.15cm]{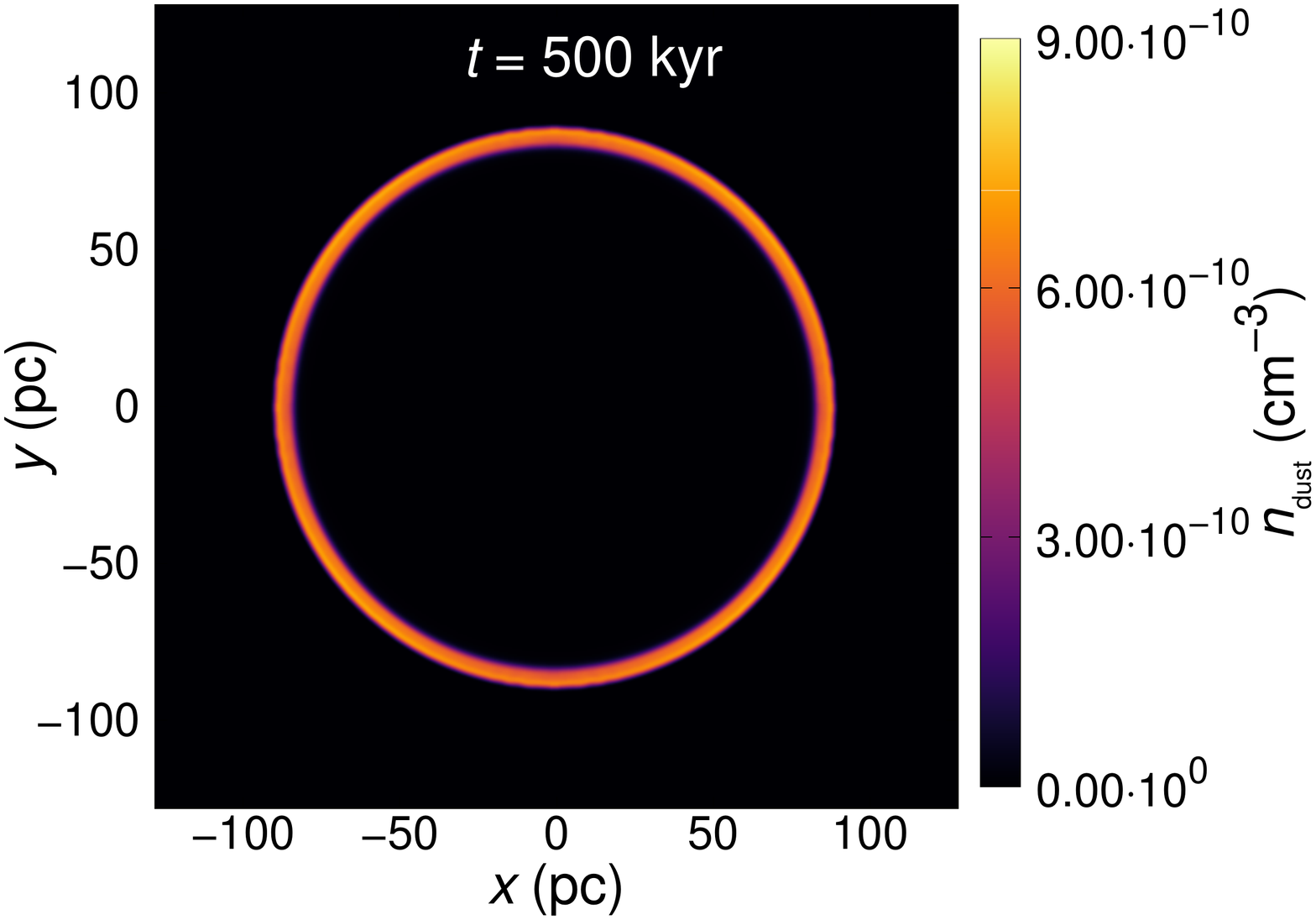}\hspace*{-0.05cm} 
 \includegraphics[trim=6.7cm 3.4cm 0.3cm 2.0cm, clip=true,page=1,height = 3.15cm]{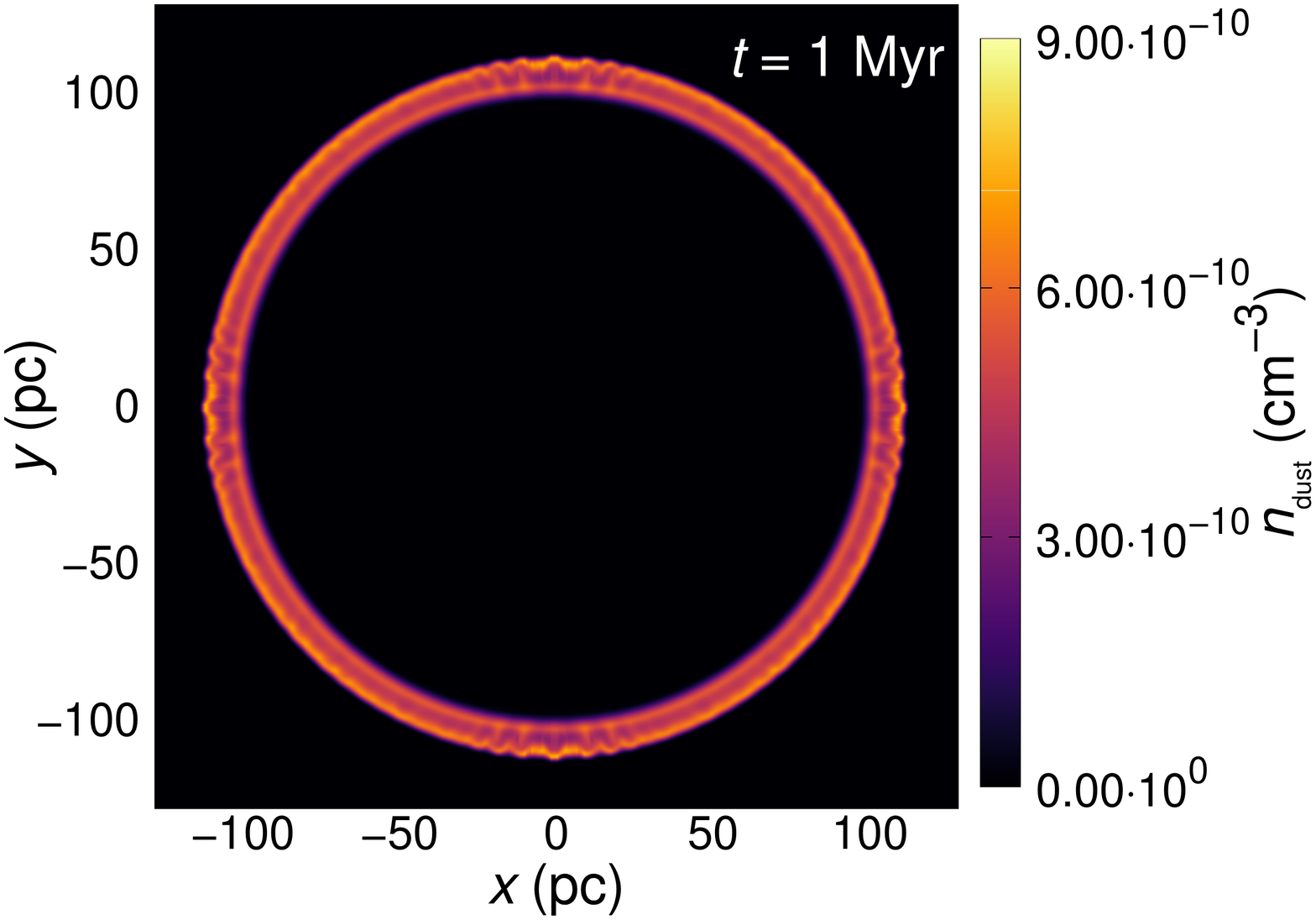}\\
 \includegraphics[trim=3.9cm 3.4cm 7.3cm 2.0cm, clip=true,page=2,height = 3.15cm]{Pics/Pics_A/Density_1_00041.pdf}\hspace*{-0.05cm} 
 \includegraphics[trim=6.7cm 3.4cm 7.3cm 2.0cm, clip=true,page=2,height = 3.15cm]{Pics/Pics_A/Density_1_00201.pdf}\hspace*{-0.05cm} 
 \includegraphics[trim=6.7cm 3.4cm 7.3cm 2.0cm, clip=true,page=2,height = 3.15cm]{Pics/Pics_A/Density_1_00801.pdf}\hspace*{-0.05cm} 
 \includegraphics[trim=6.7cm 3.4cm 7.3cm 2.0cm, clip=true,page=2,height = 3.15cm]{Pics/Pics_A/Density_1_02001.pdf}\hspace*{-0.05cm} 
 \includegraphics[trim=6.7cm 3.4cm 0.3cm 2.0cm, clip=true,page=2,height = 3.15cm]{Pics/Pics_A/Density_1_04000.pdf}\\
 \includegraphics[trim=3.9cm 3.4cm 7.3cm 2.0cm, clip=true,page=3,height = 3.15cm]{Pics/Pics_A/Density_1_00041.pdf}\hspace*{-0.05cm} 
 \includegraphics[trim=6.7cm 3.4cm 7.3cm 2.0cm, clip=true,page=3,height = 3.15cm]{Pics/Pics_A/Density_1_00201.pdf}\hspace*{-0.05cm} 
 \includegraphics[trim=6.7cm 3.4cm 7.3cm 2.0cm, clip=true,page=3,height = 3.15cm]{Pics/Pics_A/Density_1_00801.pdf}\hspace*{-0.05cm}
 \includegraphics[trim=6.7cm 3.4cm 7.3cm 2.0cm, clip=true,page=3,height = 3.15cm]{Pics/Pics_A/Density_1_02001.pdf}\hspace*{-0.05cm} 
 \includegraphics[trim=6.7cm 3.4cm 0.3cm 2.0cm, clip=true,page=3,height = 3.15cm]{Pics/Pics_A/Density_1_04000.pdf}\\
 \includegraphics[trim=3.9cm 1.3cm 7.3cm 2.0cm, clip=true,page=4,height = 3.57cm]{Pics/Pics_A/Density_1_00041.pdf}\hspace*{-0.05cm} 
 \includegraphics[trim=6.7cm 1.3cm 7.3cm 2.0cm, clip=true,page=4,height = 3.57cm]{Pics/Pics_A/Density_1_00201.pdf}\hspace*{-0.05cm} 
 \includegraphics[trim=6.7cm 1.3cm 7.3cm 2.0cm, clip=true,page=4,height = 3.57cm]{Pics/Pics_A/Density_1_00801.pdf}\hspace*{-0.05cm}  
 \includegraphics[trim=6.7cm 1.3cm 7.3cm 2.0cm, clip=true,page=4,height = 3.57cm]{Pics/Pics_A/Density_1_02001.pdf}\hspace*{-0.05cm} 
 \includegraphics[trim=6.7cm 1.3cm 0.3cm 2.0cm, clip=true,page=4,height = 3.57cm]{Pics/Pics_A/Density_1_04000.pdf}\\
    \hspace*{-0.3cm}\includegraphics[trim=1.6cm 1.4cm 8.6cm 2.0cm, clip=true,page=1,height = 3.6cm]{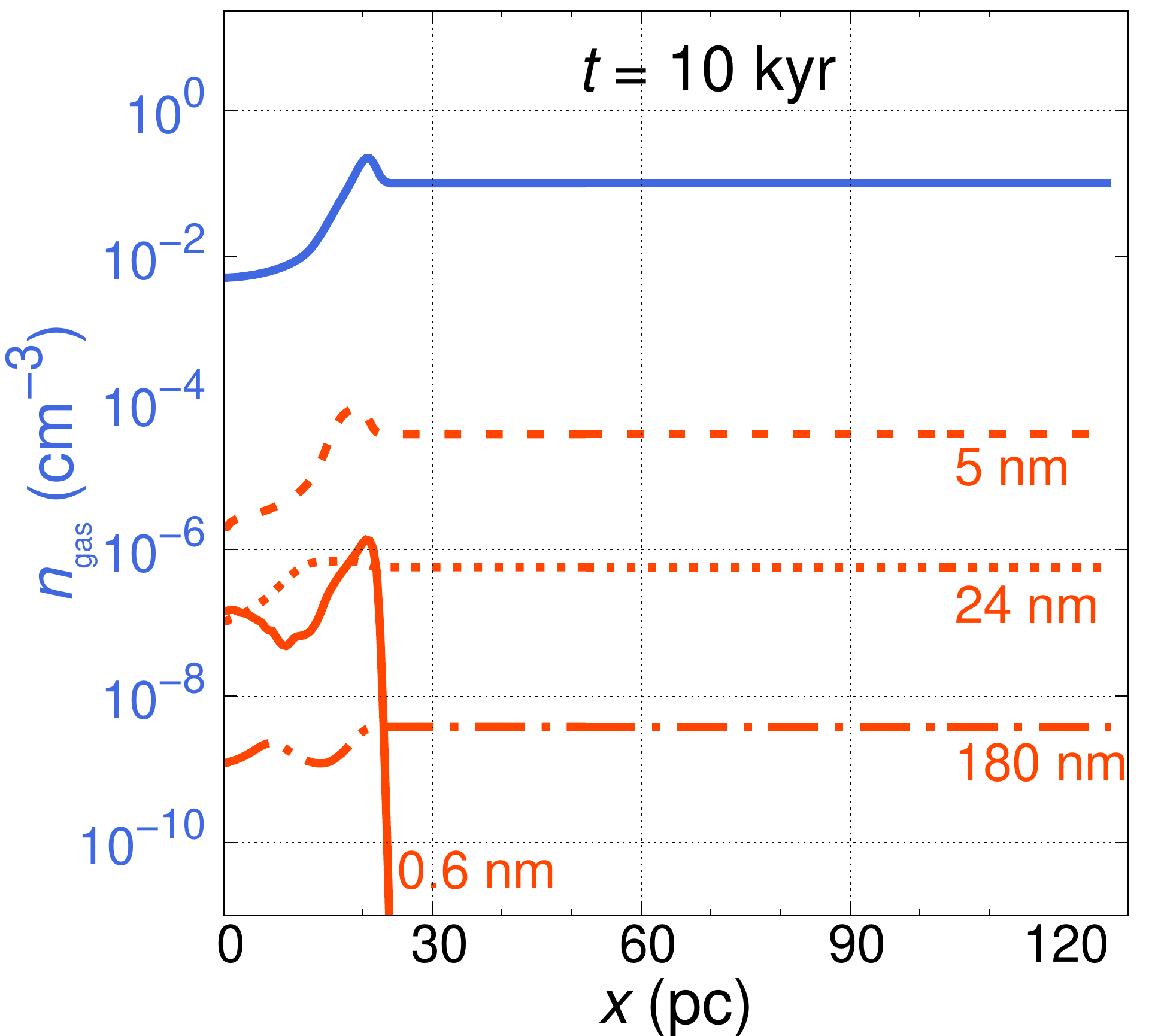}\hspace*{-0.1cm} 
  \includegraphics[trim=5.2cm 1.4cm 8.6cm 2.0cm, clip=true,page=2,height = 3.6cm]{Pics/Pics_A/Profile_dust.pdf}\hspace*{-0.1cm} 
  \includegraphics[trim=5.2cm 1.4cm 8.6cm 2.0cm, clip=true,page=3,height = 3.6cm]{Pics/Pics_A/Profile_dust.pdf}\hspace*{-0.1cm} 
  \includegraphics[trim=5.2cm 1.4cm 8.6cm 2.0cm, clip=true,page=4,height = 3.6cm]{Pics/Pics_A/Profile_dust.pdf}\hspace*{-0.1cm} 
  \includegraphics[trim=5.2cm 1.4cm 3.2cm 2.0cm, clip=true,page=5,height = 3.6cm]{Pics/Pics_A/Profile_dust.pdf}
  \caption{Temporal evolution of the spatial dust density for simulation A ($n_{\rm gas,0}=0.1\,\textrm{cm}^{-3}$, no turbulence, transport + sputtering + grain-grain collisions). The first, second, third, and fourth row shows the distribution of 0.6, 5, 24, 180$\,$nm grains, respectively. The colour scale is fixed for each row. The fifths row shows the radial profiles of the gas density (blue) and the number density of four dust grain sizes (red).}
   \label{fig_A} 
  \end{figure*}   
  
\begin{figure*}
 \includegraphics[trim=3.9cm 3.4cm 7.3cm 2.0cm, clip=true,page=1,height = 3.15cm]{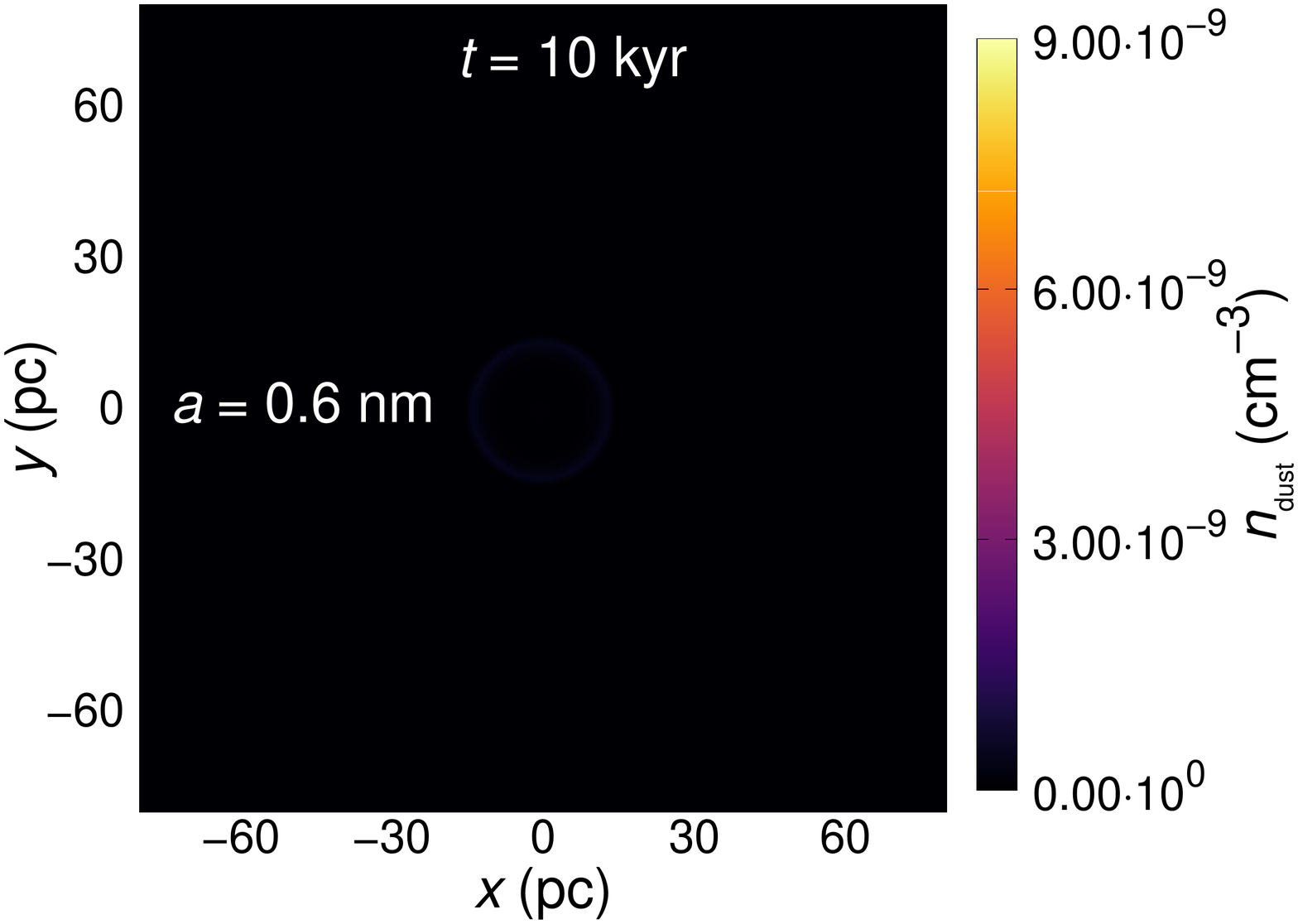}\hspace*{-0.05cm}
 \includegraphics[trim=6.7cm 3.4cm 7.3cm 2.0cm, clip=true,page=1,height = 3.15cm]{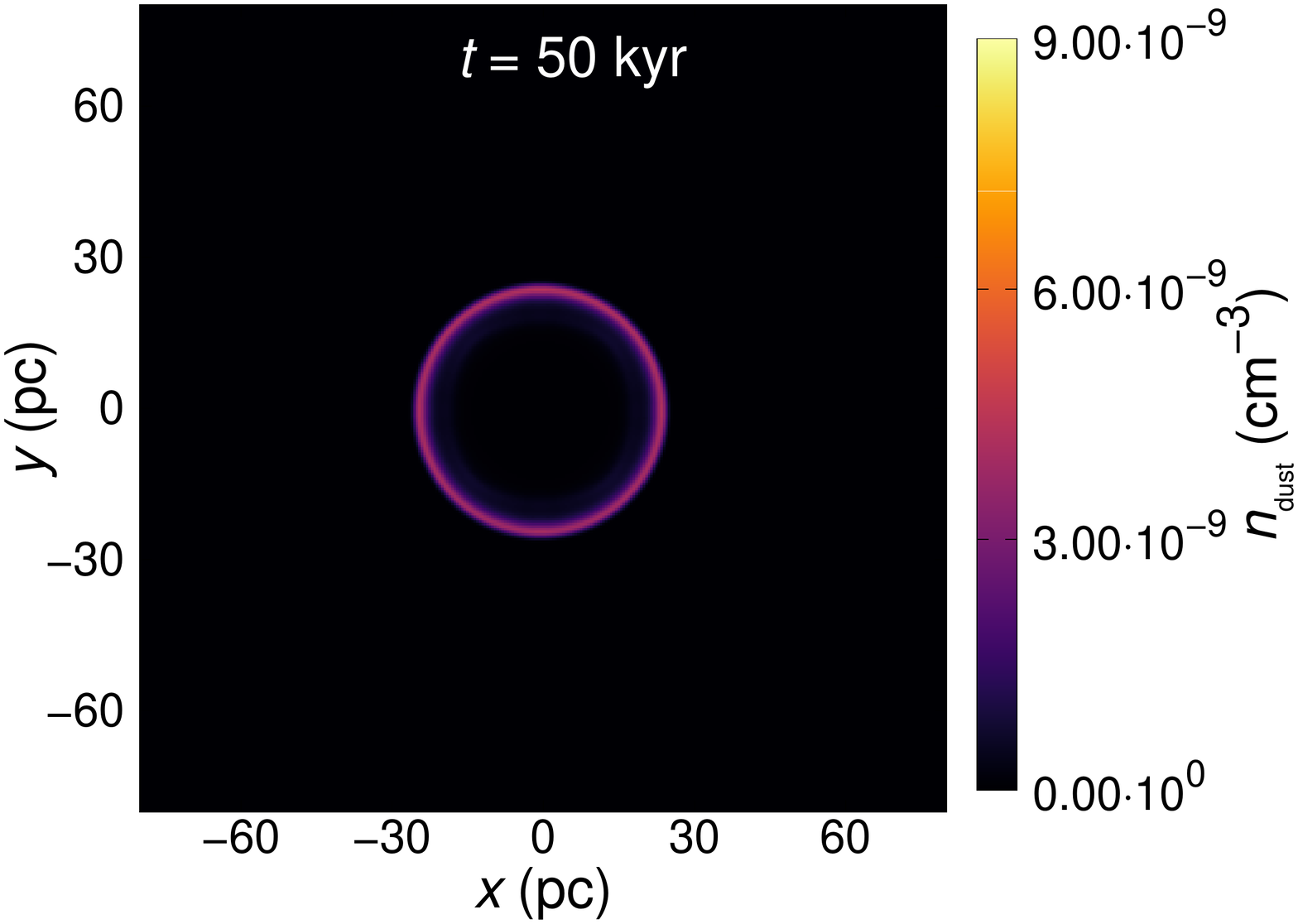}\hspace*{-0.05cm}
 \includegraphics[trim=6.7cm 3.4cm 7.3cm 2.0cm, clip=true,page=1,height = 3.15cm]{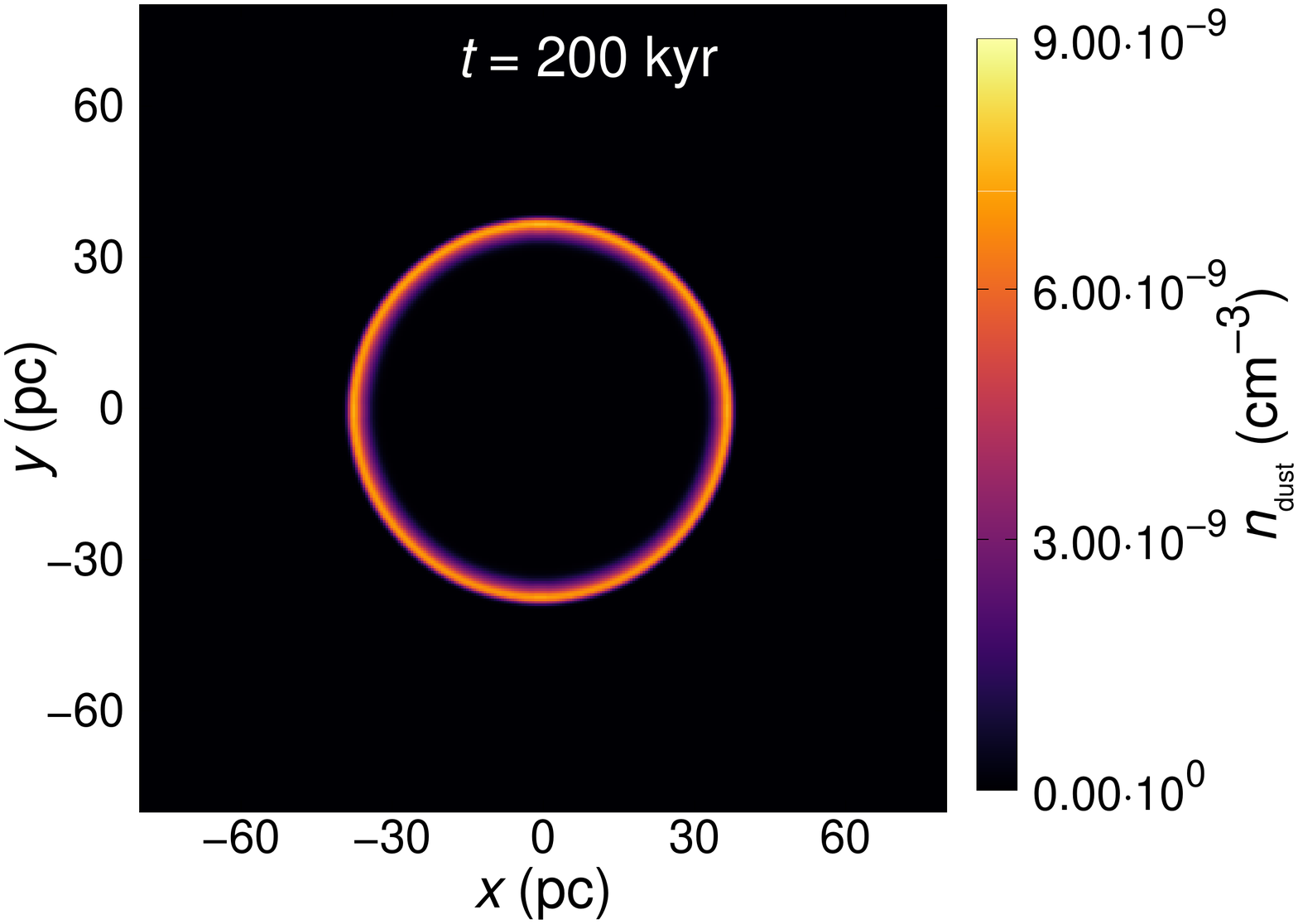}\hspace*{-0.05cm}
 \includegraphics[trim=6.7cm 3.4cm 7.3cm 2.0cm, clip=true,page=1,height = 3.15cm]{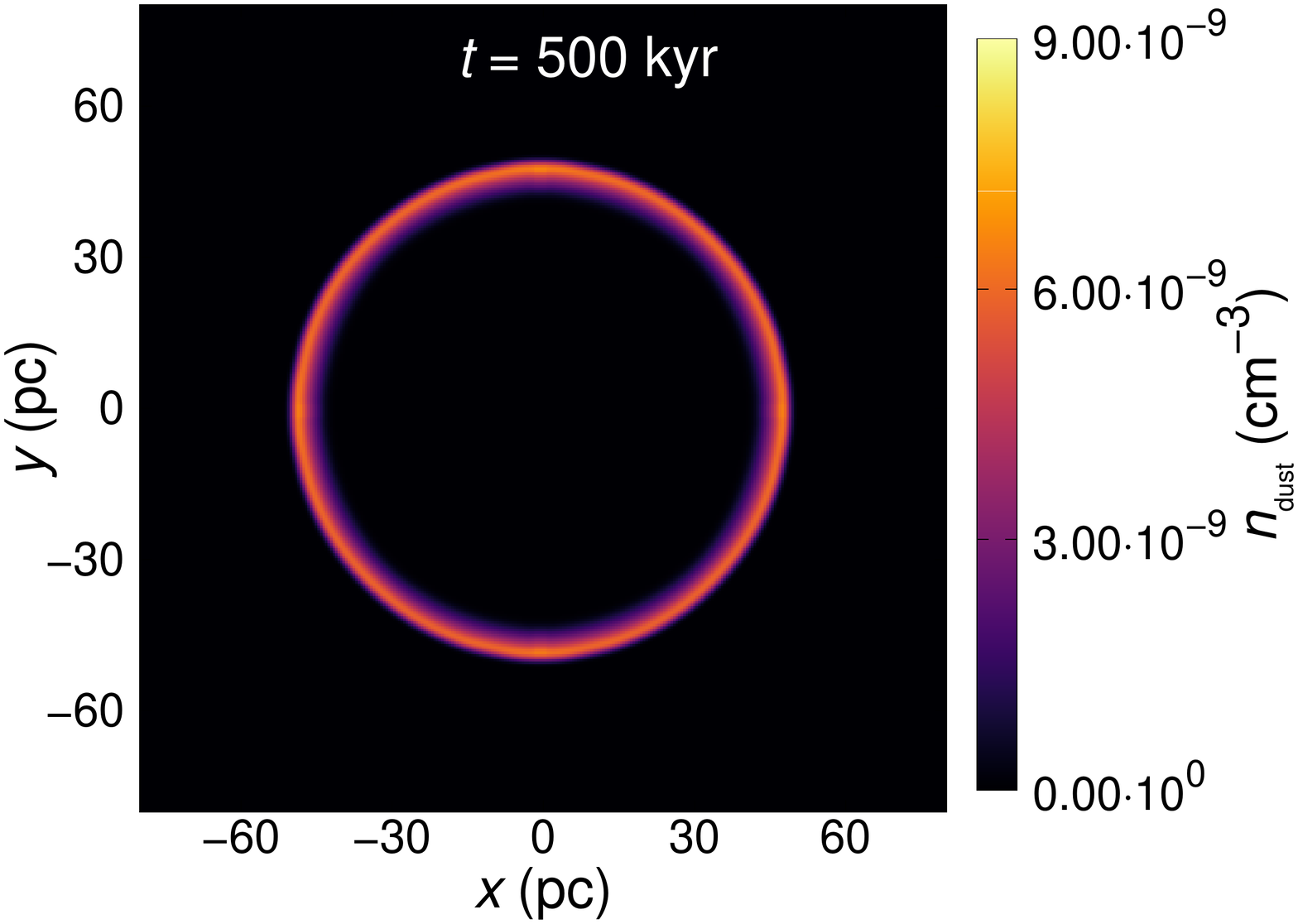}\hspace*{-0.05cm} 
 \includegraphics[trim=6.7cm 3.4cm 0.3cm 2.0cm, clip=true,page=1,height = 3.15cm]{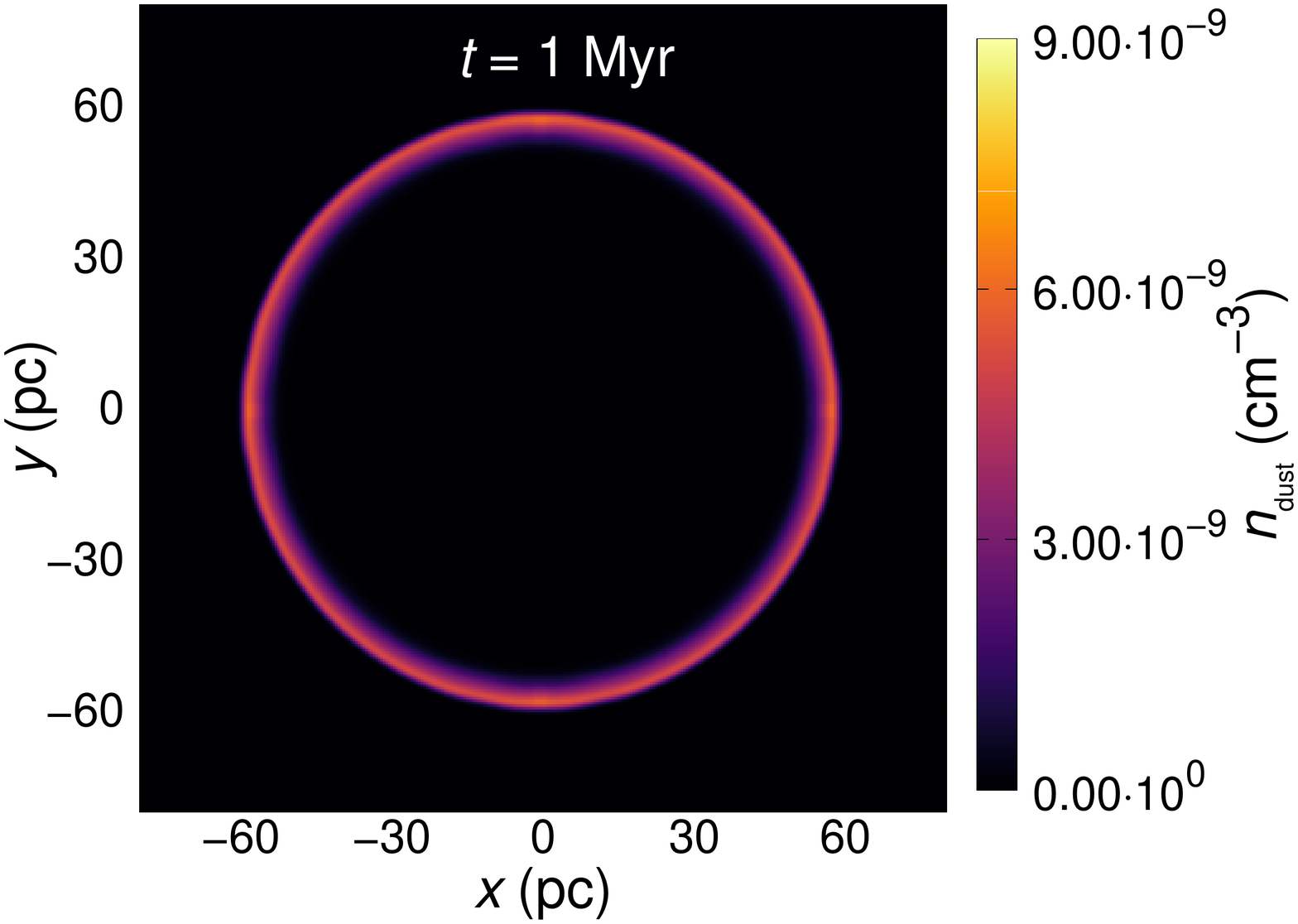}\\
 \includegraphics[trim=3.9cm 3.4cm 7.3cm 2.0cm, clip=true,page=2,height = 3.15cm]{Pics/Pics_B/Density_1_00041.pdf}\hspace*{-0.05cm} 
 \includegraphics[trim=6.7cm 3.4cm 7.3cm 2.0cm, clip=true,page=2,height = 3.15cm]{Pics/Pics_B/Density_1_00201.pdf}\hspace*{-0.05cm} 
 \includegraphics[trim=6.7cm 3.4cm 7.3cm 2.0cm, clip=true,page=2,height = 3.15cm]{Pics/Pics_B/Density_1_00801.pdf}\hspace*{-0.05cm} 
 \includegraphics[trim=6.7cm 3.4cm 7.3cm 2.0cm, clip=true,page=2,height = 3.15cm]{Pics/Pics_B/Density_1_02001.pdf}\hspace*{-0.05cm} 
 \includegraphics[trim=6.7cm 3.4cm 0.3cm 2.0cm, clip=true,page=2,height = 3.15cm]{Pics/Pics_B/Density_1_04000.pdf}\\
 \includegraphics[trim=3.9cm 3.4cm 7.3cm 2.0cm, clip=true,page=3,height = 3.15cm]{Pics/Pics_B/Density_1_00041.pdf}\hspace*{-0.05cm} 
 \includegraphics[trim=6.7cm 3.4cm 7.3cm 2.0cm, clip=true,page=3,height = 3.15cm]{Pics/Pics_B/Density_1_00201.pdf}\hspace*{-0.05cm} 
 \includegraphics[trim=6.7cm 3.4cm 7.3cm 2.0cm, clip=true,page=3,height = 3.15cm]{Pics/Pics_B/Density_1_00801.pdf}\hspace*{-0.05cm}
 \includegraphics[trim=6.7cm 3.4cm 7.3cm 2.0cm, clip=true,page=3,height = 3.15cm]{Pics/Pics_B/Density_1_02001.pdf}\hspace*{-0.05cm} 
 \includegraphics[trim=6.7cm 3.4cm 0.3cm 2.0cm, clip=true,page=3,height = 3.15cm]{Pics/Pics_B/Density_1_04000.pdf}\\
 \includegraphics[trim=3.9cm 1.3cm 7.3cm 2.0cm, clip=true,page=4,height = 3.57cm]{Pics/Pics_B/Density_1_00041.pdf}\hspace*{-0.05cm} 
 \includegraphics[trim=6.7cm 1.3cm 7.3cm 2.0cm, clip=true,page=4,height = 3.57cm]{Pics/Pics_B/Density_1_00201.pdf}\hspace*{-0.05cm} 
 \includegraphics[trim=6.7cm 1.3cm 7.3cm 2.0cm, clip=true,page=4,height = 3.57cm]{Pics/Pics_B/Density_1_00801.pdf}\hspace*{-0.05cm}  
 \includegraphics[trim=6.7cm 1.3cm 7.3cm 2.0cm, clip=true,page=4,height = 3.57cm]{Pics/Pics_B/Density_1_02001.pdf}\hspace*{-0.05cm} 
 \includegraphics[trim=6.7cm 1.3cm 0.3cm 2.0cm, clip=true,page=4,height = 3.57cm]{Pics/Pics_B/Density_1_04000.pdf}\\
    \hspace*{-0.3cm}\includegraphics[trim=1.6cm 1.4cm 8.6cm 2.0cm, clip=true,page=1,height = 3.6cm]{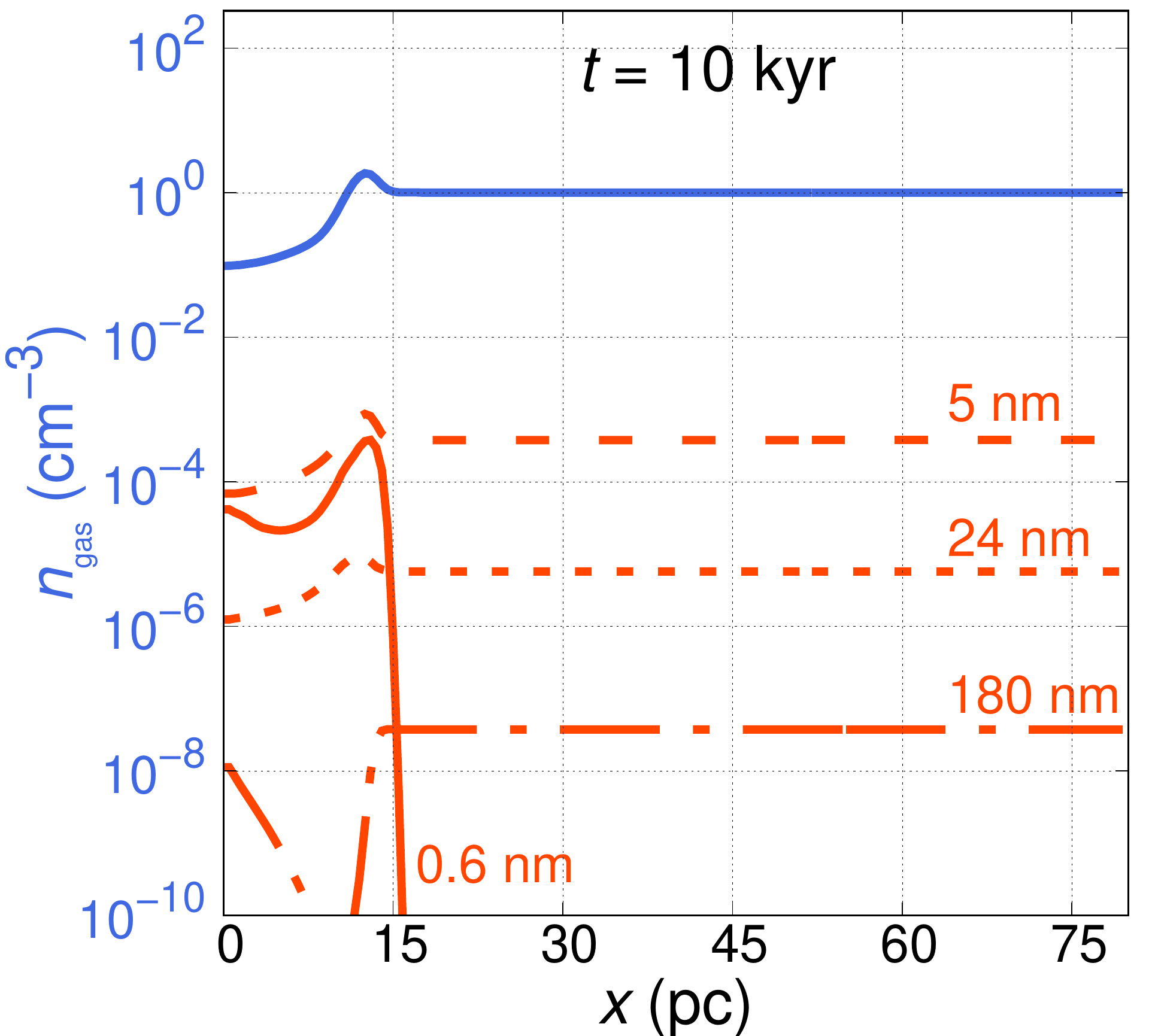}\hspace*{-0.1cm} 
  \includegraphics[trim=5.2cm 1.4cm 8.6cm 2.0cm, clip=true,page=2,height = 3.6cm]{Pics/Pics_B/Profile_dust.pdf}\hspace*{-0.1cm} 
  \includegraphics[trim=5.2cm 1.4cm 8.6cm 2.0cm, clip=true,page=3,height = 3.6cm]{Pics/Pics_B/Profile_dust.pdf}\hspace*{-0.1cm} 
  \includegraphics[trim=5.2cm 1.4cm 8.6cm 2.0cm, clip=true,page=4,height = 3.6cm]{Pics/Pics_B/Profile_dust.pdf}\hspace*{-0.1cm} 
  \includegraphics[trim=5.2cm 1.4cm 3.2cm 2.0cm, clip=true,page=5,height = 3.6cm]{Pics/Pics_B/Profile_dust.pdf}
  \caption{Same as Fig.~\ref{fig_A} but for simulation B.}
   \label{fig_B} 
  \end{figure*} 
  
\begin{figure*}
 \includegraphics[trim=3.9cm 3.4cm 7.3cm 2.0cm, clip=true,page=1,height = 3.15cm]{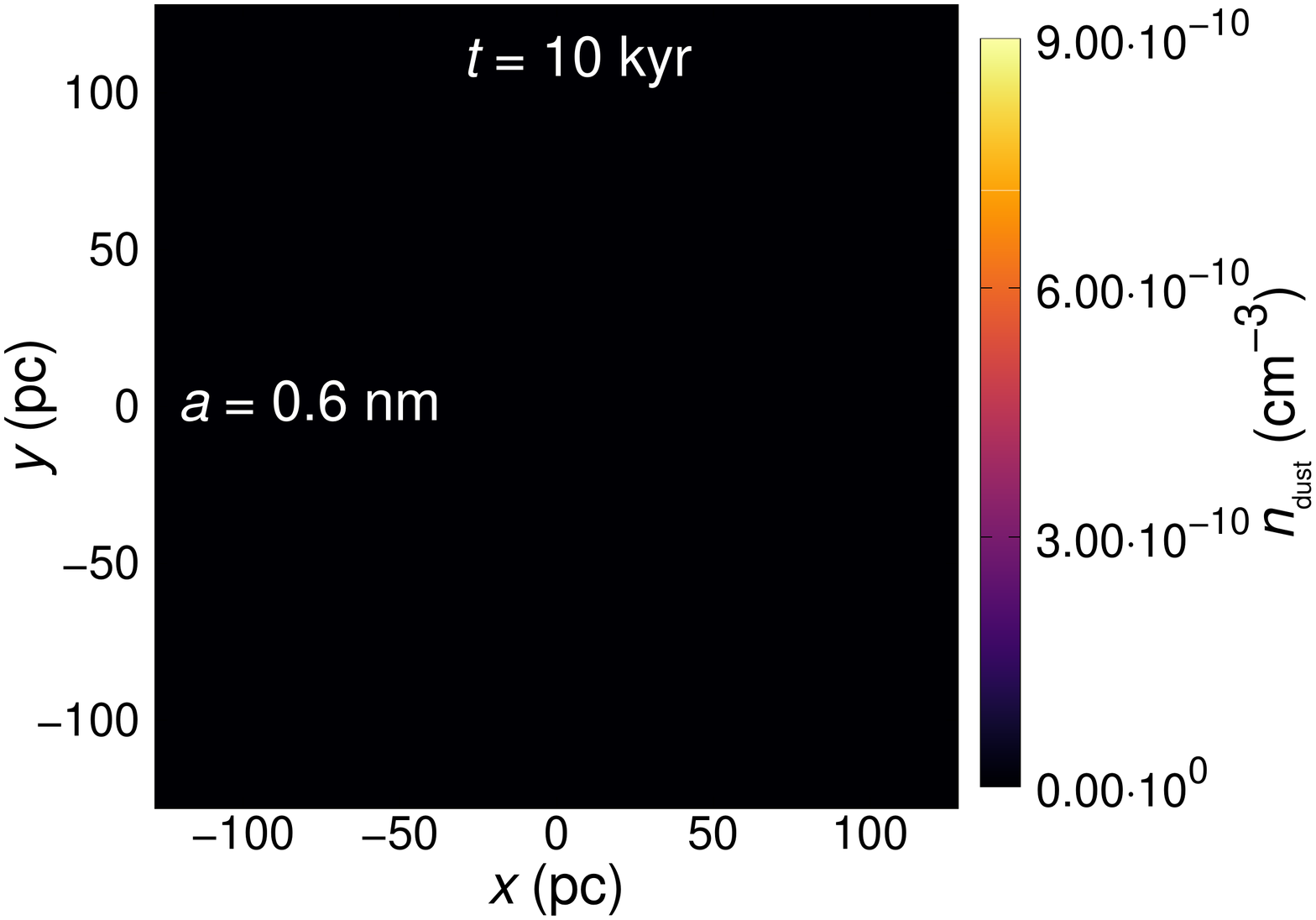}\hspace*{-0.05cm}
 \includegraphics[trim=6.7cm 3.4cm 7.3cm 2.0cm, clip=true,page=1,height = 3.15cm]{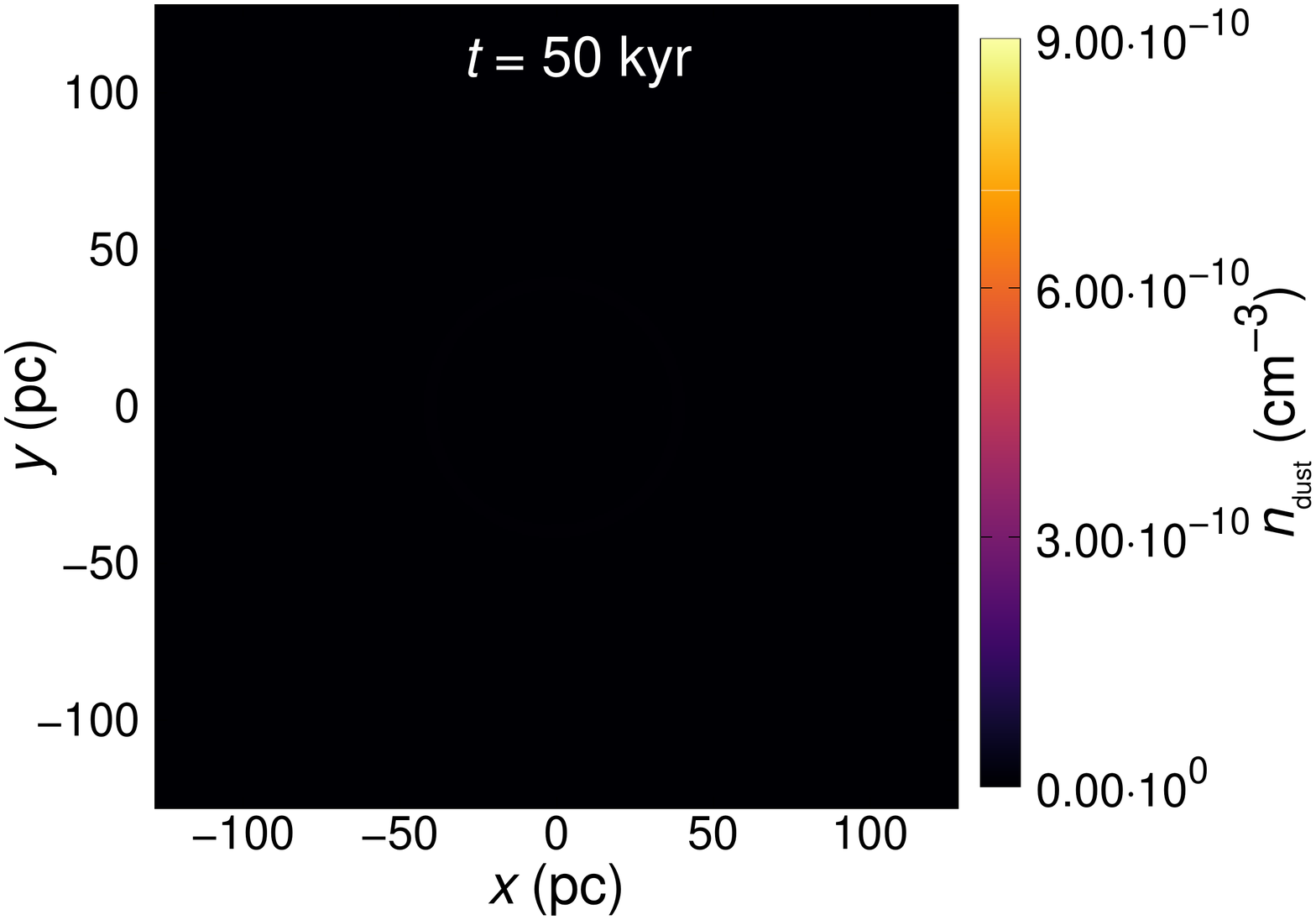}\hspace*{-0.05cm}
 \includegraphics[trim=6.7cm 3.4cm 7.3cm 2.0cm, clip=true,page=1,height = 3.15cm]{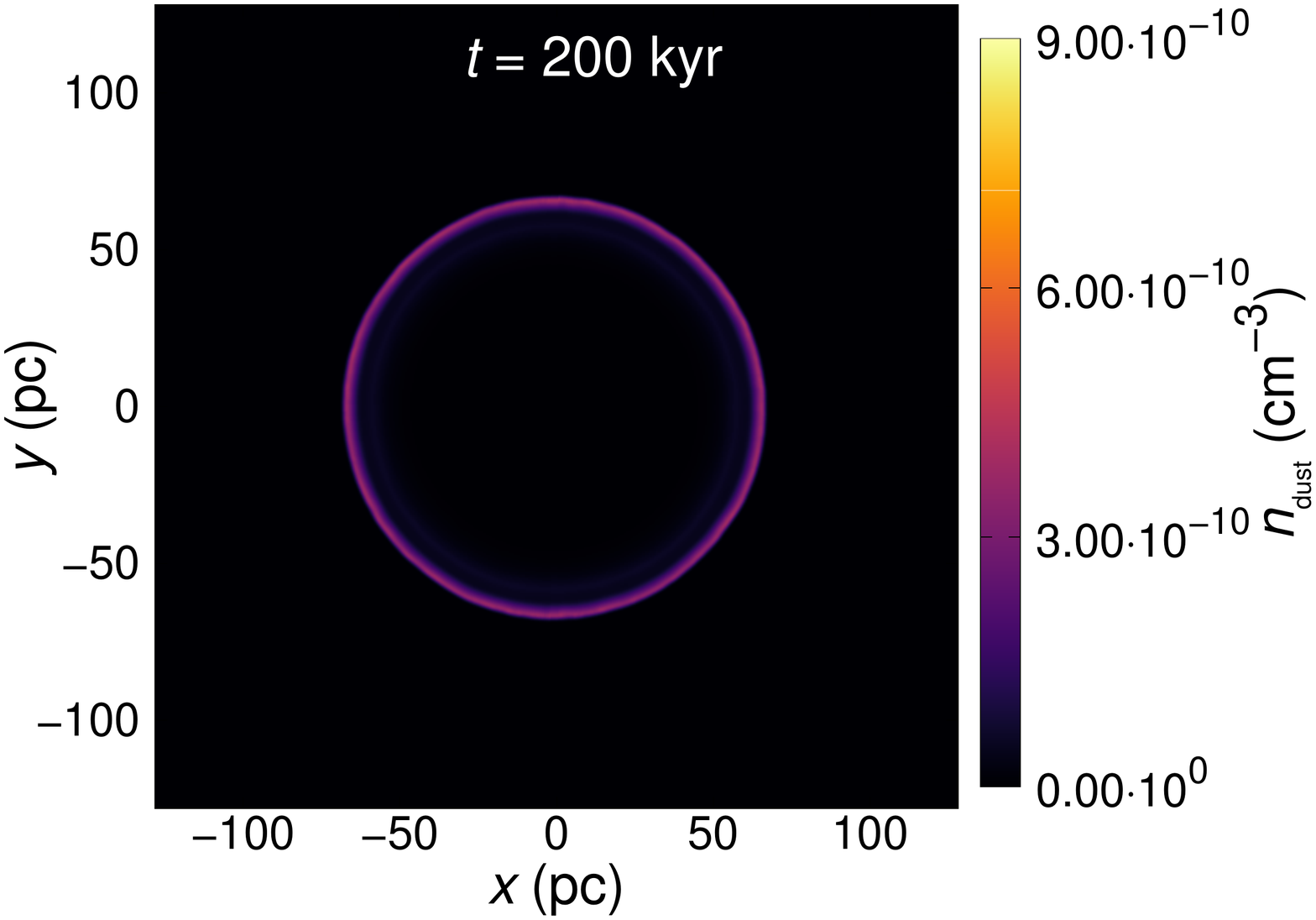}\hspace*{-0.05cm}
 \includegraphics[trim=6.7cm 3.4cm 7.3cm 2.0cm, clip=true,page=1,height = 3.15cm]{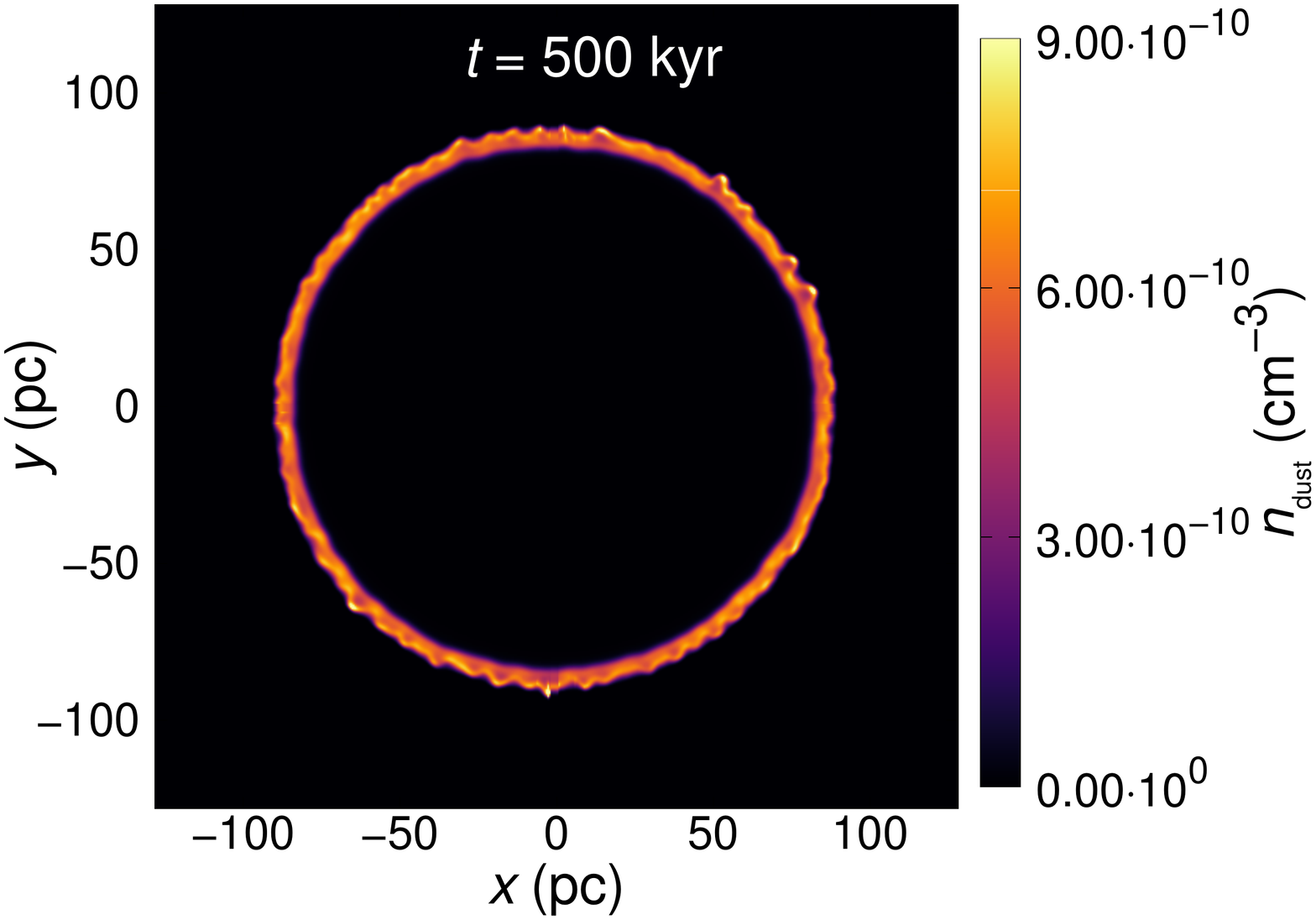}\hspace*{-0.05cm} 
 \includegraphics[trim=6.7cm 3.4cm 0.3cm 2.0cm, clip=true,page=1,height = 3.15cm]{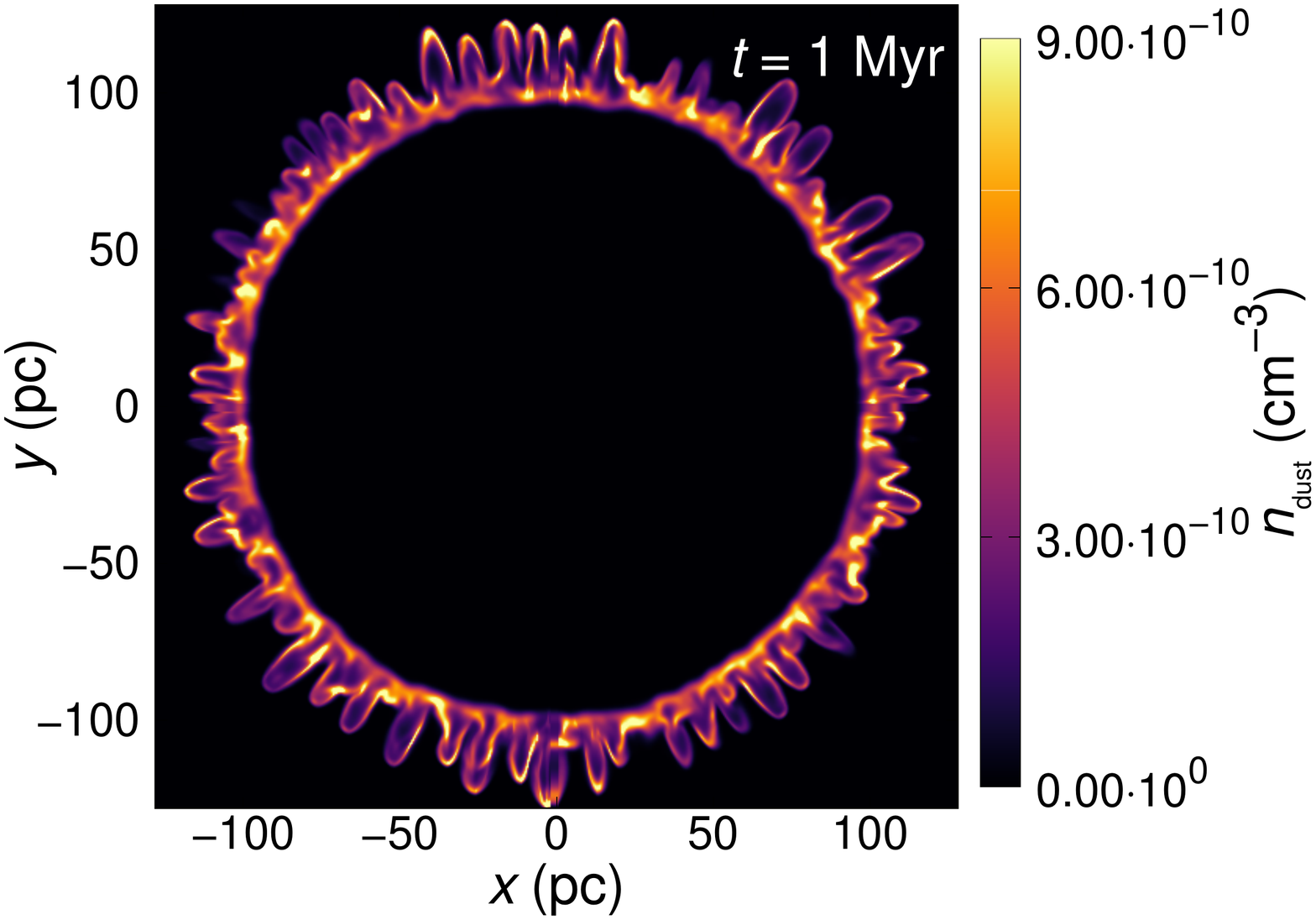}\\
 \includegraphics[trim=3.9cm 3.4cm 7.3cm 2.0cm, clip=true,page=2,height = 3.15cm]{Pics/Pics_C/Density_1_00041.pdf}\hspace*{-0.05cm} 
 \includegraphics[trim=6.7cm 3.4cm 7.3cm 2.0cm, clip=true,page=2,height = 3.15cm]{Pics/Pics_C/Density_1_00201.pdf}\hspace*{-0.05cm} 
 \includegraphics[trim=6.7cm 3.4cm 7.3cm 2.0cm, clip=true,page=2,height = 3.15cm]{Pics/Pics_C/Density_1_00801.pdf}\hspace*{-0.05cm} 
 \includegraphics[trim=6.7cm 3.4cm 7.3cm 2.0cm, clip=true,page=2,height = 3.15cm]{Pics/Pics_C/Density_1_02001.pdf}\hspace*{-0.05cm} 
 \includegraphics[trim=6.7cm 3.4cm 0.3cm 2.0cm, clip=true,page=2,height = 3.15cm]{Pics/Pics_C/Density_1_04000.pdf}\\
 \includegraphics[trim=3.9cm 3.4cm 7.3cm 2.0cm, clip=true,page=3,height = 3.15cm]{Pics/Pics_C/Density_1_00041.pdf}\hspace*{-0.05cm} 
 \includegraphics[trim=6.7cm 3.4cm 7.3cm 2.0cm, clip=true,page=3,height = 3.15cm]{Pics/Pics_C/Density_1_00201.pdf}\hspace*{-0.05cm} 
 \includegraphics[trim=6.7cm 3.4cm 7.3cm 2.0cm, clip=true,page=3,height = 3.15cm]{Pics/Pics_C/Density_1_00801.pdf}\hspace*{-0.05cm}
 \includegraphics[trim=6.7cm 3.4cm 7.3cm 2.0cm, clip=true,page=3,height = 3.15cm]{Pics/Pics_C/Density_1_02001.pdf}\hspace*{-0.05cm} 
 \includegraphics[trim=6.7cm 3.4cm 0.3cm 2.0cm, clip=true,page=3,height = 3.15cm]{Pics/Pics_C/Density_1_04000.pdf}\\
 \includegraphics[trim=3.9cm 1.3cm 7.3cm 2.0cm, clip=true,page=4,height = 3.57cm]{Pics/Pics_C/Density_1_00041.pdf}\hspace*{-0.05cm} 
 \includegraphics[trim=6.7cm 1.3cm 7.3cm 2.0cm, clip=true,page=4,height = 3.57cm]{Pics/Pics_C/Density_1_00201.pdf}\hspace*{-0.05cm} 
 \includegraphics[trim=6.7cm 1.3cm 7.3cm 2.0cm, clip=true,page=4,height = 3.57cm]{Pics/Pics_C/Density_1_00801.pdf}\hspace*{-0.05cm}  
 \includegraphics[trim=6.7cm 1.3cm 7.3cm 2.0cm, clip=true,page=4,height = 3.57cm]{Pics/Pics_C/Density_1_02001.pdf}\hspace*{-0.05cm} 
 \includegraphics[trim=6.7cm 1.3cm 0.3cm 2.0cm, clip=true,page=4,height = 3.57cm]{Pics/Pics_C/Density_1_04000.pdf}\\
  \hspace*{-0.3cm}\includegraphics[trim=1.6cm 1.4cm 8.6cm 2.0cm, clip=true,page=1,height = 3.6cm]{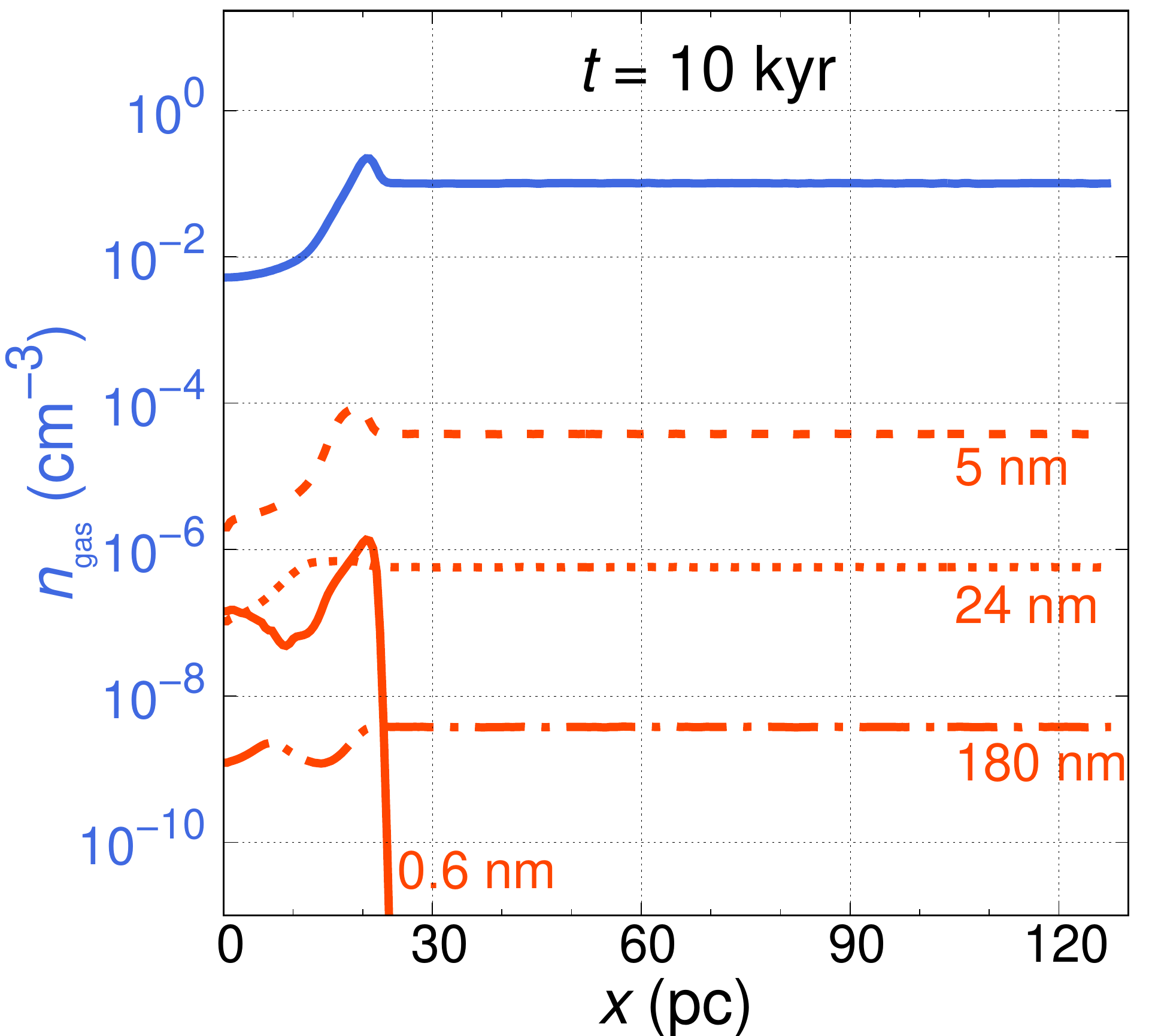}\hspace*{-0.1cm} 
  \includegraphics[trim=5.2cm 1.4cm 8.6cm 2.0cm, clip=true,page=2,height = 3.6cm]{Pics/Pics_C/Profile_dust.pdf}\hspace*{-0.1cm} 
  \includegraphics[trim=5.2cm 1.4cm 8.6cm 2.0cm, clip=true,page=3,height = 3.6cm]{Pics/Pics_C/Profile_dust.pdf}\hspace*{-0.1cm} 
  \includegraphics[trim=5.2cm 1.4cm 8.6cm 2.0cm, clip=true,page=4,height = 3.6cm]{Pics/Pics_C/Profile_dust.pdf}\hspace*{-0.1cm} 
  \includegraphics[trim=5.2cm 1.4cm 3.2cm 2.0cm, clip=true,page=5,height = 3.6cm]{Pics/Pics_C/Profile_dust.pdf}
  \caption{Same as Fig.~\ref{fig_A} but for simulation C.}
   \label{fig_C} 
  \end{figure*}

\begin{figure*}
 \includegraphics[trim=3.9cm 3.4cm 7.3cm 2.0cm, clip=true,page=1,height = 3.15cm]{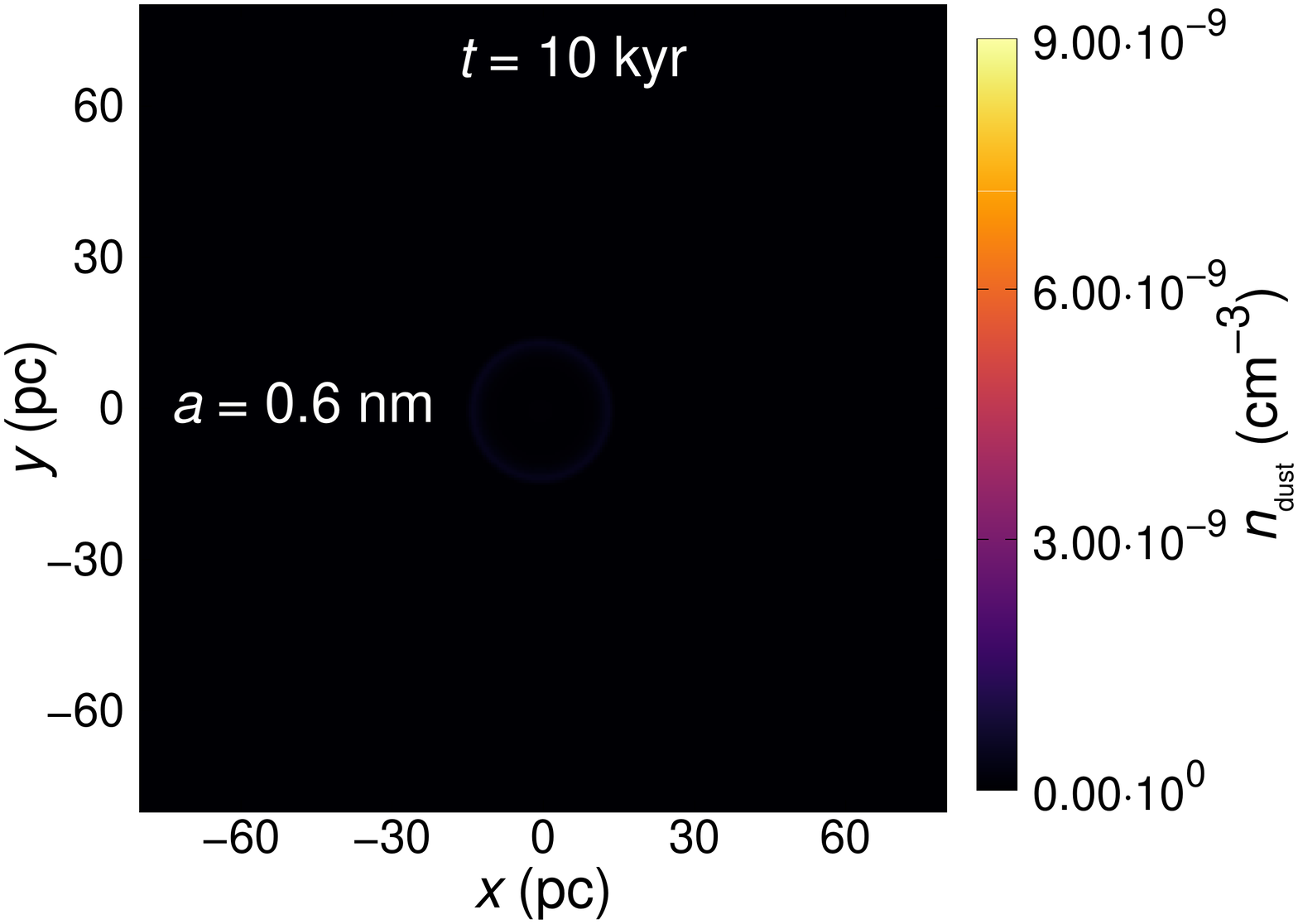}\hspace*{-0.05cm}
 \includegraphics[trim=6.7cm 3.4cm 7.3cm 2.0cm, clip=true,page=1,height = 3.15cm]{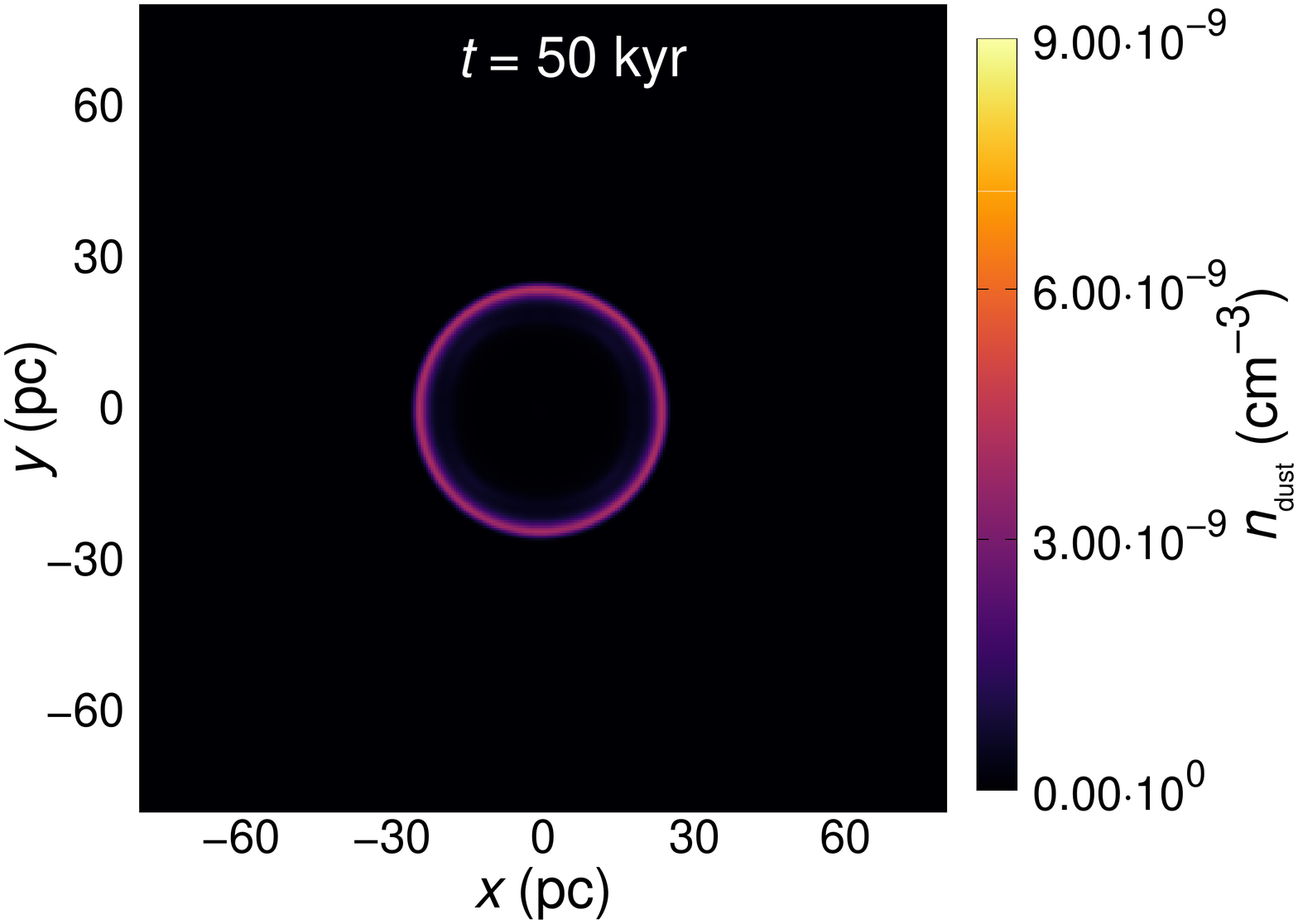}\hspace*{-0.05cm}
 \includegraphics[trim=6.7cm 3.4cm 7.3cm 2.0cm, clip=true,page=1,height = 3.15cm]{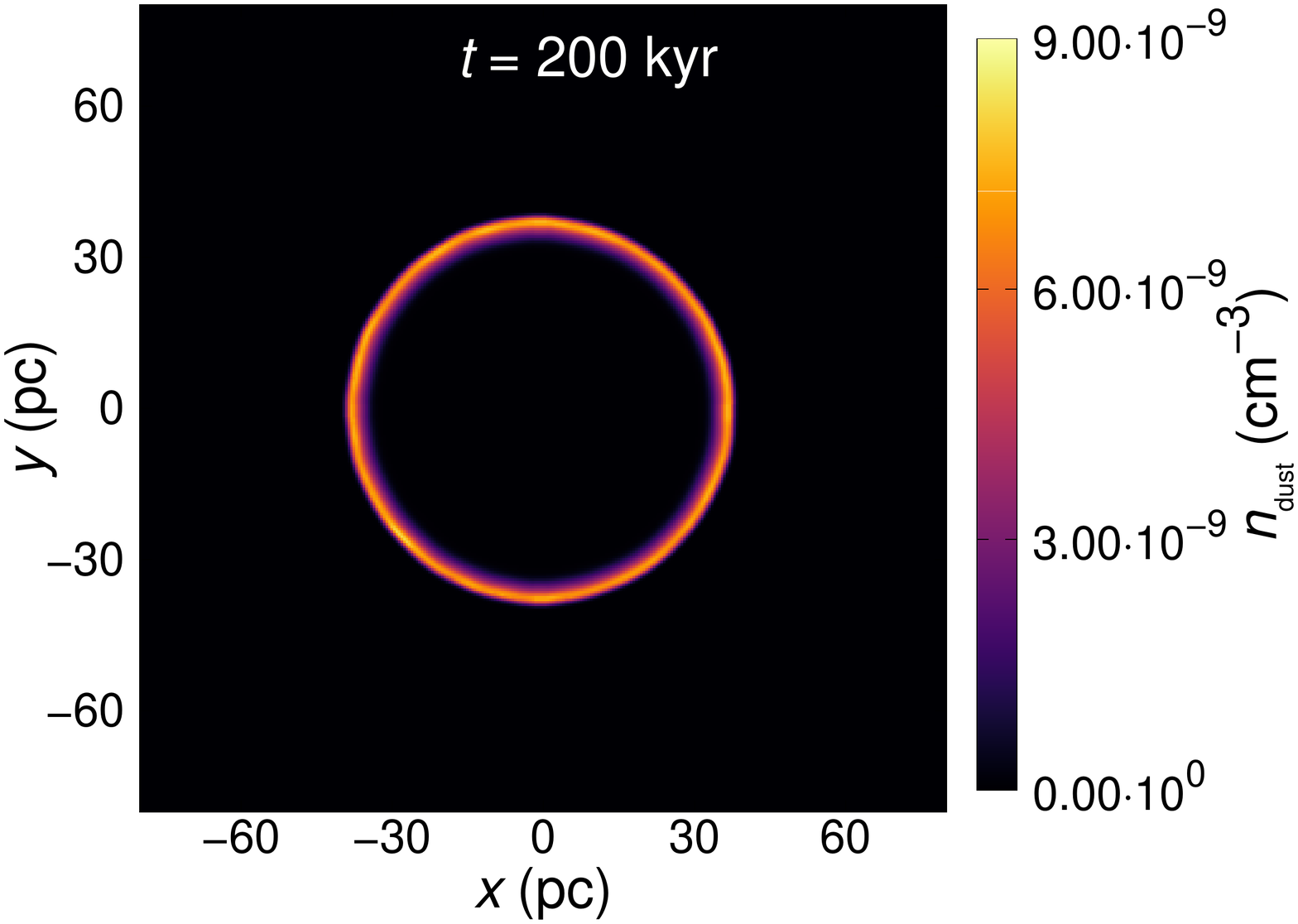}\hspace*{-0.05cm}
 \includegraphics[trim=6.7cm 3.4cm 7.3cm 2.0cm, clip=true,page=1,height = 3.15cm]{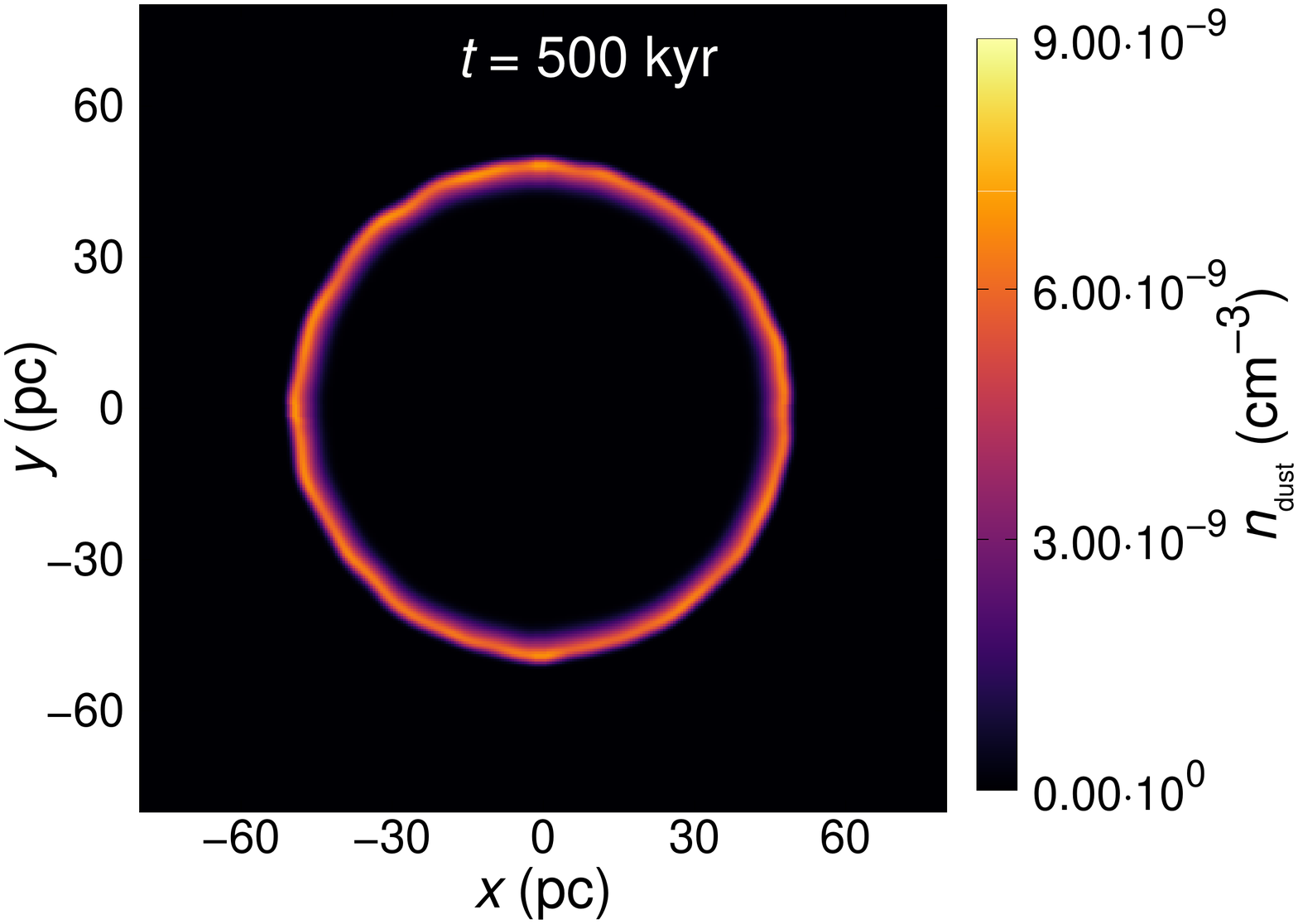}\hspace*{-0.05cm} 
 \includegraphics[trim=6.7cm 3.4cm 0.3cm 2.0cm, clip=true,page=1,height = 3.15cm]{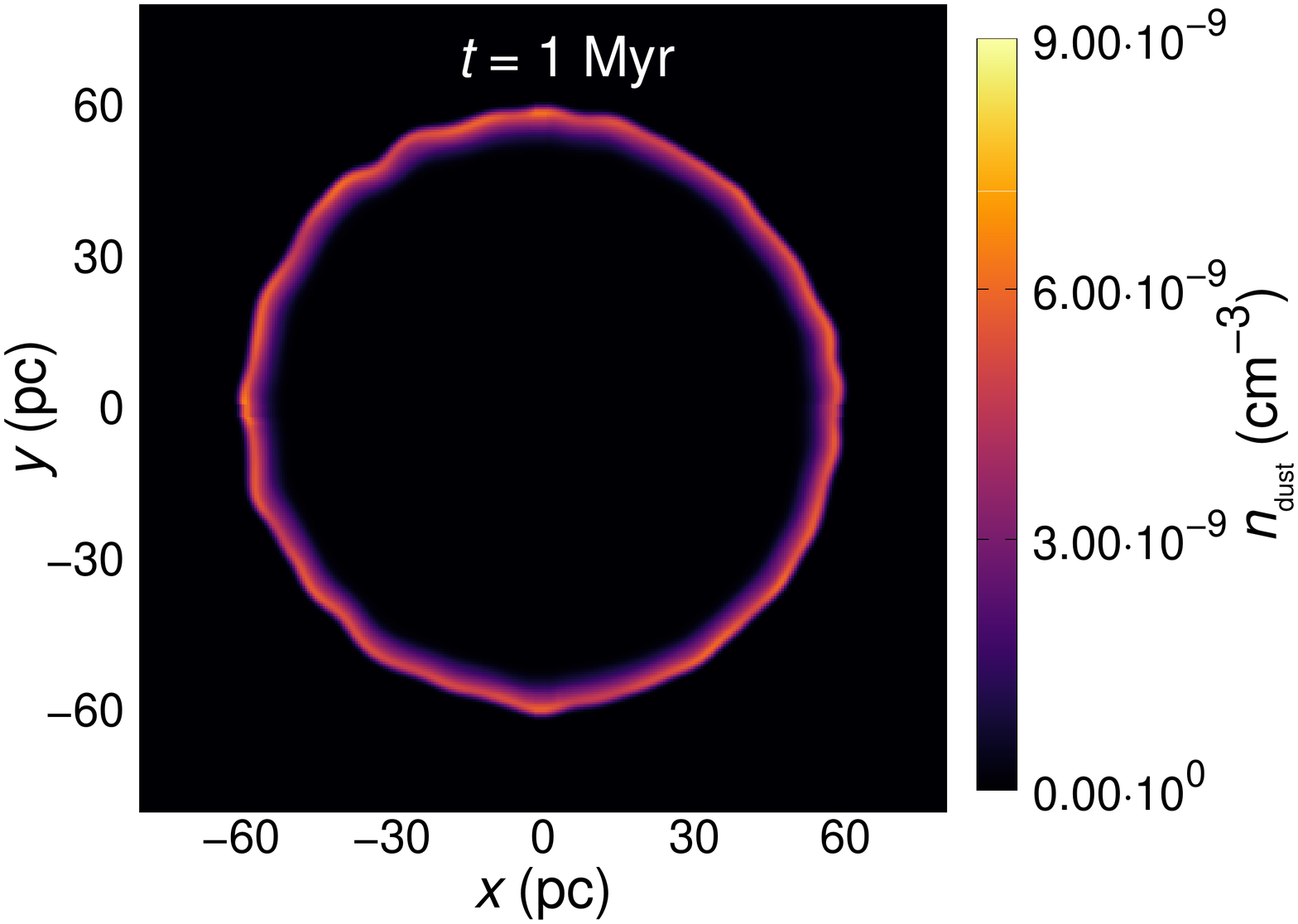}\\
 \includegraphics[trim=3.9cm 3.4cm 7.3cm 2.0cm, clip=true,page=2,height = 3.15cm]{Pics/Pics_D/Density_1_00041.pdf}\hspace*{-0.05cm} 
 \includegraphics[trim=6.7cm 3.4cm 7.3cm 2.0cm, clip=true,page=2,height = 3.15cm]{Pics/Pics_D/Density_1_00201.pdf}\hspace*{-0.05cm} 
 \includegraphics[trim=6.7cm 3.4cm 7.3cm 2.0cm, clip=true,page=2,height = 3.15cm]{Pics/Pics_D/Density_1_00801.pdf}\hspace*{-0.05cm} 
 \includegraphics[trim=6.7cm 3.4cm 7.3cm 2.0cm, clip=true,page=2,height = 3.15cm]{Pics/Pics_D/Density_1_02001.pdf}\hspace*{-0.05cm} 
 \includegraphics[trim=6.7cm 3.4cm 0.3cm 2.0cm, clip=true,page=2,height = 3.15cm]{Pics/Pics_D/Density_1_04000.pdf}\\
 \includegraphics[trim=3.9cm 3.4cm 7.3cm 2.0cm, clip=true,page=3,height = 3.15cm]{Pics/Pics_D/Density_1_00041.pdf}\hspace*{-0.05cm} 
 \includegraphics[trim=6.7cm 3.4cm 7.3cm 2.0cm, clip=true,page=3,height = 3.15cm]{Pics/Pics_D/Density_1_00201.pdf}\hspace*{-0.05cm} 
 \includegraphics[trim=6.7cm 3.4cm 7.3cm 2.0cm, clip=true,page=3,height = 3.15cm]{Pics/Pics_D/Density_1_00801.pdf}\hspace*{-0.05cm}
 \includegraphics[trim=6.7cm 3.4cm 7.3cm 2.0cm, clip=true,page=3,height = 3.15cm]{Pics/Pics_D/Density_1_02001.pdf}\hspace*{-0.05cm} 
 \includegraphics[trim=6.7cm 3.4cm 0.3cm 2.0cm, clip=true,page=3,height = 3.15cm]{Pics/Pics_D/Density_1_04000.pdf}\\
 \includegraphics[trim=3.9cm 1.3cm 7.3cm 2.0cm, clip=true,page=4,height = 3.57cm]{Pics/Pics_D/Density_1_00041.pdf}\hspace*{-0.05cm} 
 \includegraphics[trim=6.7cm 1.3cm 7.3cm 2.0cm, clip=true,page=4,height = 3.57cm]{Pics/Pics_D/Density_1_00201.pdf}\hspace*{-0.05cm} 
 \includegraphics[trim=6.7cm 1.3cm 7.3cm 2.0cm, clip=true,page=4,height = 3.57cm]{Pics/Pics_D/Density_1_00801.pdf}\hspace*{-0.05cm}  
 \includegraphics[trim=6.7cm 1.3cm 7.3cm 2.0cm, clip=true,page=4,height = 3.57cm]{Pics/Pics_D/Density_1_02001.pdf}\hspace*{-0.05cm} 
 \includegraphics[trim=6.7cm 1.3cm 0.3cm 2.0cm, clip=true,page=4,height = 3.57cm]{Pics/Pics_D/Density_1_04000.pdf}\\
  \hspace*{-0.3cm}\includegraphics[trim=1.6cm 1.4cm 8.6cm 2.0cm, clip=true,page=1,height = 3.6cm]{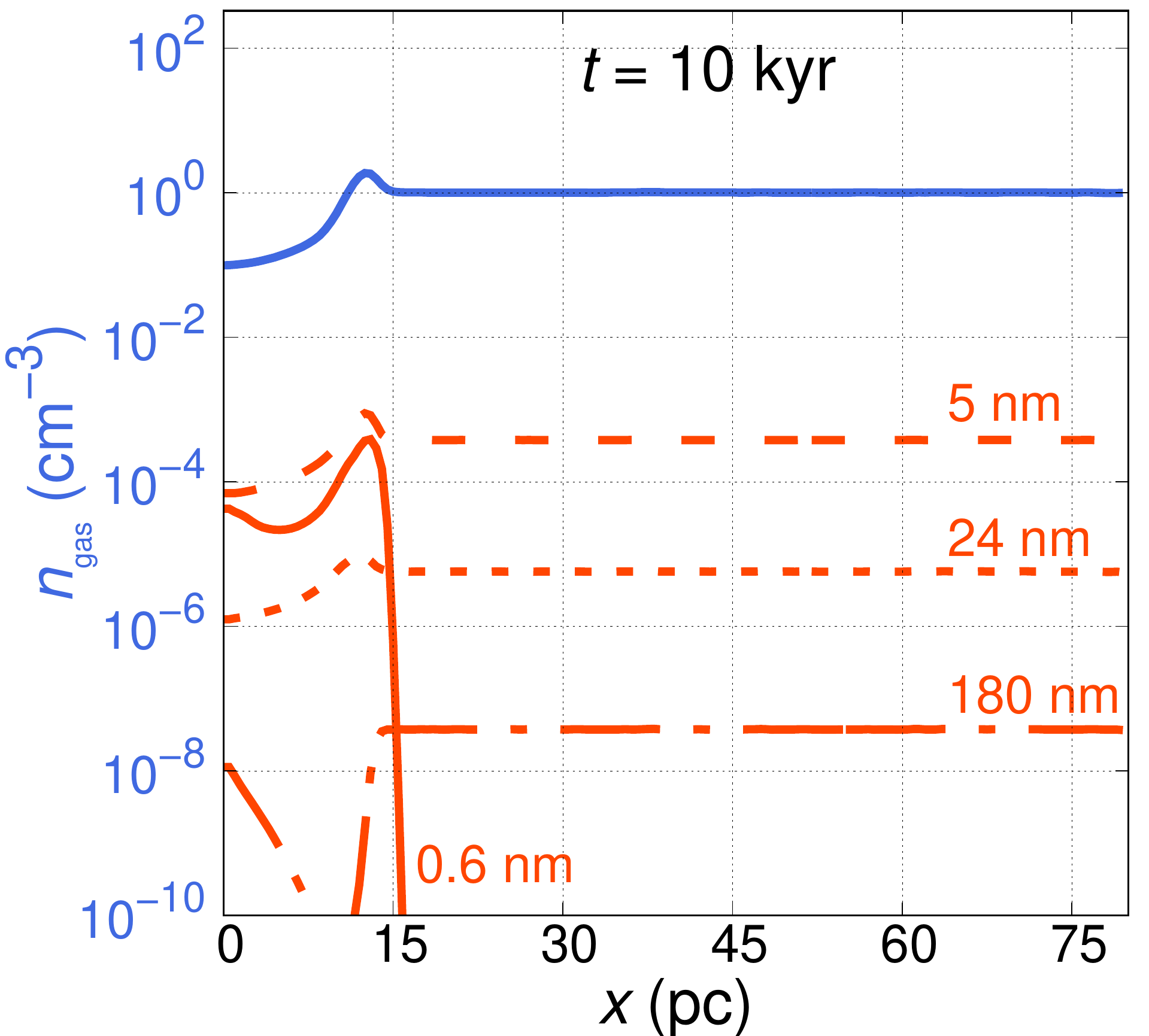}\hspace*{-0.1cm} 
  \includegraphics[trim=5.2cm 1.4cm 8.6cm 2.0cm, clip=true,page=2,height = 3.6cm]{Pics/Pics_D/Profile_dust.pdf}\hspace*{-0.1cm} 
  \includegraphics[trim=5.2cm 1.4cm 8.6cm 2.0cm, clip=true,page=3,height = 3.6cm]{Pics/Pics_D/Profile_dust.pdf}\hspace*{-0.1cm} 
  \includegraphics[trim=5.2cm 1.4cm 8.6cm 2.0cm, clip=true,page=4,height = 3.6cm]{Pics/Pics_D/Profile_dust.pdf}\hspace*{-0.1cm} 
  \includegraphics[trim=5.2cm 1.4cm 3.2cm 2.0cm, clip=true,page=5,height = 3.6cm]{Pics/Pics_D/Profile_dust.pdf}
  \caption{Same as Fig.~\ref{fig_A} but for simulation D.}
   \label{fig_D} 
  \end{figure*}

      \begin{figure*}
 \includegraphics[trim=2.5cm 1.5cm 2.2cm 2.3cm, clip=true,page=1,height = 6cm]{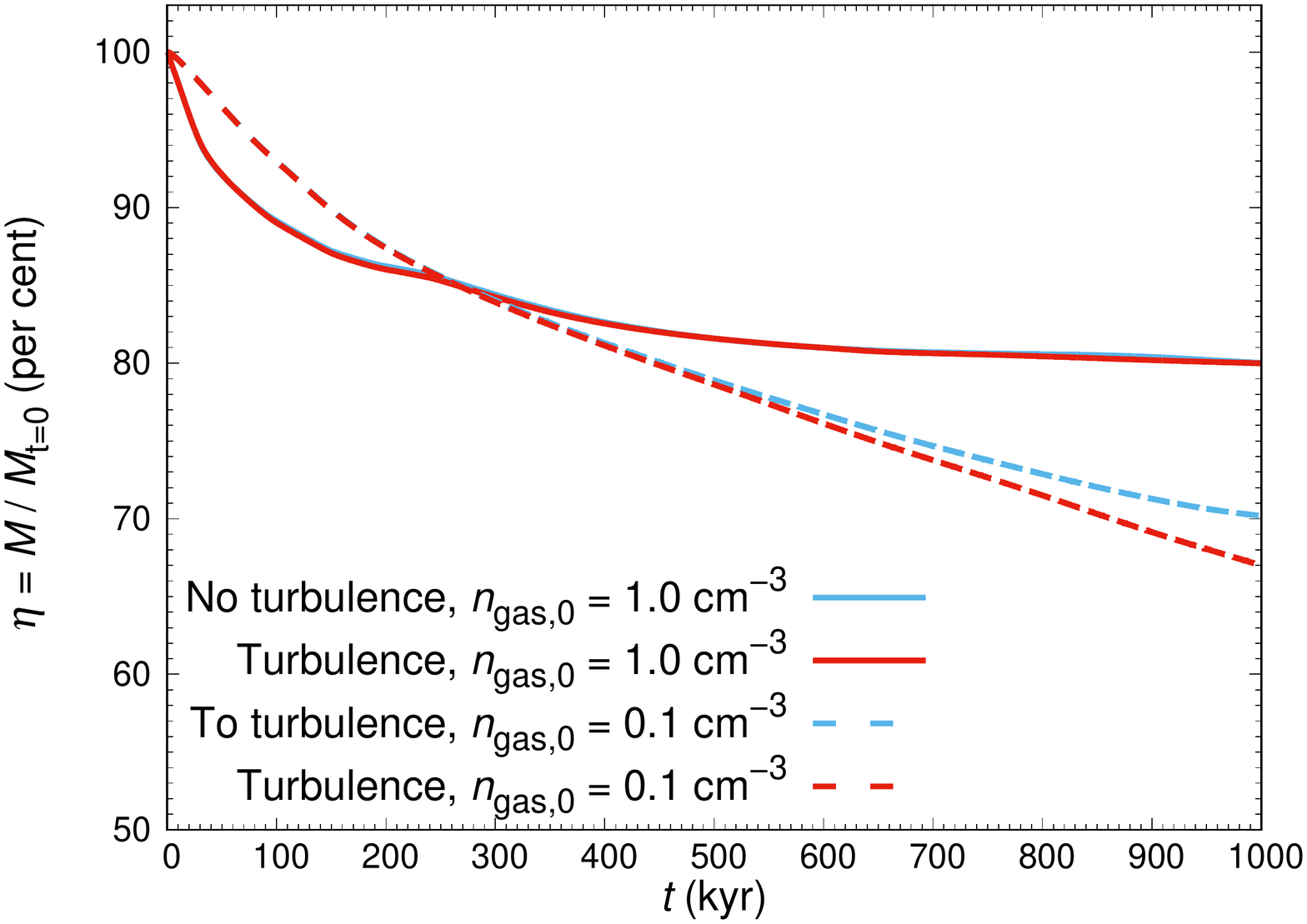}\hspace*{-0.1cm}
 \includegraphics[trim=2.5cm 1.5cm 2.2cm 2.3cm, clip=true,page=2,height = 6cm]{Pics/Pics_A/Evolution_total.pdf}\\ 
  \caption{Evolution of the total dust mass within a distance of $128\,$pc (simulation A and C) or $70\,$pc (simulation B and D) from the explosion centre, respectively. \textit{Left:} Dust survival rate $\eta = M/M_{\rm t=0}$ as a function of time. \textit{Right:} Destruction rate $\textrm{d}M/\textrm{d}t$ as a function of time.}
  \label{fig_dustmass_evolution}  
  \end{figure*}

  \begin{figure*}
 \includegraphics[trim=2.4cm 1.5cm 3.9cm 2.45cm, clip=true,page=1,height = 6cm, page=1 ]{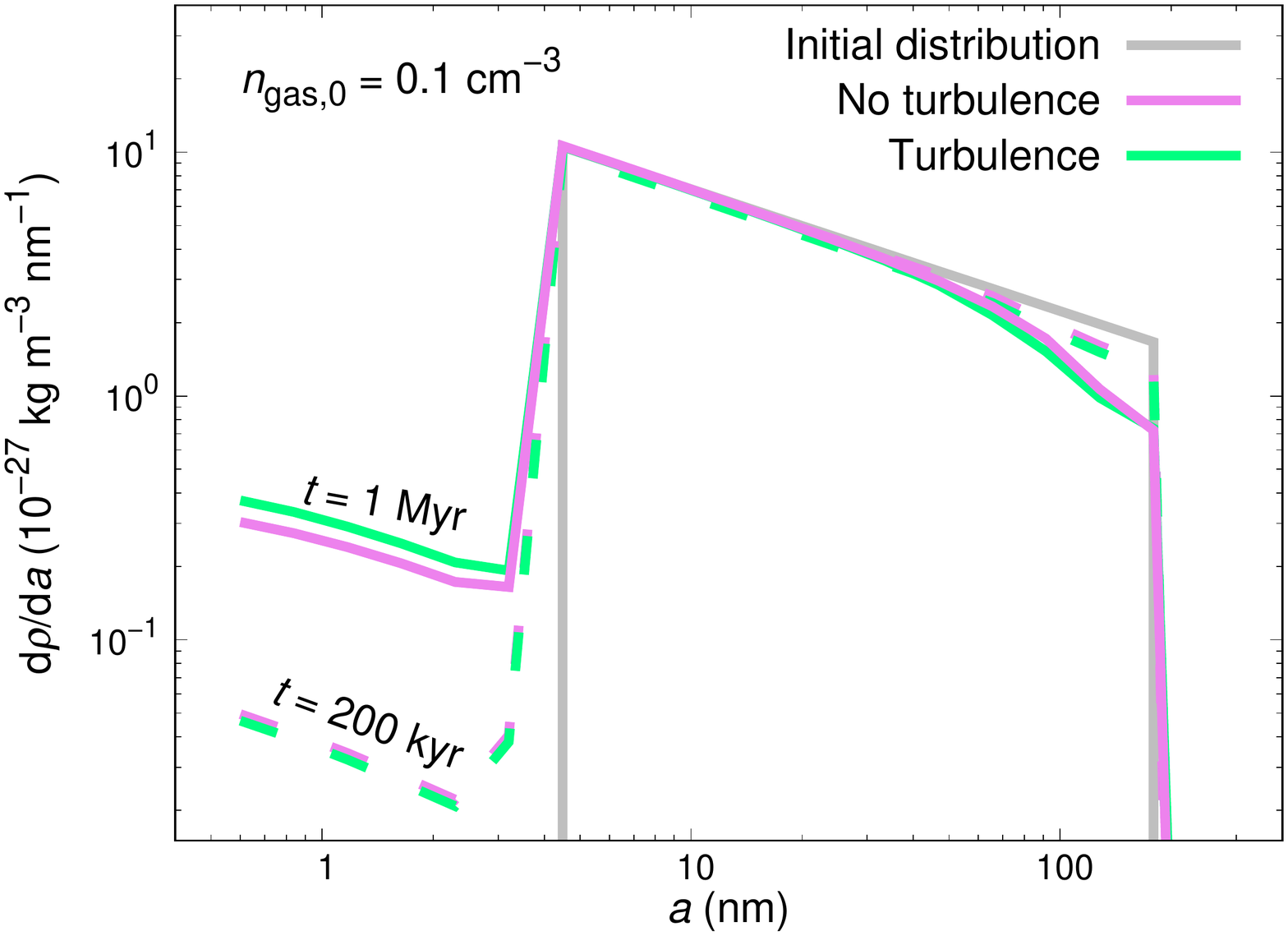}\hspace*{1.1cm}
 \includegraphics[trim=2.4cm 1.5cm 3.9cm 2.45cm, clip=true,page=1,height = 6cm, page=2 ]{Pics/Pics_A/Particlenumbers_04000_compare.pdf}\\ 
  \caption{Dust mass density per grain size for $n_{\textrm{gas}}=0.1\,\textrm{cm}^{-3}$ (\textit{left}) and $1.0\,\textrm{cm}^{-3}$ (\textit{right}). The initial dust mass density ($t=0$) is shown as grey solid line and the final distribution with (without) turbulence as green (violet) solid line. The dashed lines show the distribution at $t=\unit[200]{kyr}$.}
  \label{fig_ABCD_sizedistribution}  
  \end{figure*}
  
  
     \begin{figure}
 \includegraphics[trim=1.5cm 1.5cm 2.2cm 2.3cm, clip=true,page=4,height = 6cm]{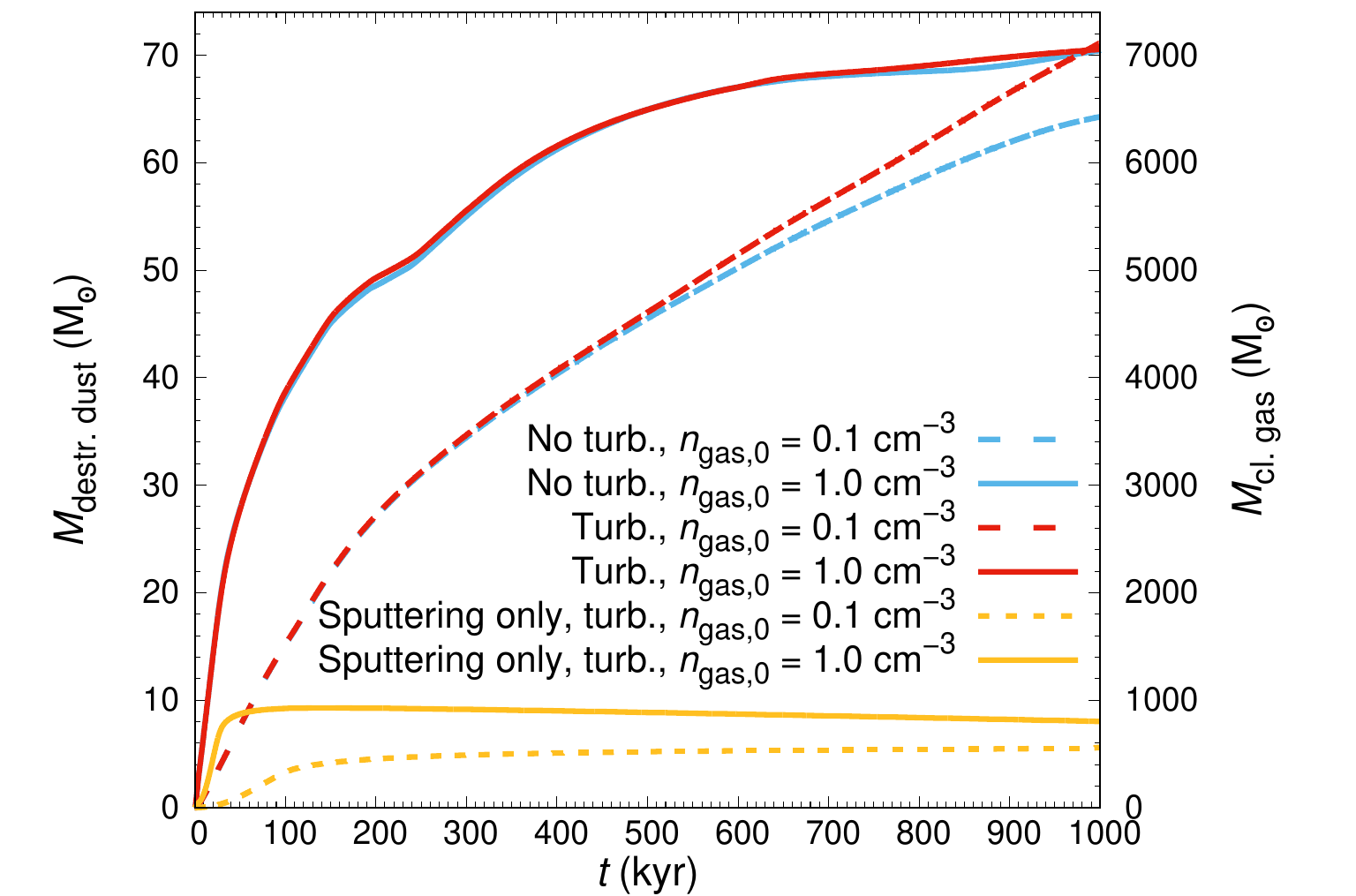}\\ 
  \caption{Destroyed dust masses for simulation A, B, C, and D (red and blue lines). Additionally, we show the results of turbulence simulations C~and~D when only sputtering as destruction process is considered (yellow lines).}
  \label{fig_sputtering}  
  \end{figure}

         \begin{figure}
 \includegraphics[trim=2.5cm 1.5cm 2.2cm 2.3cm, clip=true,page=6,height = 6cm]{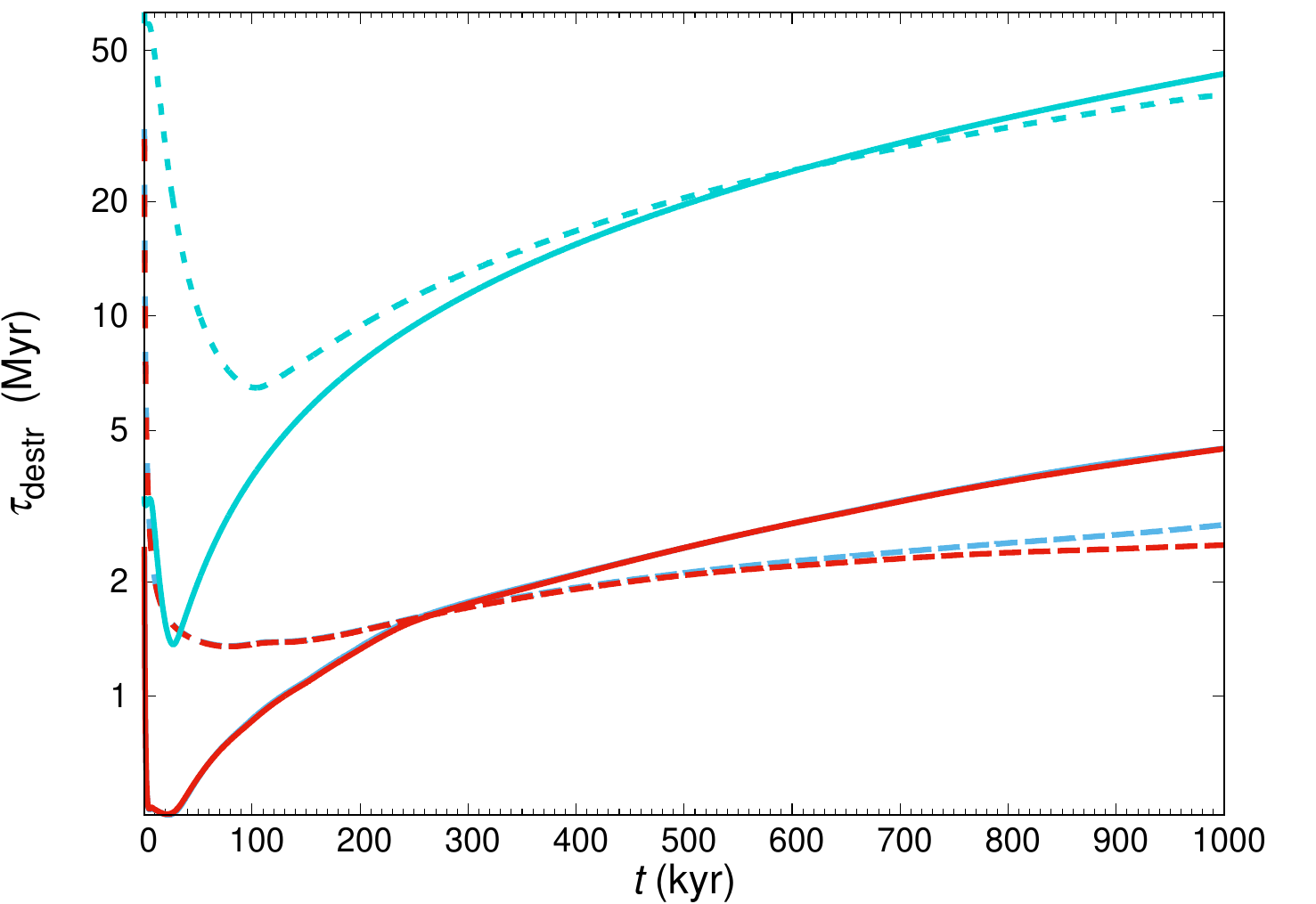}\\ 
  \caption{Dust destruction time-scale $\tau_{\rm destr}=-t/\ln{(\eta)}$. Line colours and types are the same as in Fig.~\ref{fig_sputtering}.}
  \label{fig_time-scale}  
  \end{figure}

         \begin{figure}
 \includegraphics[trim=2.5cm 1.5cm 2.2cm 2.1cm, clip=true,page=1,height = 6cm]{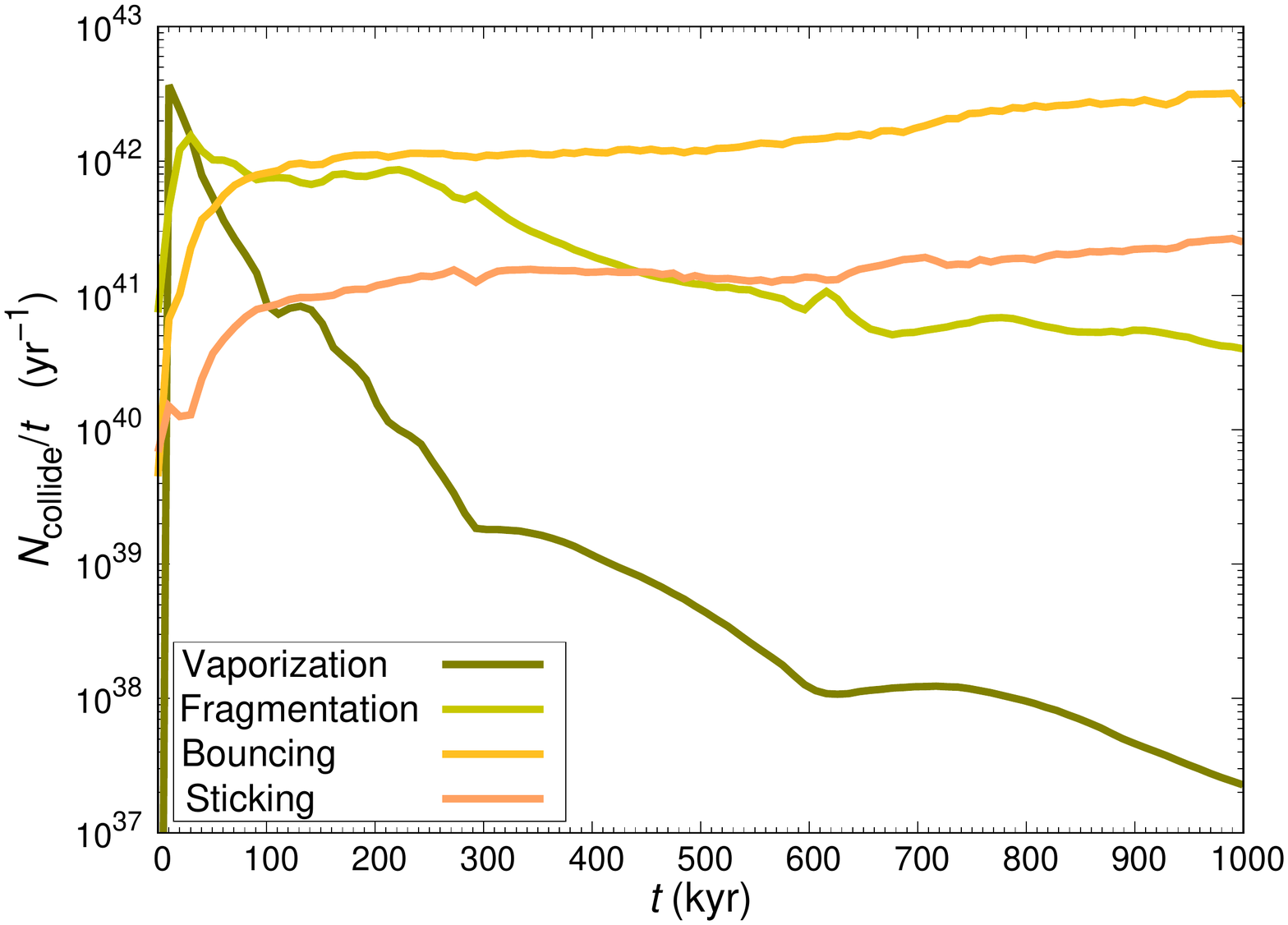}\\[-0.2cm] 
  \caption{Number of dust grains per time which vaporise, fragment, bounce, or stick together, for simulation~D ($n_{\rm gas,0}=1.0\,\textrm{cm}^{-3}$, turbulence).}
  \label{fig_collisions}  
  \end{figure}
 
  
\subsection{Dust processing induced by the shock wave}
\label{sec:dustproc}

In Figs. \ref{fig_A} -- \ref{fig_D} we show the resultant evolution of the dust component from the post-processing runs with \textsc{\textsc{Paperboats}} for all four simulations. The overall evolution is rather similar regardless of the different initial conditions.

We determine the dust mass survival rate $\eta = M/M_{\rm t=0}$ after
$\unit[1]{Myr}$ within a radius of $\unit[128]{pc}$ for the low-density case
($n_{\rm gas,0}=0.1\,\textrm{cm}^{-3}$) and within a radius of $\unit[70]{pc}$
for the high-density case ($n_{\rm gas,0}=1.0\,\textrm{cm}^{-3}$).
The radii are extended to encompass the secondary
instabilities 
beyond the shell when background turbulence is considered. The dust survival rates  for $n_{\rm gas,0} = 1.0$~cm$^{-3}$ are higher than for $n_{\rm gas,0} = 0.1$~cm$^{-3}$ (Table~\ref{List_hydrosimulations_res}, Fig.~\ref{fig_dustmass_evolution}). The higher gas density causes a slower evolution and thus smaller gas and dust velocities which reduces the destruction. On the other hand, the dust survival rates for $\Delta_\text{\rm gd}=100$ (Fig.~\ref{fig_dustmass_evolution}) are between 67 and 80\%  and for $\Delta_\text{\rm gd}=10$ (Fig.~\ref{fig_DM_evolution}) between 43 and 55\%. This gives an indication of the importance of the gas-to-dust 
ratio
$\Delta_\text{\rm gd}$ and implies that the dust-mass fraction being destroyed increases with the amount of dust initially present in the ISM (when the gas density is unchanged). A very dusty ISM will lose more of its dust due to destruction by SNe than a dust-poor ISM, conceptually hypothesised in, e.g., \citet{Mattsson14a}. We note that this may play an important role in how the dust components of galaxies evolve and suggests a different prescription for dust destruction than the commonly used model introduced by \cite{Tielens94} and \cite{Jones94,Jones96} which is based on the work by \citet{McKee89}. 

\subsubsection{Creation of nano-sized grains}
The top row of Figs. \ref{fig_A} -- \ref{fig_D} show the formation and evolution of the  $0.6\,$nm  grains. Due to the initial minimum dust grain radius of $\unit[5]{nm}$, smaller grains are not present at the beginning and have to be formed due to sputtering and grain-grain collisions. These small grains are well-coupled to the gas and accumulate in a dust shell which is in conjunction with
the shell of shocked gas. 

Nano-sized grains are of fundamental importance for the rate of dust-mass destruction as they are much more easily destroyed by sputtering. This effect is even greater if the gas-to-dust mass ratio is lower than usual ($\Delta_\text{\rm gd}=10$ instead of 100), which results in an elevated grain-grain collision rate.

\subsubsection{Intermediate-sized grains}
For grains of intermediate sizes ($a\sim 10\,$nm), all simulations display a 
distinct
 dust shell forming and efficient dust cleansing inside these shells. The low-density simulations (A and C) also show a secondary shell forming (see, in particular, the evolution of $a=24\,$nm grains in Figs.~\ref{fig_A} and \ref{fig_C}), which is due to a reverse shock that appears early on and then ``bounces'' at the origin and propagates outwards again. A secondary shell is clearly visible at $t = 200$~kyr in Figs.~\ref{fig_A} and~\ref{fig_C}. These secondary shells (and reverse shocks in the dust) do not seem to affect the dust processing much, which is reasonable given that the density in the secondary shell is lower compared to the primary shell formed by the forward shock.

\subsubsection{Destruction of large grains}
The largest grains in our simulations ($a\gtrsim 100\,$nm) are mainly destroyed by the SN shock as it propagates. However, the destruction is not due to direct ion-sputtering of these large grains, but rather a result of destructive grain-grain collisions, i.e., grain shattering and vaporisation. As outlined in \cite{Kirchschlager2019}, grain-grain collisions and sputtering are synergistic processes. The destruction of grain mass is then the result of sputtering of the small splinters created by the shattering of large grains. 

In simulation A (low density), large grains are effectively destroyed as the shock propagates outwards (see evolution for $a= 180\,$nm in Fig.~\ref{fig_A}). In simulation B (high density), on the other hand, the SN ``bubble'' is not only cleansed from large dust grains, but also large grains are accumulated in a dust shell building up in conjunction with the shell of shocked gas (see evolution for $a= 180\,$nm in Fig.~\ref{fig_B}).

\subsubsection{Evolution of the grain-size distribution}
As large grains are partly destroyed by shattering and nano-sized grains formed in that process are sputtered away by the ionised gas, we expect some evolution of the grain-size distribution. Fig. \ref{fig_ABCD_sizedistribution} shows the resultant grain-size distribution at $t=0,\, 200$, and $1000$~kyr in case of $n_{\rm gas,0} = 0.1$~cm$^{-3}$ (simulations A and C; left panel) and $n_{\rm gas,0} = 1.0$~cm$^{-3}$ (simulations B and D; right panel). Normalised to the gas density $n_{\rm gas,0}$, the typical number of nano-sized grains in the low-density cases after $1\,$Myr is more than twice of that in the high-density cases. Furthermore, large grains are also shattered at a higher rate if the overall matter density is low ($n_{\rm gas,0} = 0.1$~cm$^{-3}$), which is reflected in the shape of the grain-size distribution. This is completely in line with what we argued above: dust is destroyed by destructive grain-grain collisions combined with ion sputtering. The large grains are thus depleted while tiny grains are created. The shape of the grain-size distribution in the intermediate size range is not much affected, however. The slope of the grain-size distribution of the debris from grain-grain collisions is similar to that of the MRN-like distribution we use for the initial ISM dust component. Thus, the shape of the grain-size distribution is retained in the intermediate size range (roughly $a\sim 5\dots 50\,$nm), while some evolution occurs in the small-grain end as well as in the large-grain end of the grain-size distribution. 

Although the main effect on the large-grain end of the grain-size distribution is depletion due to shattering, some grains will actually grow and become bigger than the initial maximum grain size ($a_{\rm max}=250\,$nm), which can be discerned in both panels of Fig. \ref{fig_ABCD_sizedistribution}. The net growth is very small, however, and the growth processes (coagulation of grains and accretion of molecules onto grains) must be regarded as very inefficient and negligible in the first approximation.

\subsubsection{Impact of weak ISM turbulence}
Comparing the high-density cases (simulations B and D), adding ``turbulence'' has little effect on the dust processing and the formation of a dust shell, while the increased number of density structures seen in the low-density case (simulation C) leads to a broader and less distinct dust shell.  During the first $\sim 500$~kyr we see essentially no difference in the overall dust destruction rate between simulations A and C (Fig. \ref{fig_dustmass_evolution}), which is mainly due to the fact that $u_{\rm s}/u_{\rm t,\,max}\gg 1$, where $u_{\rm s}$ is the expansion velocity of the shell and $u_{\rm t,\,max}$ is the maximum velocity of the background ``turbulence''. As the expansion slows down and $u_{\rm s}/u_{\rm t,\,max}\sim 1$, the energy variance in the background becomes significant at the same time as instabilities can form. After $\sim 500$~kyr the dust destruction rate is somewhat higher when turbulence is considered.
 

\section{Destroyed dust masses and destruction time-scales}\label{sec:compare}
In order to emphasise the importance of grain-grain collisions for the destroyed dust masses as well as to compare our destruction rates with previous studies, we also conduct ``turbulence'' simulations~C and~D without grain-grain collisions and consider only thermal and non-thermal sputtering as well as gas accretion (Section~\ref{sec:spu_onl}). The combined destruction by sputtering and grain-grain collisions is discussed in Section~\ref{sec:spu_and_gg}. The results for the dust mass survival rates, the total destroyed dust masses and gas masses cleared of dust of all simulations are summarised in Table~\ref{List_hydrosimulations_res}.


  \begin{table}
 \centering
 \caption{Results of simulations {A-D}: Dust mass survival rate $\eta = M/M_{\rm t=0}$, the total destroyed dust masses $M_\textrm{\rm destr}$, gas masses cleared of dust $M_{\rm cl.\,gas}$. and dust destruction time-scale $\tau_{\rm destr}$.}
 \begin{tabular}{c c c c c}
 \hline\hline
 Index&$\eta\,[\%]$&$M_{\rm destr.\,dust}\,[\textrm{M}_\odot]$&$M_\textrm{\rm cl.\,gas}\,[\textrm{M}_\odot]$&$\tau_{\rm destr}\,[\textrm{Myr}]$\\\hline 
 A&$70.2$&$64.7$&6470&$\phantom{0}2.8$\\ 
 B&$80.0$&$70.9$&7090&$\phantom{0}4.5$\\  
 C&$67.0$&$71.6$&7160&$\phantom{0}2.5$\\  
 D&$80.0$&$71.0$&7100&$\phantom{0}4.5$\\ 
 C (only sputt.)&$97.4$&$\phantom{0}5.6$&\phantom{0}560&$38.0$\\  
 D (only sputt.)&$97.7$&$\phantom{0}8.0$&\phantom{0}800&$43.0$\\ 
 \hline   
 \end{tabular}
 \label{List_hydrosimulations_res}
 \end{table}  


\subsection{Destruction by sputtering only}
\label{sec:spu_onl}
When the grain-grain collision rate is forced to zero, the net dust destruction is reduced by a factor of ten (see Fig.~\ref{fig_sputtering}, yellow lines). After $\unit[{\sim}100]{kyr}$ (low-density) or $\unit[{\sim}30]{kyr}$ (high-density), the gas temperature in the shocked shell drops below $\unit[10^5]{K}$ while the relative velocity between gas and dust is not higher than $\unit[200]{km/s}$. Following e.g. \cite{Goodson2016}, sputtering becomes insignificant at those gas temperatures and velocities, and the slope of the destruction curve flattens. For the low-density case, the dust destruction saturates, while the higher gas density in simulation D even causes a slight dust mass growth due to gas accretion after $\unit[145]{kyr}$. The destroyed dust mass after $\unit[1]{Myr}$ is $M_{\rm destr.\,dust} = \unit[5.6]{M_\odot}$ for $n_{\rm gas,0} = 0.1$~cm$^{-3}$ (dust survival rate $\eta=97.4\,\%$ within $\unit[128]{pc}$) and $M_{\rm destr.\,dust} = \unit[8.0]{M_\odot}$ for $n_{\rm gas,0} = 1.0$~cm$^{-3}$ ($\eta=97.7\,\%$ within $\unit[70]{pc}$). The gas mass cleared of dust, defined as $M_{\rm cl.\,gas} =  \Delta_{\rm gd} \times M_{\rm destr.\,dust}$ (e.g.~\citealt{Hu2019}), amounts to $M_{\rm cl.\,gas} = \unit[560]{M_\odot}$ for $n_{\rm gas,0} = 0.1$~cm$^{-3}$ and $M_{\rm cl.\,gas} =\unit[800]{M_\odot}$ for $n_{\rm gas,0} = 1.0$~cm$^{-3}$. 

We can also calculate the dust destruction time-scale\footnote{We note that this time-scale describes the destruction of dust within a certain radius around a single SNe. It is different to other destruction time-scales frequently used to describe the dust destruction in an entire galaxy which also takes into account the number of SNe per time and volume (see e.g.\,\citealt{Slavin2015, Hu2019}).} 
\begin{equation}
\tau_{\rm destr}=\frac{-t}{\ln{(\eta)}}
\end{equation}
as a function of time $t$ (Fig.~\ref{fig_time-scale}). Considering only sputtering as destruction process, $\tau_{\rm destr}$ amounts to $\unit[{\sim}40]{Myr}$ for the low and the high-density cases after $\unit[1]{Myr}$.

We compare our results to previous studies that mainly model dust sputtering in single SN blast waves in a homogeneous ISM. Most of them assumed a steady-state shock model, whose applicability to realistic time-dependent shocks is uncertain. 
\cite{Hu2019} have also investigated the destruction of interstellar dust via thermal and non-thermal sputtering in SN shocks using 3D hydrodynamic simulations. However, they neglected grain-grain collisions and kept the grain-size distribution fixed. Assuming an initial dust-to-gas mass ratio of $\Delta_{\rm gd}=200$, they found a total destroyed silicate dust mass of $M_{\rm destr.\,dust} = \unit[9.95]{M_\odot}$ for $n_{\rm gas,0} = 0.1$~cm$^{-3}$, corresponding to $M_{\rm cl.\,gas} = \unit[1990]{M_\odot}$ of cleared gas. For $n_{\rm gas,0} = 1.0$~cm$^{-3}$ the blast wave destroys $\unit[6.85]{M_\odot}$ of silicate dust, corresponding to $M_{\rm cl.\,gas} = \unit[1370]{M_\odot}$. The gas masses estimated by \cite{Hu2019} agree within a factor of $1.7-3.5$ with our results. The main difference is their consideration of multiple SN explosions resembling the solar-neighbourhood environment which causes a multiphase ISM. 

\cite{Slavin2015} conducted 1D hydrodynamic simulations of an SN expansion. They followed the evolution of the grain-size distribution and also included a treatment for the magnetic pressure support, which can suppress non-thermal sputtering. For the gas mass cleared of dust, they obtained $M_{\rm cl.\,gas} = \unit[1990]{M_\odot}$ for  $n_{\rm gas,0} = 0.25$~cm$^{-3}$. 

\cite{Martinez2019} evaluated  the impact of SN explosions on the evolution of dust grains when the explosion occurs within wind-driven bubbles. They used 3D hydrodynamic simulations and studied the destruction of dust generated within the ejecta and of the dust initially present in the ambient ISM. Considering a tight coupling between gas and dust, they had to neglect grain-grain collisions and non-thermal sputtering. In the case of a wind-driven shell model 
and a surrounding ISM gas density of $n_{\rm gas,0} = 1.0$~cm$^{-3}$, only   $\unit[{\sim}0.02]{M_\odot}$ of the ejecta dust and $\unit[{\sim}0.45]{M_\odot}$ of the ambient dust is destroyed by thermal sputtering, and the cleared gas mass is $M_{\rm cl.\,gas} {\sim} \unit[47]{M_\odot}$. On the other hand, when the ambient medium is initially homogeneous ($n_{\rm gas,0} = 1.0$~cm$^{-3}$), $\unit[0.34]{M_\odot}$ of the ejecta dust and $\unit[1.2]{M_\odot}$ of the ambient dust is destroyed within $\unit[{\sim}6100]{yr}$ after the SN explosion. These small dust masses are a result of the much shorter evolution time compared to our study. In Fig.~\ref{fig_sputtering_100kyr} in the appendix, we show the first $\unit[100]{kyr}$ of Fig.~\ref{fig_sputtering} on a logarithmic scale. We can see that our sputtering simulations for $n_{\rm gas,0} = 1.0$~cm$^{-3}$ give a destroyed dust mass of $\unit[{\sim}0.7]{M_\odot}$ after $\unit[{\sim}6100]{yr}$ ($M_{\rm cl.\,gas}=\unit[70]{M_\odot}$), which agrees within a factor of $1.5$ with the dust mass derived by \cite{Martinez2019}.
 
In summary, we can conclude that our results for the sputtered dust masses and the gas masses cleared of dust are consistent with masses derived in previous studies.  

\subsection{Destruction by grain-grain collisions and sputtering}
\label{sec:spu_and_gg}
When taking grain-grain collisions into account, dust destruction is more efficient (Table~\ref{List_hydrosimulations}). For the low-density case without turbulence (simulation~A), $\unit[64.7]{M_\odot}$ of dust is destroyed and a gas mass of $\unit[6470]{M_\odot}$ is cleared of dust. This is related to a dust survival rate of $\eta=70.2\%$ and the destruction time-scale amounts to $\tau_{\rm destr}=\unit[2.8]{Myr}$. For the other three simulations B-D, the destruction is even $10\,\%$ higher and results in $M_{\rm destr.\,dust} \sim \unit[70]{M_\odot}$ ($M_{\rm cl.\,gas}\sim\unit[7000]{M_\odot}$). The survival rate is $\eta=80\%$ ($\tau_{\rm destr}=\unit[4.5]{Myr}$) for the two high-density cases (simulation B and D), and  $\eta=67\%$  ($\tau_{\rm destr}=\unit[2.5]{Myr}$) for the low-density case with turbulence (simulation C).

The destroyed dust masses are one order of magnitude larger and the destruction time-scales one order of magnitude smaller compared to dust destruction by pure sputtering (Section~\ref{sec:spu_onl}). As mentioned above, the destruction is evoked by the direct destruction in grain-grain collisions (catastrophic fragmentation, vaporisation) but also by the proceeding sputtering of the small fragments. The masses of $\unit[64-70]{M_\odot}$ of dust destroyed by a single SN blast wave in a homogeneous medium are a big challenge for SNe as potential dust factories as well as for infrared observations of the Large Magellanic Cloud which showed that an average mass of $\unit[1.2-11.2]{M_\odot}$ of dust is removed per SN (\citealt{Lakicevic2015}).

The significance of grain-grain collisions is also demonstrated by the collisional time-scale $\tau_{\rm col}$. As outlined in \cite{Kirchschlager2019}, the collisional time-scale under the assumption of a single grain size is
\begin{equation}
  \tau_{\rm col}= \frac{4\,\Delta_{\rm gd}\rho_{\rm bulk}}{3\,m_{\rm amu}}\frac{a}{\overline{n}_{\rm gas} \,\overline{v}},\\
               \approx 5.6\frac{\left(a/{\rm nm}\right)}{\left(\overline{n}_{\rm gas}/{\rm cm}^{-3}\right) \left(\overline{v}/{\rm (km/s)}\right)}\unit[]{Myr},\label{taucol}
\end{equation}
where $a$ is the grain radius, $\overline{n}_{\rm gas}$ is the mean gas number density, $\overline{v}$ is the mean relative velocity between the grains, and $m_{\rm amu}$ and $\rho_{\rm bulk}=\unit[3.3]{{\rm g}\,{\rm cm}^{-3}}$ are the atomic mass unit and the silicate density, respectively. Assuming a typical gas density in the forward shock shell  of $\overline{n}_{\rm gas}=\unit[10]{cm^{-3}}$ for $n_{\rm gas,0}=1.0\,\textrm{cm}^{-3}$ (see bottom row in Figs.~\ref{fig_B} and~\ref{fig_D}), grains with radius $a=\unit[10]{nm}$, and a mean velocity $\overline{v}=\unit[50]{km/s}$, the collisional time-scale  is $\tau_{\rm col}\approx\unit[112]{kyr}$, which is ${\sim}11\,\%$ of the total simulation time. For $n_{\rm gas,0}=0.1\,\textrm{cm}^{-3}$, the typical gas density in the forward shock shell is $\overline{n}_{\rm gas}=\unit[1]{cm^{-3}}$ (Figs.~\ref{fig_A} and~\ref{fig_C}) and the mean velocity is $\overline{v}=\unit[100]{km/s}$, resulting in $\tau_{\rm col}\approx\unit[560]{kyr}$. In both cases, the time-scale for grain-grain collisions is 
less than the simulation time, revealing that a significant number of grains is involved in grain-grain collisions. 

The outcome of the collisions depends strongly on the collision energy that declines with evolution time as the blast wave velocity and thus the relative velocity between dust grains decrease.  Fig.~\ref{fig_collisions} shows the number rate of dust grains involved in vaporisation, fragmentation, bouncing or sticking. Vaporisation is the dominant process in the first $\unit[30]{kyr}$ only, followed by fragmentation (until $\unit[100]{kyr}$), before non-destructive processes as bouncing and coagulation take over. While the number of dust grains that are vaporised is steeply falling with time ($\propto t^{-3}$), the effect of fragmentation decreases slowly ($\propto t^{-1}$). The fragmentation rate at $\unit[1]{Myr}$ is only a factor of ${\sim}30$ below the maximum fragmentation rate at $\unit[30]{kyr}$. We note that, although the total dust mass in the ambient ISM decreases with time, the total number of dust grains can even increase as initially large grains are fragmented into smaller pieces. 
This will increase the number of dust grains involved in vaporisation, fragmentation, bouncing and sticking to higher levels.

We can conclude that grain-grain collisions have a crucial effect on the total dust survival rate. Moreover, the temporal evolution of the remnant and in particular its blast wave velocity and gas density determine the predominance of different collision processes, starting with high energetic processes as vaporisation over grain shattering down to non-destructive processes as bouncing and coagulation.


\section{Conclusions}\label{sec:conclusions}
We have conducted 3D hydrodynamic simulations of an SN blast wave propagating through the ISM for an evolution time of $\unit[1]{Myr}$. The late-stage evolution ($t\gtrsim 0.1$~Myr), which is dominated by the interaction between the forward shock and the surrounding ISM, has been of primary interest as our aim has been to study the destruction of ISM dust. We calculated the dust processing due to sputtering, accretion of atoms/molecules, and grain-grain collisions (vaporisation, fragmentation, and coagulation) in 2D slices from the output of the hydrodynamic simulations. 
 
We have considered both a low ($n_{\rm gas,0} = 0.1$~cm$^{-3}$) and a normal/high-density ISM gas density ($n_{\rm gas,0} = 1.0$~cm$^{-3}$) and explored also the impact of adding a weakly compressive turbulent background, which means we have presented four different simulations. Our conclusions based on these simulations are:
\begin{itemize}
    \item The SN blast wave creates an evacuated region around the explosion centre, surrounded by a shell structure of compressed gas. The shell has a diameter of $\unit[{\sim}110]{pc}$ in the low-density ISM and $\unit[{\sim}70]{pc}$  in the high-density ISM. 
    
    \item    Gas turbulence creates gas density fluctuations which evoke Vishniac-Ostriker-Bertschinger overstabilities and Richtmyer-Meshkov instabilities at the forward shock region. For the low-density ISM, these instabilities can extend outwards to $\unit[{\sim}125]{pc}$.    
  
    \item The dust survival rate for $n_{\rm gas,0} = 0.1$~cm$^{-3}$ is lower than for $n_{\rm gas,0} = 1.0$~cm$^{-3}$ ($67-70\,\%$ vs. $80\,\%$) as a higher density means that the blast wave slows down faster. The dust survival rate for a gas-to-dust mass ratio $\Delta_{\rm gd}=10$ is lower than for $\Delta_{\rm gd}=100$ ($43-55\,\%$ vs. $67-80\,\%$) as the larger grain number density increases the frequency of grain-grain collisions. For all studied scenarios, the largest dust survival rate is $80\,\%$ for $n_{\rm gas,0} = 1.0$~cm$^{-3}$ and  $\Delta_{\rm gd}=100$.
    
    \item For the low gas density, the destroyed dust masses are  $10\,\%$ higher when turbulence is considered. Turbulence has a negligible effect on the destroyed dust masses in the high-density case, for both gas-to-dust mass ratios $\Delta_{\rm gd}=10$ and 100.
 
    \item Taking sputtering and grain-grain collisions into account, the total destroyed dust masses for a gas-to-dust mass ratio $\Delta_{\rm gd}=100$ are between $64.7$ and  $\unit[71.6]{M_\odot}$ and the gas masses cleared of dust are $\unit[6470-7160]{M_\odot}$. Considering only sputtering in the turbulence cases, the total destroyed dust masses are   $\unit[5.6-8.0]{M_\odot}$.
    
    \item The time-scale for destruction by sputtering and grain-grain collisions within the front of the forward shock amounts to $\unit[2.5-4.5]{Myr}$. In contrast, $\unit[{\sim}40]{Myr}$ are required when only sputtering is considered.
\end{itemize}
It is well known that dust grains can form in the ejecta of core-collapse SNe and the derived dust masses are in the range of ${\sim}0.1-1$ solar masses (\citealt{Gall2014, Owen2015, Wesson2015, Bevan2016, Bevan2017,Bevan2019, Priestley2019, Priestley2020, Niculescu2021}). The dust masses destroyed in a homogeneous medium around the SN are much larger, even when only sputtering is considered. This makes SNe net dust destroyers and worsens the dust-budget crises seen in galaxies at high redshifts. Ways to reduce the amount of destroyed dust have been reported to be the extensive evacuation of the stellar environment in form of wind-blown bubbles before the explosion (\citealt{Martinez2019}), or the consideration of a more complex, turbulent multiphase ISM that mitigates the blast wave earlier (\citealt{Hu2019}). However, the impact of grain-grain collisions on these scenarios has  to be studied in the future.

\section*{Acknowledgements}
F.K. acknowledges funding from the European Research Council Grant SNDUST ERC-2015-AdG-694520.
L.M. acknowledges funding from the Swedish Research Council (Vetenskapsr\aa det), grant no. 2015-04505.
F.A.G. acknowledges support from the Academy of Finland ReSoLVE Centre of
Excellence (grant 307411) and the ERC under the EU's Horizon 2020 research and
innovation programme (Project UniSDyn, grant 818665). The hydrodynamic simulations were performed using computational resources provided by the Swedish National Infrastructure for Computing (SNIC) at the PDC Center for High Performance Computing, KTH Royal Institute of Technology in Stockholm.\\[-0.8cm]
\section*{Data Availability}

The data underlying this article will be made available upon request.



\bibliographystyle{mnras}
\bibliography{refs} 

\begin{thebibliography}{}
\makeatletter
\relax
\def\mn@urlcharsother{\let\do\@makeother \do\$\do\&\do\#\do\^\do\_\do\%\do\~}
\def\mn@doi{\begingroup\mn@urlcharsother \@ifnextchar [ {\mn@doi@}
  {\mn@doi@[]}}
\def\mn@doi@[#1]#2{\def\@tempa{#1}\ifx\@tempa\@empty \href
  {http://dx.doi.org/#2} {doi:#2}\else \href {http://dx.doi.org/#2} {#1}\fi
  \endgroup}
\def\mn@eprint#1#2{\mn@eprint@#1:#2::\@nil}
\def\mn@eprint@arXiv#1{\href {http://arxiv.org/abs/#1} {{\tt arXiv:#1}}}
\def\mn@eprint@dblp#1{\href {http://dblp.uni-trier.de/rec/bibtex/#1.xml}
  {dblp:#1}}
\def\mn@eprint@#1:#2:#3:#4\@nil{\def\@tempa {#1}\def\@tempb {#2}\def\@tempc
  {#3}\ifx \@tempc \@empty \let \@tempc \@tempb \let \@tempb \@tempa \fi \ifx
  \@tempb \@empty \def\@tempb {arXiv}\fi \@ifundefined
  {mn@eprint@\@tempb}{\@tempb:\@tempc}{\expandafter \expandafter \csname
  mn@eprint@\@tempb\endcsname \expandafter{\@tempc}}}

\bibitem[\protect\citeauthoryear{{Baines}, {Williams}  \& {Asebiomo}}{{Baines}
  et~al.}{1965}]{Baines65}
{Baines} M.~J.,  {Williams} I.~P.,   {Asebiomo} A.~S.,  1965, \mn@doi [MNRAS]
  {10.1093/mnras/130.1.63}, \href
  {https://ui.adsabs.harvard.edu/abs/1965MNRAS.130...63B} {130, 63}

\bibitem[\protect\citeauthoryear{{Barlow}}{{Barlow}}{1978}]{Barlow78}
{Barlow} M.~J.,  1978, \mn@doi [MNRAS] {10.1093/mnras/183.3.367}, \href
  {https://ui.adsabs.harvard.edu/abs/1978MNRAS.183..367B} {183, 367}

\bibitem[\protect\citeauthoryear{{Bertoldi}, {Carilli}, {Cox}, {Fan},
  {Strauss}, {Beelen}, {Omont}  \& {Zylka}}{{Bertoldi}
  et~al.}{2003}]{Bertoldi03}
{Bertoldi} F.,  {Carilli} C.~L.,  {Cox} P.,  {Fan} X.,  {Strauss} M.~A.,
  {Beelen} A.,  {Omont} A.,   {Zylka} R.,  2003, \mn@doi [A\&A]
  {10.1051/0004-6361:20030710}, \href
  {http://adsabs.harvard.edu/abs/2003A%26A...406L..55B} {406, L55}

\bibitem[\protect\citeauthoryear{{Bevan} \& {Barlow}}{{Bevan} \&
  {Barlow}}{2016}]{Bevan2016}
{Bevan} A.,  {Barlow} M.~J.,  2016, \mn@doi [\mnras] {10.1093/mnras/stv2651},
  \href {http://adsabs.harvard.edu/abs/2016MNRAS.456.1269B} {456, 1269}

\bibitem[\protect\citeauthoryear{{Bevan}, {Barlow}  \& {Milisavljevic}}{{Bevan}
  et~al.}{2017}]{Bevan2017}
{Bevan} A.,  {Barlow} M.~J.,   {Milisavljevic} D.,  2017, \mn@doi [\mnras]
  {10.1093/mnras/stw2985}, \href
  {http://adsabs.harvard.edu/abs/2017MNRAS.465.4044B} {465, 4044}

\bibitem[\protect\citeauthoryear{{Bevan} et~al.,}{{Bevan}
  et~al.}{2019}]{Bevan2019}
{Bevan} A.,  et~al., 2019, \mn@doi [\mnras] {10.1093/mnras/stz679}, 485, 5192

\bibitem[\protect\citeauthoryear{{Bocchio}, {Micelotta}, {Gautier}  \&
  {Jones}}{{Bocchio} et~al.}{2012}]{Bocchio2012}
{Bocchio} M.,  {Micelotta} E.~R.,  {Gautier} A.-L.,   {Jones} A.~P.,  2012,
  \mn@doi [\aap] {10.1051/0004-6361/201219705}, \href
  {http://adsabs.harvard.edu/abs/2012A%26A...545A.124B} {545, A124}

\bibitem[\protect\citeauthoryear{{Bocchio}, {Jones}  \& {Slavin}}{{Bocchio}
  et~al.}{2014}]{Bocchio2014}
{Bocchio} M.,  {Jones} A.~P.,   {Slavin} J.~D.,  2014, \mn@doi [\aap]
  {10.1051/0004-6361/201424368}, \href
  {http://adsabs.harvard.edu/abs/2014A%26A...570A..32B} {570, A32}

\bibitem[\protect\citeauthoryear{{Brandenburg} \& {Sarson}}{{Brandenburg} \&
  {Sarson}}{2002}]{ABGS02}
{Brandenburg} A.,  {Sarson} G.~R.,  2002, \mn@doi [\prl]
  {10.1103/PhysRevLett.88.055003}, \href
  {https://ui.adsabs.harvard.edu/abs/2002PhRvL..88e5003B} {88, 055003}

\bibitem[\protect\citeauthoryear{{Brouillette}}{{Brouillette}}{2002}]{Bro02}
{Brouillette} M.,  2002, \mn@doi [Annual Review of Fluid Mechanics]
  {10.1146/annurev.fluid.34.090101.162238}, \href
  {https://ui.adsabs.harvard.edu/abs/2002AnRFM..34..445B} {34, 445}

\bibitem[\protect\citeauthoryear{{Cioffi}, {McKee}  \& {Bertschinger}}{{Cioffi}
  et~al.}{1988}]{Cioffi88}
{Cioffi} D.~F.,  {McKee} C.~F.,   {Bertschinger} E.,  1988, \mn@doi [\apj]
  {10.1086/166834}, \href {http://adsabs.harvard.edu/abs/1988ApJ...334..252C}
  {334, 252}

\bibitem[\protect\citeauthoryear{{De Cia}, {Ledoux}, {Savaglio}, {Schady}  \&
  {Vreeswijk}}{{De Cia} et~al.}{2013}]{DeCia13}
{De Cia} A.,  {Ledoux} C.,  {Savaglio} S.,  {Schady} P.,   {Vreeswijk} P.~M.,
  2013, \mn@doi [A\&A] {10.1051/0004-6361/201321834}, \href
  {http://adsabs.harvard.edu/abs/2013A%26A...560A..88D} {560, A88}

\bibitem[\protect\citeauthoryear{{De Cia}, {Ledoux}, {Mattsson}, {Petitjean},
  {Srianand}, {Gavignaud}  \& {Jenkins}}{{De Cia} et~al.}{2016}]{DeCia16}
{De Cia} A.,  {Ledoux} C.,  {Mattsson} L.,  {Petitjean} P.,  {Srianand} R.,
  {Gavignaud} I.,   {Jenkins} E.~B.,  2016, \mn@doi [A\&A]
  {10.1051/0004-6361/201527895}, \href
  {http://adsabs.harvard.edu/abs/2016A%26A...596A..97D} {596, A97}

\bibitem[\protect\citeauthoryear{{Dolan} \& {Mathieu}}{{Dolan} \&
  {Mathieu}}{2002}]{Dolan2002}
{Dolan} C.~J.,  {Mathieu} R.~D.,  2002, \mn@doi [\aj] {10.1086/324631}, \href
  {https://ui.adsabs.harvard.edu/abs/2002AJ....123..387D} {123, 387}

\bibitem[\protect\citeauthoryear{{Draine}}{{Draine}}{1990}]{Draine90}
{Draine} B.~T.,  1990, in {Blitz} L.,  ed.,  Astronomical Society of the
  Pacific Conference Series Vol. 12, The Evolution of the Interstellar Medium.
  pp 193--205

\bibitem[\protect\citeauthoryear{{Draine} \& {Salpeter}}{{Draine} \&
  {Salpeter}}{1979}]{Draine79}
{Draine} B.~T.,  {Salpeter} E.~E.,  1979, \mn@doi [ApJ] {10.1086/157206}, \href
  {http://adsabs.harvard.edu/abs/1979ApJ...231..438D} {231, 438}

\bibitem[\protect\citeauthoryear{{Fry}, {Fields}  \& {Ellis}}{{Fry}
  et~al.}{2020}]{Fry2020}
{Fry} B.~J.,  {Fields} B.~D.,   {Ellis} J.~R.,  2020, \mn@doi [\apj]
  {10.3847/1538-4357/ab86bf}, \href
  {https://ui.adsabs.harvard.edu/abs/2020ApJ...894..109F} {894, 109}

\bibitem[\protect\citeauthoryear{{Gall}, {Andersen}  \& {Hjorth}}{{Gall}
  et~al.}{2011a}]{Gall11a}
{Gall} C.,  {Andersen} A.~C.,   {Hjorth} J.,  2011a, \mn@doi [A\&A]
  {10.1051/0004-6361/201015286}, \href
  {http://adsabs.harvard.edu/abs/2011A%26A...528A..13G} {528, A13}

\bibitem[\protect\citeauthoryear{{Gall}, {Andersen}  \& {Hjorth}}{{Gall}
  et~al.}{2011b}]{Gall11b}
{Gall} C.,  {Andersen} A.~C.,   {Hjorth} J.,  2011b, \mn@doi [A\&A]
  {10.1051/0004-6361/201015605}, \href
  {http://adsabs.harvard.edu/abs/2011A%26A...528A..14G} {528, A14}

\bibitem[\protect\citeauthoryear{{Gall} et~al.,}{{Gall}
  et~al.}{2014}]{Gall2014}
{Gall} C.,  et~al., 2014, \mn@doi [\nat] {10.1038/nature13558}, \href
  {http://adsabs.harvard.edu/abs/2014Natur.511..326G} {511, 326}

\bibitem[\protect\citeauthoryear{{Gent}, {Shukurov}, {Sarson}, {Fletcher}  \&
  {Mantere}}{{Gent} et~al.}{2013a}]{Gent:2013a}
{Gent} F.~A.,  {Shukurov} A.,  {Sarson} G.~R.,  {Fletcher} A.,   {Mantere}
  M.~J.,  2013a, \mn@doi [\mnras] {10.1093/mnrasl/sls042}, \href
  {http://adsabs.harvard.edu/abs/2013MNRAS.430L..40G} {430, L40}

\bibitem[\protect\citeauthoryear{{Gent}, {Shukurov}, {Fletcher}, {Sarson}  \&
  {Mantere}}{{Gent} et~al.}{2013b}]{Gent:2013b}
{Gent} F.~A.,  {Shukurov} A.,  {Fletcher} A.,  {Sarson} G.~R.,   {Mantere}
  M.~J.,  2013b, \mn@doi [\mnras] {10.1093/mnras/stt560}, \href
  {http://adsabs.harvard.edu/abs/2013MNRAS.432.1396G} {432, 1396}

\bibitem[\protect\citeauthoryear{{Gent}, {Mac Low}, {K{\"a}pyl{\"a}}, {Sarson}
  \& {Hollins}}{{Gent} et~al.}{2020}]{GMKSH20}
{Gent} F.~A.,  {Mac Low} M.-M.,  {K{\"a}pyl{\"a}} M.~J.,  {Sarson} G.~R.,
  {Hollins} J.~F.,  2020, \mn@doi [Geophysical and Astrophysical Fluid
  Dynamics] {10.1080/03091929.2019.1634705}, \href
  {https://ui.adsabs.harvard.edu/abs/2020GApFD.114...77G} {114, 77}

\bibitem[\protect\citeauthoryear{{Gent}, {Mac Low}, {K{\"a}pyl{\"a}}  \&
  {Singh}}{{Gent} et~al.}{2021}]{GMKS21}
{Gent} F.~A.,  {Mac Low} M.-M.,  {K{\"a}pyl{\"a}} M.~J.,   {Singh} N.~K.,
  2021, \mn@doi [\apjl] {10.3847/2041-8213/abed59}, \href
  {https://ui.adsabs.harvard.edu/abs/2021ApJ...910L..15G} {910, L15}

\bibitem[\protect\citeauthoryear{{Gomez} et~al.,}{{Gomez}
  et~al.}{2012}]{Gomez12a}
{Gomez} H.~L.,  et~al., 2012, \mn@doi [ApJ] {10.1088/0004-637X/760/1/96}, \href
  {http://adsabs.harvard.edu/abs/2012ApJ...760...96G} {760, 96}

\bibitem[\protect\citeauthoryear{{Goodson}, {Luebbers}, {Heitsch}  \&
  {Frazer}}{{Goodson} et~al.}{2016}]{Goodson2016}
{Goodson} M.~D.,  {Luebbers} I.,  {Heitsch} F.,   {Frazer} C.~C.,  2016,
  \mn@doi [\mnras] {10.1093/mnras/stw1796}, \href
  {https://ui.adsabs.harvard.edu/abs/2016MNRAS.462.2777G} {462, 2777}

\bibitem[\protect\citeauthoryear{{Haugen} \& {Brandenburg}}{{Haugen} \&
  {Brandenburg}}{2004}]{HB04}
{Haugen} N. E.~L.,  {Brandenburg} A.,  2004, \mn@doi [\pre]
  {10.1103/PhysRevE.70.036408}, \href
  {https://ui.adsabs.harvard.edu/abs/2004PhRvE..70c6408H} {70, 036408}

\bibitem[\protect\citeauthoryear{{Hirashita} \& {Yan}}{{Hirashita} \&
  {Yan}}{2009}]{Hirashita09}
{Hirashita} H.,  {Yan} H.,  2009, \mn@doi [MNRAS]
  {10.1111/j.1365-2966.2009.14405.x}, \href
  {http://adsabs.harvard.edu/abs/2009MNRAS.394.1061H} {394, 1061}

\bibitem[\protect\citeauthoryear{{Hu}, {Zhukovska}, {Somerville}  \&
  {Naab}}{{Hu} et~al.}{2019}]{Hu2019}
{Hu} C.-Y.,  {Zhukovska} S.,  {Somerville} R.~S.,   {Naab} T.,  2019, \mn@doi
  [\mnras] {10.1093/mnras/stz1481}, \href
  {https://ui.adsabs.harvard.edu/abs/2019MNRAS.487.3252H} {487, 3252}

\bibitem[\protect\citeauthoryear{{Jones} \& {Nuth}}{{Jones} \&
  {Nuth}}{2011}]{Jones11}
{Jones} A.~P.,  {Nuth} J.~A.,  2011, \mn@doi [A\&A]
  {10.1051/0004-6361/201014440}, \href
  {http://adsabs.harvard.edu/abs/2011A%26A...530A..44J} {530, A44}

\bibitem[\protect\citeauthoryear{{Jones}, {Tielens}, {Hollenbach}  \&
  {McKee}}{{Jones} et~al.}{1994}]{Jones94}
{Jones} A.~P.,  {Tielens} A.~G.~G.~M.,  {Hollenbach} D.~J.,   {McKee} C.~F.,
  1994, \mn@doi [ApJ] {10.1086/174689}, \href
  {http://adsabs.harvard.edu/abs/1994ApJ...433..797J} {433, 797}

\bibitem[\protect\citeauthoryear{{Jones}, {Tielens}  \& {Hollenbach}}{{Jones}
  et~al.}{1996}]{Jones96}
{Jones} A.~P.,  {Tielens} A.~G.~G.~M.,   {Hollenbach} D.~J.,  1996, \mn@doi
  [ApJ] {10.1086/177823}, \href
  {http://adsabs.harvard.edu/abs/1996ApJ...469..740J} {469, 740}

\bibitem[\protect\citeauthoryear{{Kirchschlager}, {Schmidt}, {Barlow},
  {Fogerty}, {Bevan}  \& {Priestley}}{{Kirchschlager}
  et~al.}{2019}]{Kirchschlager2019}
{Kirchschlager} F.,  {Schmidt} F.~D.,  {Barlow} M.~J.,  {Fogerty} E.~L.,
  {Bevan} A.,   {Priestley} F.~D.,  2019, \mn@doi [\mnras]
  {10.1093/mnras/stz2399}, \href
  {https://ui.adsabs.harvard.edu/abs/2019MNRAS.489.4465K} {489, 4465}

\bibitem[\protect\citeauthoryear{{Kirchschlager}, {Barlow}  \&
  {Schmidt}}{{Kirchschlager} et~al.}{2020}]{Kirchschlager2020}
{Kirchschlager} F.,  {Barlow} M.~J.,   {Schmidt} F.~D.,  2020, \mn@doi [\apj]
  {10.3847/1538-4357/ab7db8}, \href
  {https://ui.adsabs.harvard.edu/abs/2020ApJ...893...70K} {893, 70}

\bibitem[\protect\citeauthoryear{{Kolmogorov}}{{Kolmogorov}}{1941}]{Kolmogorov41}
{Kolmogorov} A.,  1941, Akademiia Nauk SSSR Doklady, \href
  {http://adsabs.harvard.edu/abs/1941DoSSR..30..301K} {30, 301}

\bibitem[\protect\citeauthoryear{{Kuo} \& {Hirashita}}{{Kuo} \&
  {Hirashita}}{2012}]{Kuo12}
{Kuo} T.-M.,  {Hirashita} H.,  2012, \mn@doi [MNRAS]
  {10.1111/j.1745-3933.2012.01282.x}, \href
  {http://adsabs.harvard.edu/abs/2012MNRAS.424L..34K} {424, L34}

\bibitem[\protect\citeauthoryear{{Laki{\'c}evi{\'c}}
  et~al.,}{{Laki{\'c}evi{\'c}} et~al.}{2015}]{Lakicevic2015}
{Laki{\'c}evi{\'c}} M.,  et~al., 2015, \mn@doi [\apj]
  {10.1088/0004-637X/799/1/50}, \href
  {https://ui.adsabs.harvard.edu/abs/2015ApJ...799...50L} {799, 50}

\bibitem[\protect\citeauthoryear{{Mart{\'\i}nez-Gonz{\'a}lez}, {W{\"u}nsch},
  {Palou{\v{s}}}, {Mu{\~n}oz-Tu{\~n}{\'o}n}, {Silich}  \&
  {Tenorio-Tagle}}{{Mart{\'\i}nez-Gonz{\'a}lez} et~al.}{2018}]{Martinez2018}
{Mart{\'\i}nez-Gonz{\'a}lez} S.,  {W{\"u}nsch} R.,  {Palou{\v{s}}} J.,
  {Mu{\~n}oz-Tu{\~n}{\'o}n} C.,  {Silich} S.,   {Tenorio-Tagle} G.,  2018,
  \mn@doi [\apj] {10.3847/1538-4357/aadb88}, \href
  {https://ui.adsabs.harvard.edu/abs/2018ApJ...866...40M} {866, 40}

\bibitem[\protect\citeauthoryear{{Mart{\'\i}nez-Gonz{\'a}lez}, {W{\"u}nsch},
  {Silich}, {Tenorio-Tagle}, {Palou{\v{s}}}  \&
  {Ferrara}}{{Mart{\'\i}nez-Gonz{\'a}lez} et~al.}{2019}]{Martinez2019}
{Mart{\'\i}nez-Gonz{\'a}lez} S.,  {W{\"u}nsch} R.,  {Silich} S.,
  {Tenorio-Tagle} G.,  {Palou{\v{s}}} J.,   {Ferrara} A.,  2019, \mn@doi [\apj]
  {10.3847/1538-4357/ab571b}, \href
  {https://ui.adsabs.harvard.edu/abs/2019ApJ...887..198M} {887, 198}

\bibitem[\protect\citeauthoryear{{Mathis}, {Rumpl}  \& {Nordsieck}}{{Mathis}
  et~al.}{1977}]{Mathis77}
{Mathis} J.~S.,  {Rumpl} W.,   {Nordsieck} K.~H.,  1977, \mn@doi [ApJ]
  {10.1086/155591}, \href {http://adsabs.harvard.edu/abs/1977ApJ...217..425M}
  {217, 425}

\bibitem[\protect\citeauthoryear{{Matsuura} et~al.,}{{Matsuura}
  et~al.}{2009}]{Matsuura09}
{Matsuura} M.,  et~al., 2009, \mn@doi [MNRAS]
  {10.1111/j.1365-2966.2009.14743.x}, \href
  {http://adsabs.harvard.edu/abs/2009MNRAS.396..918M} {396, 918}

\bibitem[\protect\citeauthoryear{{Mattsson}}{{Mattsson}}{2011}]{Mattsson11b}
{Mattsson} L.,  2011, \mn@doi [MNRAS] {10.1111/j.1365-2966.2011.18447.x}, \href
  {http://adsabs.harvard.edu/abs/2011MNRAS.414..781M} {414, 781}

\bibitem[\protect\citeauthoryear{{Mattsson}}{{Mattsson}}{2016}]{Mattsson16}
{Mattsson} L.,  2016, \mn@doi [\planss] {10.1016/j.pss.2016.05.002}, \href
  {http://adsabs.harvard.edu/abs/2016P%26SS..133..107M} {133, 107}

\bibitem[\protect\citeauthoryear{{Mattsson} \& {Andersen}}{{Mattsson} \&
  {Andersen}}{2012}]{Mattsson12b}
{Mattsson} L.,  {Andersen} A.~C.,  2012, \mn@doi [MNRAS]
  {10.1111/j.1365-2966.2012.20574.x}, \href
  {http://adsabs.harvard.edu/abs/2012MNRAS.423...38M} {423, 38}

\bibitem[\protect\citeauthoryear{{Mattsson}, {Andersen}  \&
  {Munkhammar}}{{Mattsson} et~al.}{2012}]{Mattsson12a}
{Mattsson} L.,  {Andersen} A.~C.,   {Munkhammar} J.~D.,  2012, \mn@doi [MNRAS]
  {10.1111/j.1365-2966.2012.20575.x}, \href
  {http://adsabs.harvard.edu/abs/2012MNRAS.423...26M} {423, 26}

\bibitem[\protect\citeauthoryear{{Mattsson}, {De Cia}, {Andersen}  \&
  {Zafar}}{{Mattsson} et~al.}{2014a}]{Mattsson14a}
{Mattsson} L.,  {De Cia} A.,  {Andersen} A.~C.,   {Zafar} T.,  2014a, \mn@doi
  [MNRAS] {10.1093/mnras/stu370}, \href
  {http://adsabs.harvard.edu/abs/2014MNRAS.440.1562M} {440, 1562}

\bibitem[\protect\citeauthoryear{{Mattsson} et~al.,}{{Mattsson}
  et~al.}{2014b}]{Mattsson14b}
{Mattsson} L.,  et~al., 2014b, \mn@doi [MNRAS] {10.1093/mnras/stu1228}, \href
  {http://adsabs.harvard.edu/abs/2014MNRAS.444..797M} {444, 797}

\bibitem[\protect\citeauthoryear{{McKee}}{{McKee}}{1989}]{McKee89}
{McKee} C.,  1989, in {Allamandola} L.~J.,  {Tielens} A.~G.~G.~M.,  eds,  IAU
  Symposium Vol. 135, Interstellar Dust. p.~431

\bibitem[\protect\citeauthoryear{{Micha{\l}owski}, {Murphy}, {Hjorth},
  {Watson}, {Gall}  \& {Dunlop}}{{Micha{\l}owski}
  et~al.}{2010a}]{Michalowski10b}
{Micha{\l}owski} M.~J.,  {Murphy} E.~J.,  {Hjorth} J.,  {Watson} D.,  {Gall}
  C.,   {Dunlop} J.~S.,  2010a, \mn@doi [A\&A] {10.1051/0004-6361/201014902},
  \href {http://adsabs.harvard.edu/abs/2010A%26A...522A..15M} {522, A15}

\bibitem[\protect\citeauthoryear{{Micha{\l}owski}, {Watson}  \&
  {Hjorth}}{{Micha{\l}owski} et~al.}{2010b}]{Michalowski10a}
{Micha{\l}owski} M.~J.,  {Watson} D.,   {Hjorth} J.,  2010b, \mn@doi [ApJ]
  {10.1088/0004-637X/712/2/942}, \href
  {http://adsabs.harvard.edu/abs/2010ApJ...712..942M} {712, 942}

\bibitem[\protect\citeauthoryear{{Niculescu-Duvaz}, {Barlow}, {Bevan},
  {Milisavljevic}  \& {De Looze}}{{Niculescu-Duvaz}
  et~al.}{2021}]{Niculescu2021}
{Niculescu-Duvaz} M.,  {Barlow} M.~J.,  {Bevan} A.,  {Milisavljevic} D.,   {De
  Looze} I.,  2021, \mn@doi [\mnras] {10.1093/mnras/stab932}, \href
  {https://ui.adsabs.harvard.edu/abs/2021MNRAS.504.2133N} {504, 2133}

\bibitem[\protect\citeauthoryear{{Ostriker} \& {McKee}}{{Ostriker} \&
  {McKee}}{1988}]{Ostriker88}
{Ostriker} J.~P.,  {McKee} C.~F.,  1988, \mn@doi [\rmp]
  {10.1103/RevModPhys.60.1}, \href
  {http://adsabs.harvard.edu/abs/1988RvMP...60....1O} {60, 1}

\bibitem[\protect\citeauthoryear{{Owen} \& {Barlow}}{{Owen} \&
  {Barlow}}{2015}]{Owen2015}
{Owen} P.~J.,  {Barlow} M.~J.,  2015, \mn@doi [\apj]
  {10.1088/0004-637X/801/2/141}, \href
  {http://adsabs.harvard.edu/abs/2015ApJ...801..141O} {801, 141}

\bibitem[\protect\citeauthoryear{{Pencil Code Collaboration} et~al.,}{{Pencil
  Code Collaboration} et~al.}{2021}]{Pencil-JOSS}
{Pencil Code Collaboration} et~al., 2021, \mn@doi [The Journal of Open Source
  Software] {10.21105/joss.02807}, \href
  {https://ui.adsabs.harvard.edu/abs/2021JOSS....6.2807P} {6, 2807}

\bibitem[\protect\citeauthoryear{{Priestley}, {Barlow}  \& {De
  Looze}}{{Priestley} et~al.}{2019}]{Priestley2019}
{Priestley} F.~D.,  {Barlow} M.~J.,   {De Looze} I.,  2019, \mn@doi [\mnras]
  {10.1093/mnras/stz414}, \href
  {http://adsabs.harvard.edu/abs/2019MNRAS.485..440P} {485, 440}

\bibitem[\protect\citeauthoryear{{Priestley}, {Bevan}, {Barlow}  \& {De
  Looze}}{{Priestley} et~al.}{2020}]{Priestley2020}
{Priestley} F.~D.,  {Bevan} A.,  {Barlow} M.~J.,   {De Looze} I.,  2020,
  \mn@doi [\mnras] {10.1093/mnras/staa2121}, \href
  {https://ui.adsabs.harvard.edu/abs/2020MNRAS.497.2227P} {497, 2227}

\bibitem[\protect\citeauthoryear{{Sarazin} \& {White}}{{Sarazin} \&
  {White}}{1987}]{Sarazin:1987}
{Sarazin} C.~L.,  {White} III R.~E.,  1987, \mn@doi [\apj] {10.1086/165522},
  \href {http://adsabs.harvard.edu/abs/1987ApJ...320...32S} {320, 32}

\bibitem[\protect\citeauthoryear{{Sedov}}{{Sedov}}{1959}]{Sedov59}
{Sedov} L.~I.,  1959, {Similarity and Dimensional Methods in Mechanics}.
New York: Academic Press

\bibitem[\protect\citeauthoryear{{Shull}}{{Shull}}{1978}]{Shull1978}
{Shull} J.~M.,  1978, \mn@doi [\apj] {10.1086/156666}, \href
  {https://ui.adsabs.harvard.edu/abs/1978ApJ...226..858S} {226, 858}

\bibitem[\protect\citeauthoryear{{Slavin}, {Jones}  \& {Tielens}}{{Slavin}
  et~al.}{2004}]{Slavin04}
{Slavin} J.~D.,  {Jones} A.~P.,   {Tielens} A.~G.~G.~M.,  2004, \mn@doi [ApJ]
  {10.1086/423834}, \href {http://adsabs.harvard.edu/abs/2004ApJ...614..796S}
  {614, 796}

\bibitem[\protect\citeauthoryear{{Slavin}, {Dwek}  \& {Jones}}{{Slavin}
  et~al.}{2015}]{Slavin2015}
{Slavin} J.~D.,  {Dwek} E.,   {Jones} A.~P.,  2015, \mn@doi [\apj]
  {10.1088/0004-637X/803/1/7}, \href
  {https://ui.adsabs.harvard.edu/abs/2015ApJ...803....7S} {803, 7}

\bibitem[\protect\citeauthoryear{{Slavin}, {Dwek}, {Mac Low}  \&
  {Hill}}{{Slavin} et~al.}{2020}]{Slavin2020}
{Slavin} J.~D.,  {Dwek} E.,  {Mac Low} M.-M.,   {Hill} A.~S.,  2020, \mn@doi
  [\apj] {10.3847/1538-4357/abb5a4}, \href
  {https://ui.adsabs.harvard.edu/abs/2020ApJ...902..135S} {902, 135}

\bibitem[\protect\citeauthoryear{{Taylor}}{{Taylor}}{1950}]{Taylor50}
{Taylor} G.,  1950, \mn@doi [Royal Society of London Proceedings Series A]
  {10.1098/rspa.1950.0049}, \href
  {http://ukads.nottingham.ac.uk/abs/1950RSPSA.201..159T} {201, 159}

\bibitem[\protect\citeauthoryear{{Testi} et~al.,}{{Testi}
  et~al.}{2014}]{Testi2014}
{Testi} L.,  et~al., 2014, in {Beuther} H.,  {Klessen} R.~S.,  {Dullemond}
  C.~P.,   {Henning} T.,  eds, Protostars and Planets VI. p.~339 (\mn@eprint
  {arXiv} {1402.1354}), \mn@doi{10.2458/azu\_uapress\_9780816531240-ch015}

\bibitem[\protect\citeauthoryear{{Tielens}, {McKee}, {Seab}  \&
  {Hollenbach}}{{Tielens} et~al.}{1994}]{Tielens94}
{Tielens} A.~G.~G.~M.,  {McKee} C.~F.,  {Seab} C.~G.,   {Hollenbach} D.~J.,
  1994, \mn@doi [ApJ] {10.1086/174488}, \href
  {http://adsabs.harvard.edu/abs/1994ApJ...431..321T} {431, 321}

\bibitem[\protect\citeauthoryear{{Vishniac}}{{Vishniac}}{1983}]{Vishniac83}
{Vishniac} E.~T.,  1983, \mn@doi [\apj] {10.1086/161433}, \href
  {http://adsabs.harvard.edu/abs/1983ApJ...274..152V} {274, 152}

\bibitem[\protect\citeauthoryear{{Vishniac}, {Ostriker}  \&
  {Bertschinger}}{{Vishniac} et~al.}{1985}]{VOB85}
{Vishniac} E.~T.,  {Ostriker} J.~P.,   {Bertschinger} E.,  1985, \mn@doi [\apj]
  {10.1086/163079}, \href {http://adsabs.harvard.edu/abs/1985ApJ...291..399V}
  {291, 399}

\bibitem[\protect\citeauthoryear{{Watson}, {Christensen}, {Knudsen}, {Richard},
  {Gallazzi}  \& {Micha{\l}owski}}{{Watson} et~al.}{2015}]{Watson15}
{Watson} D.,  {Christensen} L.,  {Knudsen} K.~K.,  {Richard} J.,  {Gallazzi}
  A.,   {Micha{\l}owski} M.~J.,  2015, \mn@doi [\nat] {10.1038/nature14164},
  \href {http://adsabs.harvard.edu/abs/2015Natur.519..327W} {519, 327}

\bibitem[\protect\citeauthoryear{{Wesson}, {Barlow}, {Matsuura}  \&
  {Ercolano}}{{Wesson} et~al.}{2015}]{Wesson2015}
{Wesson} R.,  {Barlow} M.~J.,  {Matsuura} M.,   {Ercolano} B.,  2015, \mn@doi
  [\mnras] {10.1093/mnras/stu2250}, \href
  {https://ui.adsabs.harvard.edu/abs/2015MNRAS.446.2089W} {446, 2089}

\bibitem[\protect\citeauthoryear{{Wolfire}, {Hollenbach}, {McKee}, {Tielens}
  \& {Bakes}}{{Wolfire} et~al.}{1995}]{Wolfire:1995}
{Wolfire} M.~G.,  {Hollenbach} D.,  {McKee} C.~F.,  {Tielens} A.~G.~G.~M.,
  {Bakes} E.~L.~O.,  1995, \mn@doi [\apj] {10.1086/175510}, \href
  {http://adsabs.harvard.edu/abs/1995ApJ...443..152W} {443, 152}

\bibitem[\protect\citeauthoryear{{Zhukovska}, {Gail}  \&
  {Trieloff}}{{Zhukovska} et~al.}{2008}]{Zhukovska08}
{Zhukovska} S.,  {Gail} H.-P.,   {Trieloff} M.,  2008, \mn@doi [A\&A]
  {10.1051/0004-6361:20077789}, \href
  {http://adsabs.harvard.edu/abs/2008A%26A...479..453Z} {479, 453}

\makeatother
\end{thebibliography}

\appendix
\section{Dust processing for $\Delta_{\rm gd}=10$.}
\label{apx:delta10}
\begin{figure*}
 \includegraphics[trim=3.9cm 3.4cm 7.3cm 2.0cm, clip=true,page=1,height = 3.15cm]{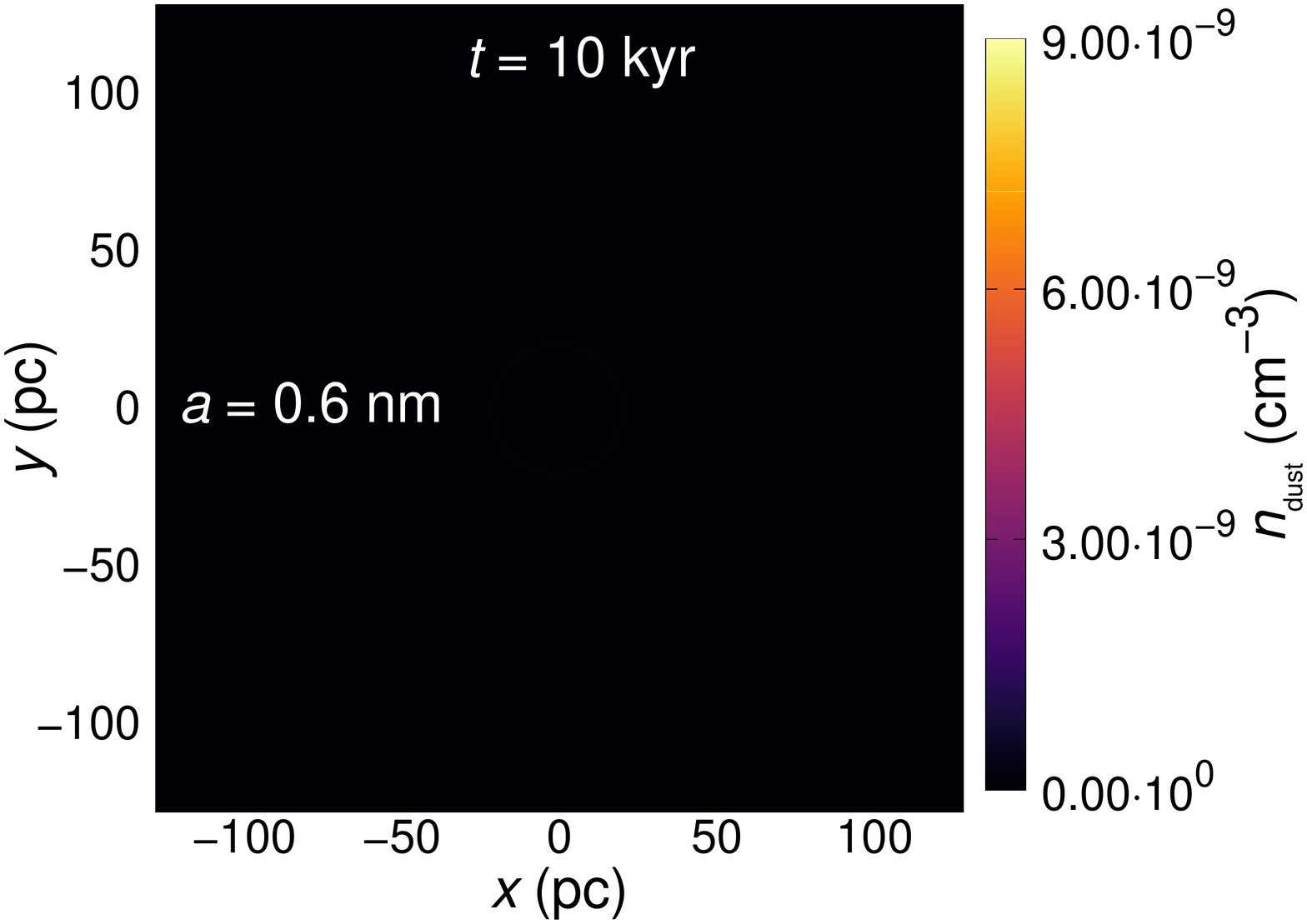}\hspace*{-0.05cm}
 \includegraphics[trim=6.7cm 3.4cm 7.3cm 2.0cm, clip=true,page=1,height = 3.15cm]{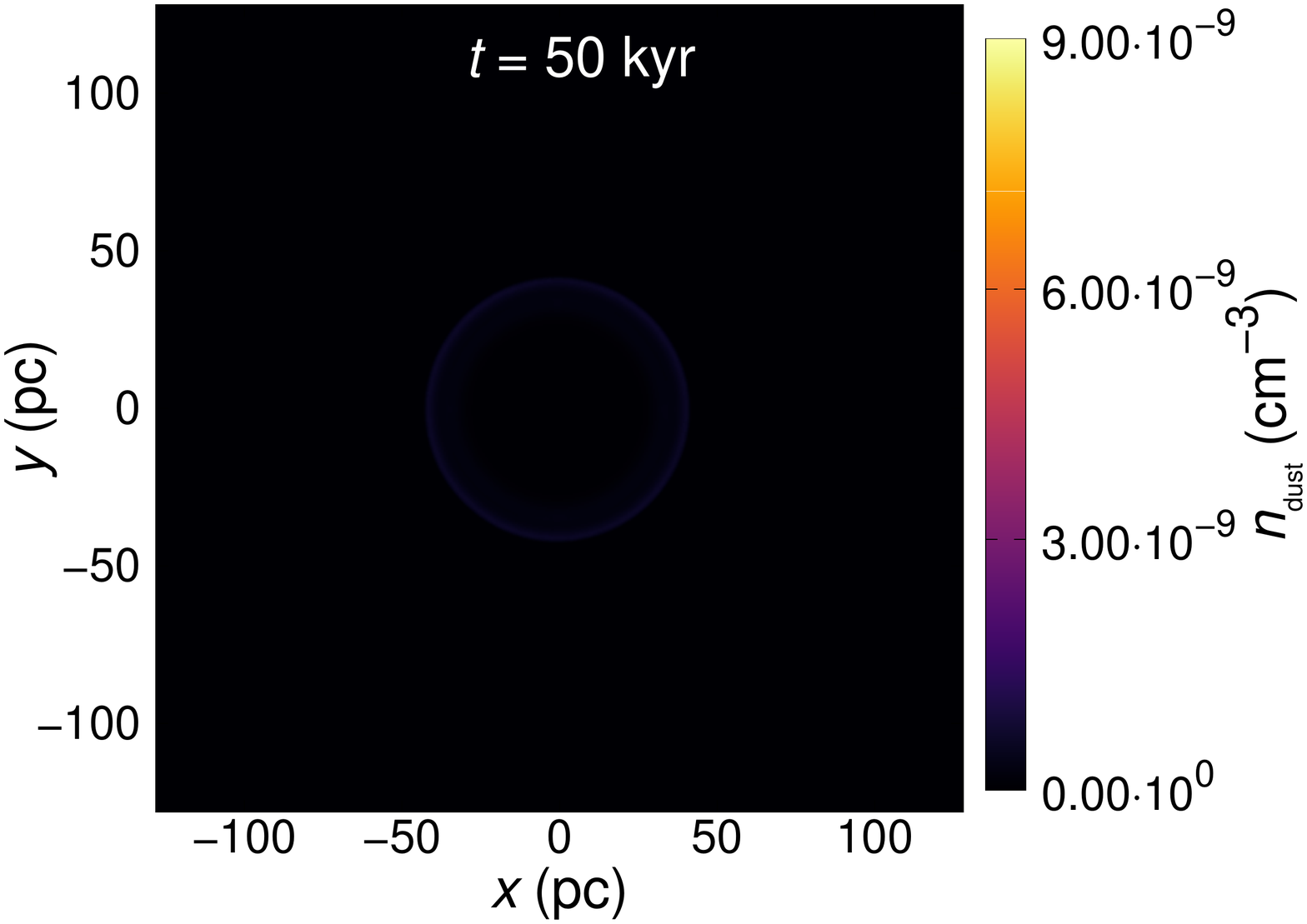}\hspace*{-0.05cm}
 \includegraphics[trim=6.7cm 3.4cm 7.3cm 2.0cm, clip=true,page=1,height = 3.15cm]{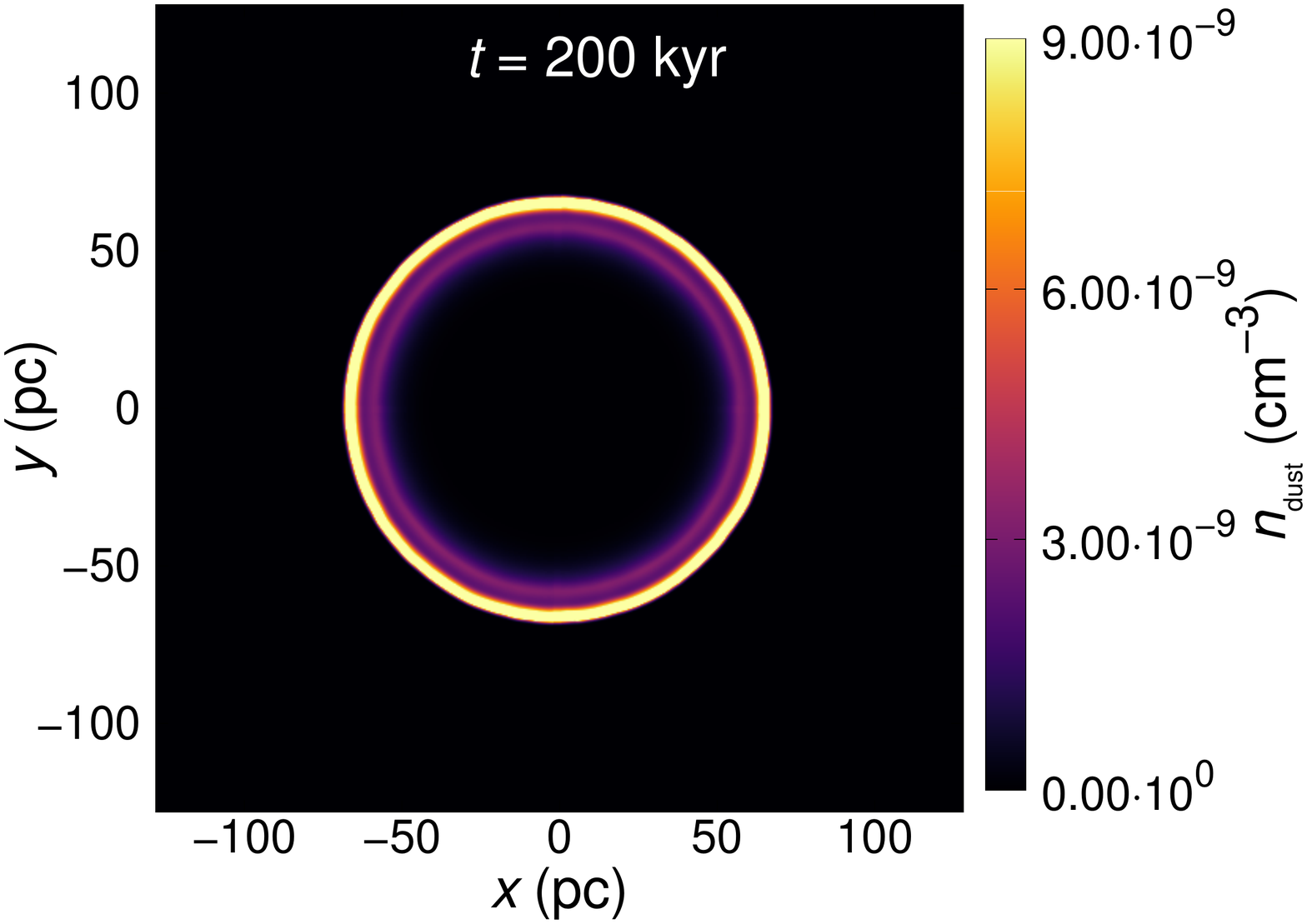}\hspace*{-0.05cm}
 \includegraphics[trim=6.7cm 3.4cm 7.3cm 2.0cm, clip=true,page=1,height = 3.15cm]{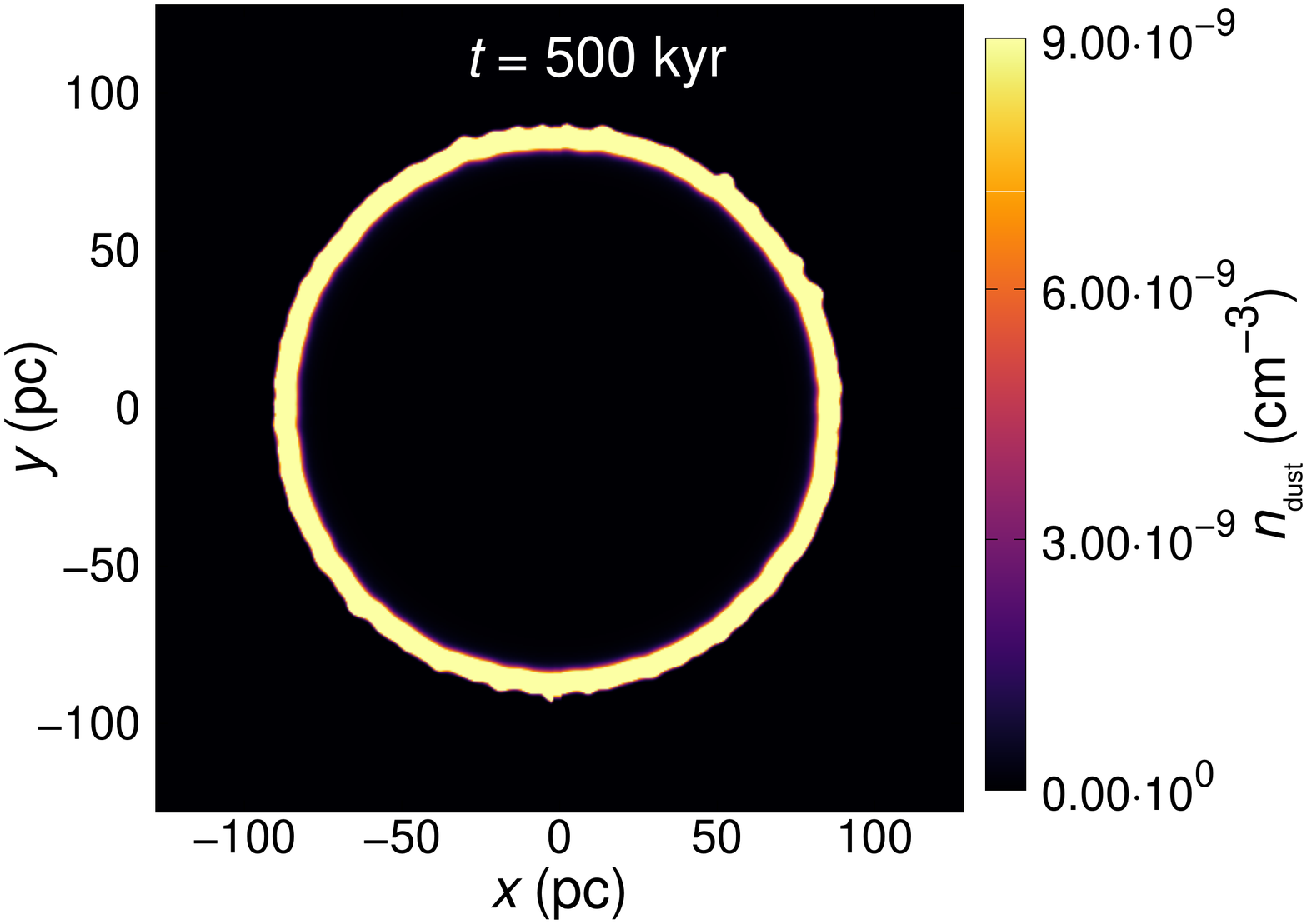}\hspace*{-0.05cm} 
 \includegraphics[trim=6.7cm 3.4cm 0.3cm 2.0cm, clip=true,page=1,height = 3.15cm]{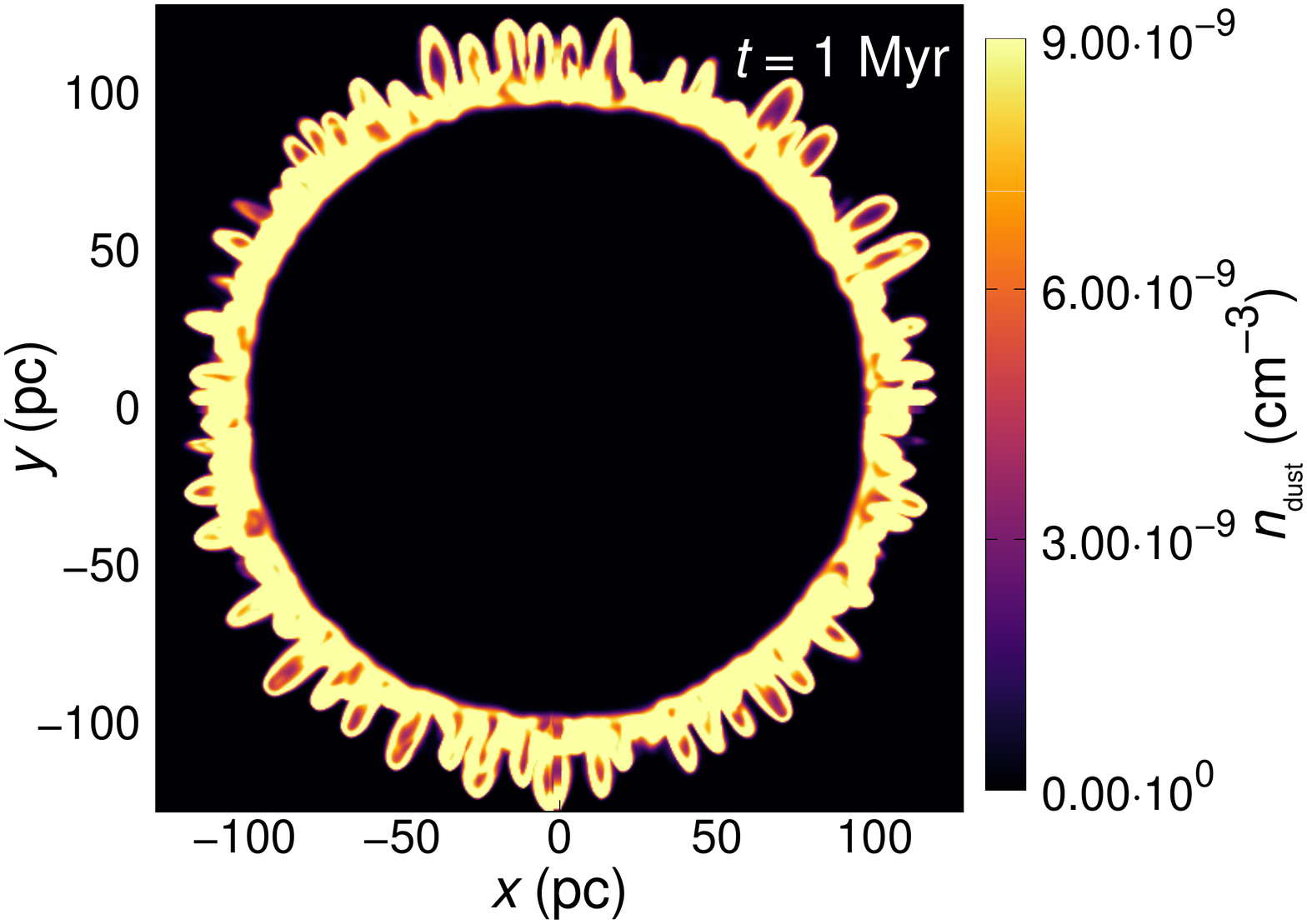}\\
 \includegraphics[trim=3.9cm 3.4cm 7.3cm 2.0cm, clip=true,page=2,height = 3.15cm]{Pics/Pics_C2/Density_1_00041.pdf}\hspace*{-0.05cm} 
 \includegraphics[trim=6.7cm 3.4cm 7.3cm 2.0cm, clip=true,page=2,height = 3.15cm]{Pics/Pics_C2/Density_1_00201.pdf}\hspace*{-0.05cm} 
 \includegraphics[trim=6.7cm 3.4cm 7.3cm 2.0cm, clip=true,page=2,height = 3.15cm]{Pics/Pics_C2/Density_1_00801.pdf}\hspace*{-0.05cm} 
 \includegraphics[trim=6.7cm 3.4cm 7.3cm 2.0cm, clip=true,page=2,height = 3.15cm]{Pics/Pics_C2/Density_1_02001.pdf}\hspace*{-0.05cm} 
 \includegraphics[trim=6.7cm 3.4cm 0.3cm 2.0cm, clip=true,page=2,height = 3.15cm]{Pics/Pics_C2/Density_1_04000.pdf}\\
 \includegraphics[trim=3.9cm 3.4cm 7.3cm 2.0cm, clip=true,page=3,height = 3.15cm]{Pics/Pics_C2/Density_1_00041.pdf}\hspace*{-0.05cm} 
 \includegraphics[trim=6.7cm 3.4cm 7.3cm 2.0cm, clip=true,page=3,height = 3.15cm]{Pics/Pics_C2/Density_1_00201.pdf}\hspace*{-0.05cm} 
 \includegraphics[trim=6.7cm 3.4cm 7.3cm 2.0cm, clip=true,page=3,height = 3.15cm]{Pics/Pics_C2/Density_1_00801.pdf}\hspace*{-0.05cm}
 \includegraphics[trim=6.7cm 3.4cm 7.3cm 2.0cm, clip=true,page=3,height = 3.15cm]{Pics/Pics_C2/Density_1_02001.pdf}\hspace*{-0.05cm} 
 \includegraphics[trim=6.7cm 3.4cm 0.3cm 2.0cm, clip=true,page=3,height = 3.15cm]{Pics/Pics_C2/Density_1_04000.pdf}\\
 \includegraphics[trim=3.9cm 1.3cm 7.3cm 2.0cm, clip=true,page=4,height = 3.57cm]{Pics/Pics_C2/Density_1_00041.pdf}\hspace*{-0.05cm} 
 \includegraphics[trim=6.7cm 1.3cm 7.3cm 2.0cm, clip=true,page=4,height = 3.57cm]{Pics/Pics_C2/Density_1_00201.pdf}\hspace*{-0.05cm} 
 \includegraphics[trim=6.7cm 1.3cm 7.3cm 2.0cm, clip=true,page=4,height = 3.57cm]{Pics/Pics_C2/Density_1_00801.pdf}\hspace*{-0.05cm}  
 \includegraphics[trim=6.7cm 1.3cm 7.3cm 2.0cm, clip=true,page=4,height = 3.57cm]{Pics/Pics_C2/Density_1_02001.pdf}\hspace*{-0.05cm} 
 \includegraphics[trim=6.7cm 1.3cm 0.3cm 2.0cm, clip=true,page=4,height = 3.57cm]{Pics/Pics_C2/Density_1_04000.pdf}\\
  \hspace*{-0.3cm}\includegraphics[trim=1.6cm 1.4cm 8.6cm 2.0cm, clip=true,page=1,height = 3.6cm]{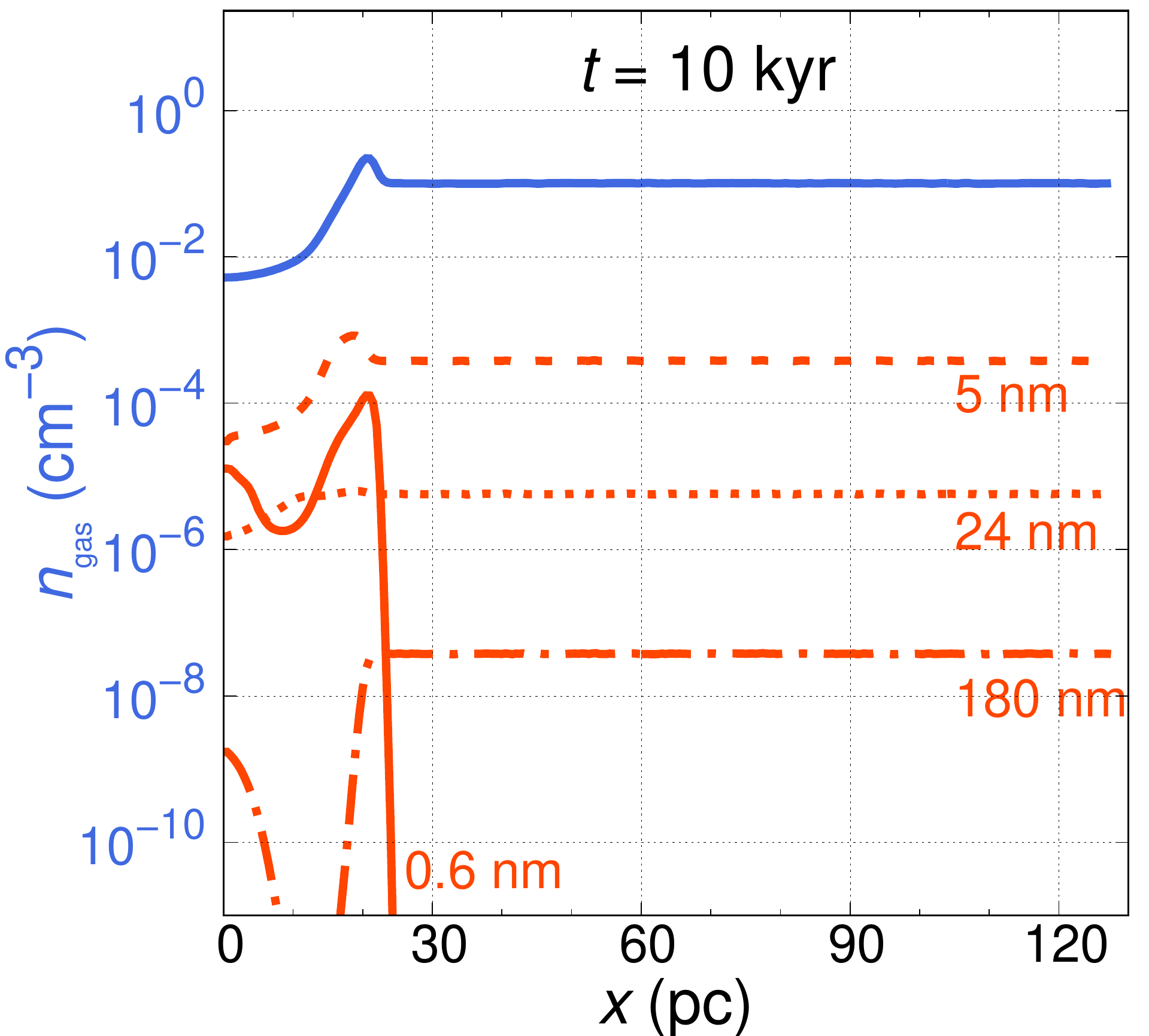}\hspace*{-0.1cm} 
  \includegraphics[trim=5.2cm 1.4cm 8.6cm 2.0cm, clip=true,page=2,height = 3.6cm]{Pics/Pics_C2/Profile_dust.pdf}\hspace*{-0.1cm} 
  \includegraphics[trim=5.2cm 1.4cm 8.6cm 2.0cm, clip=true,page=3,height = 3.6cm]{Pics/Pics_C2/Profile_dust.pdf}\hspace*{-0.1cm} 
  \includegraphics[trim=5.2cm 1.4cm 8.6cm 2.0cm, clip=true,page=4,height = 3.6cm]{Pics/Pics_C2/Profile_dust.pdf}\hspace*{-0.1cm} 
  \includegraphics[trim=5.2cm 1.4cm 3.2cm 2.0cm, clip=true,page=5,height = 3.6cm]{Pics/Pics_C2/Profile_dust.pdf}\\[-0.2cm]
 \caption{Same as Fig.~\ref{fig_C}  (simulation~C, $n_{\rm gas,0} = 1.0$~g~cm$^{-3}$, turbulence)  but for the gas-to-dust mass ratio $\Delta_{\rm gd}=10$.}
   \label{fig_C2} 
  \end{figure*}

\begin{figure*}
 \includegraphics[trim=2.5cm 1.5cm 2.2cm 2.3cm, clip=true,page=1,height = 6cm]{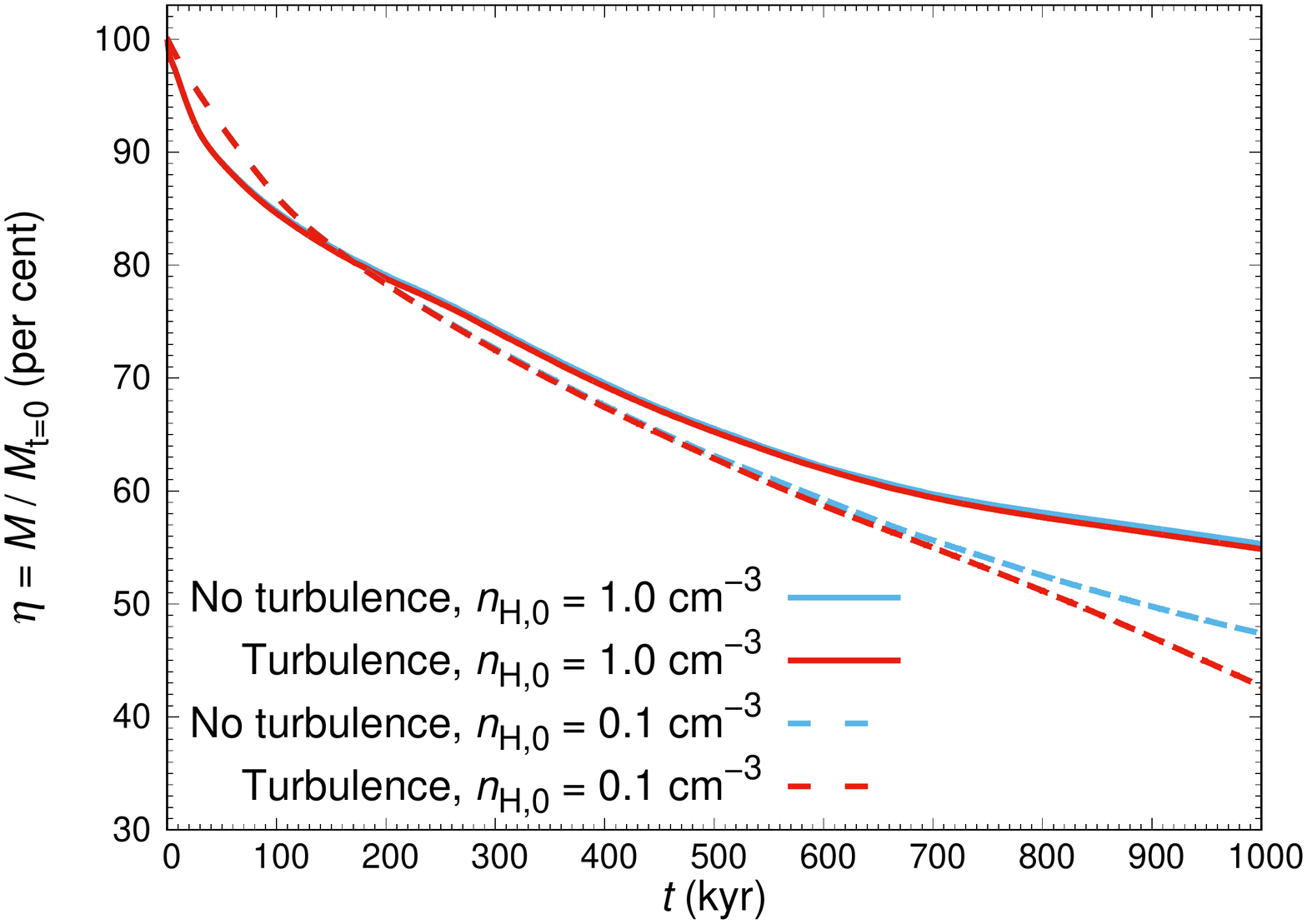}\hspace*{-0.1cm}
 \includegraphics[trim=2.5cm 1.5cm 2.2cm 2.3cm, clip=true,page=2,height = 6cm]{Pics/Pics_A2/Evolution_total.pdf}\\[-0.2cm]
  \caption{Same as Fig.~\ref{fig_dustmass_evolution} but for the gas-to-dust mass ratio $\Delta_{\rm gd}=10$.}
  \label{fig_DM_evolution}  
  \end{figure*}

\begin{figure*}
 \includegraphics[trim=2.4cm 1.5cm 3.9cm 2.45cm, clip=true,page=1,height = 6cm, page=1 ]{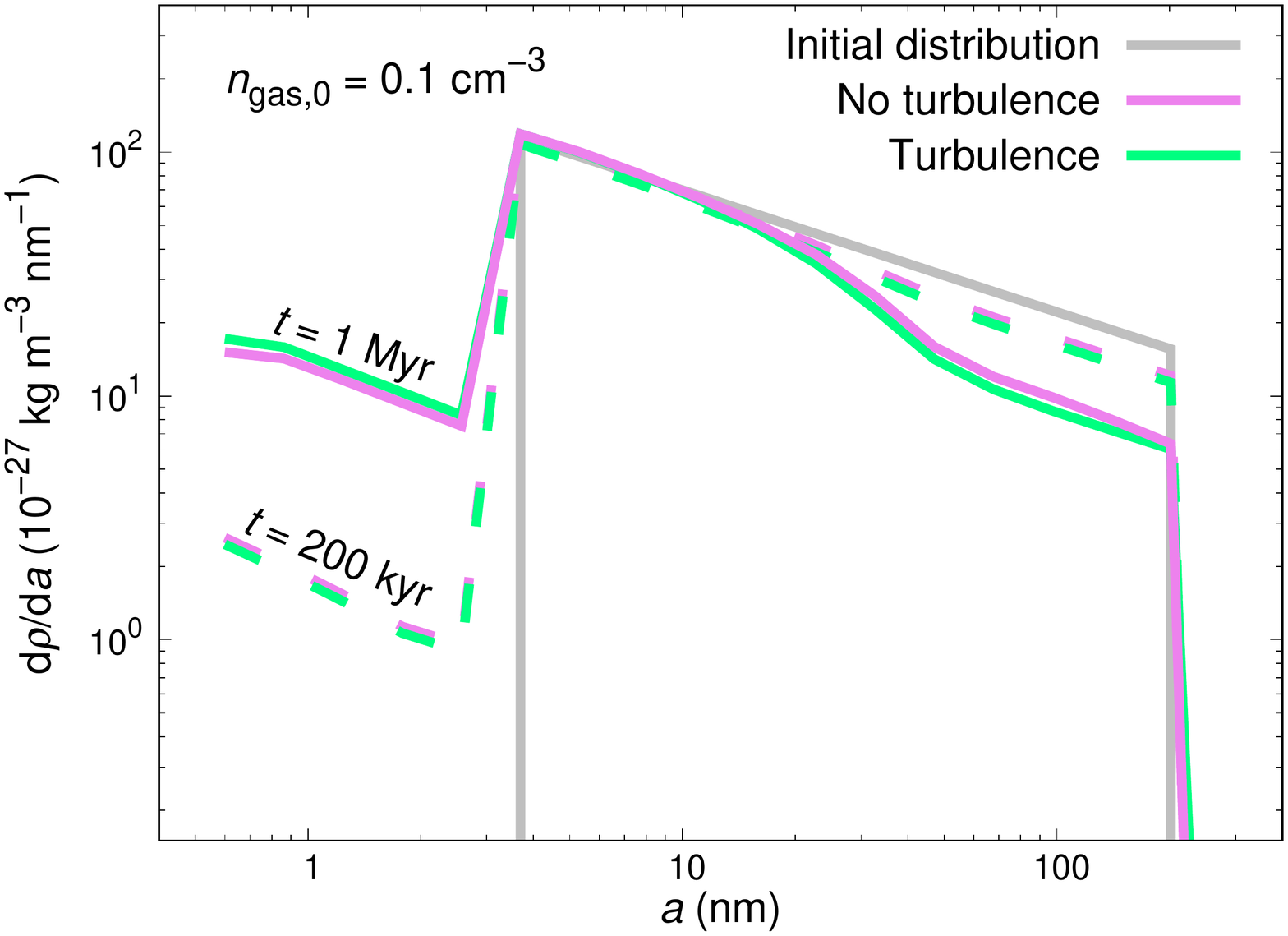}\hspace*{0.2cm}
 \includegraphics[trim=2.4cm 1.5cm 3.9cm 2.45cm, clip=true,page=1,height = 6cm, page=2 ]{Pics/Pics_A2/Particlenumbers_04000_compare.pdf}\\[-0.2cm]
  \caption{Same as Fig.~\ref{fig_ABCD_sizedistribution}  but for the gas-to-dust mass ratio $\Delta_{\rm gd}=10$.}
  \label{fig_A2C2_osizedistribution}  
  \end{figure*}

 The amount of destroyed dust and in particular the efficiency of grain-grain collisions depend strongly on the number density of dust grains and thus on the gas-to-dust mass ratio $\Delta_{\rm gd}$. We conducted the post-processing simulations A-D using \textsc{Paperboats} for \mbox{$\Delta_{\rm gd}=10$.}  The dust evolution for simulation~C ($n_{\rm gas,0} = 1.0$~g~cm$^{-3}$, turbulence) is shown in Fig.~\ref{fig_C2} and the survival rates and grain-size distributions for all four set-ups in Figs.~\ref{fig_DM_evolution} and  \ref{fig_A2C2_osizedistribution}.

\section{Turbulence-like initial velocity field}
\label{apx:faketurb}
\begin{figure*}
  \resizebox{\hsize}{!}{
      \includegraphics{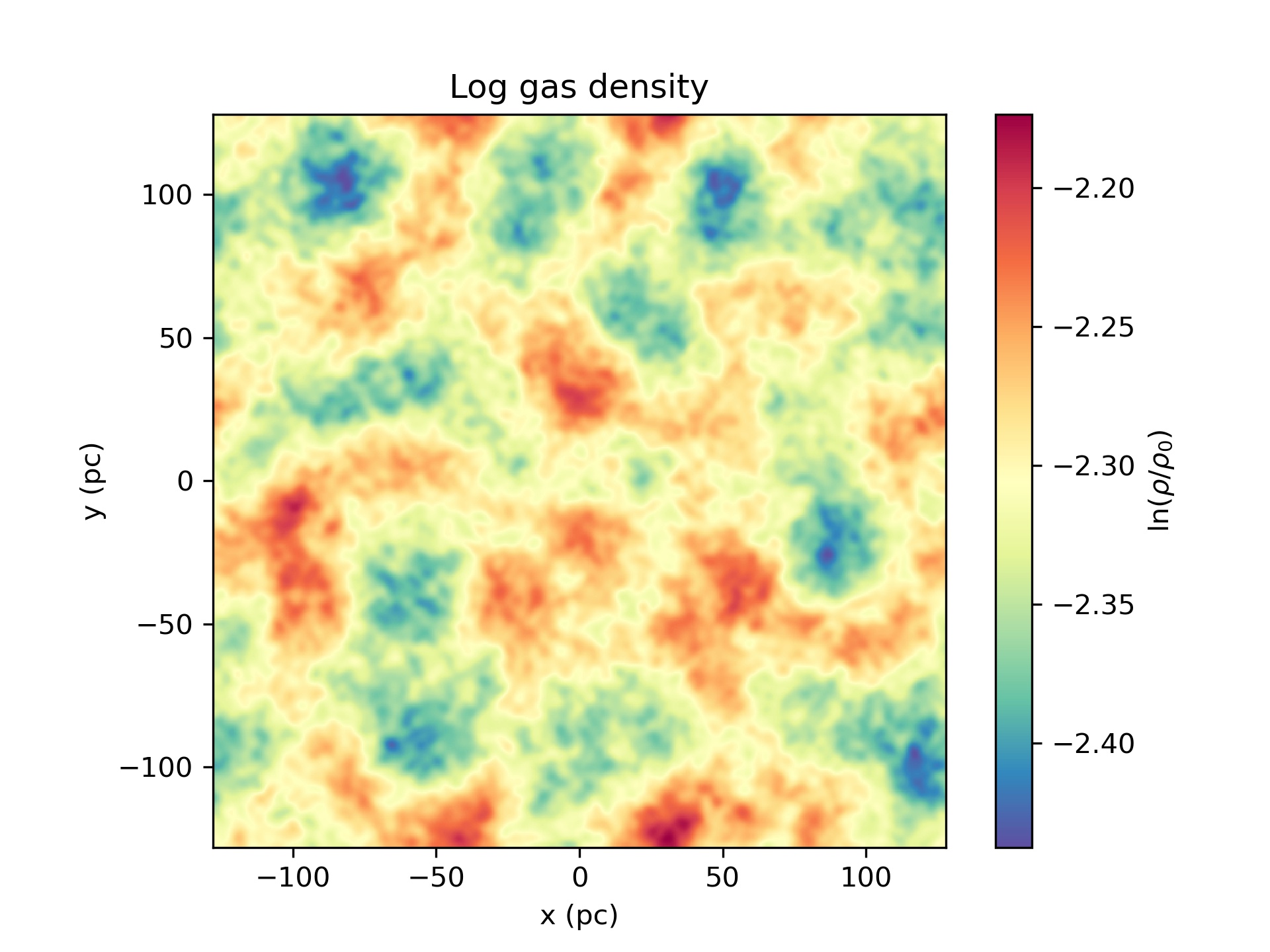}
      \includegraphics{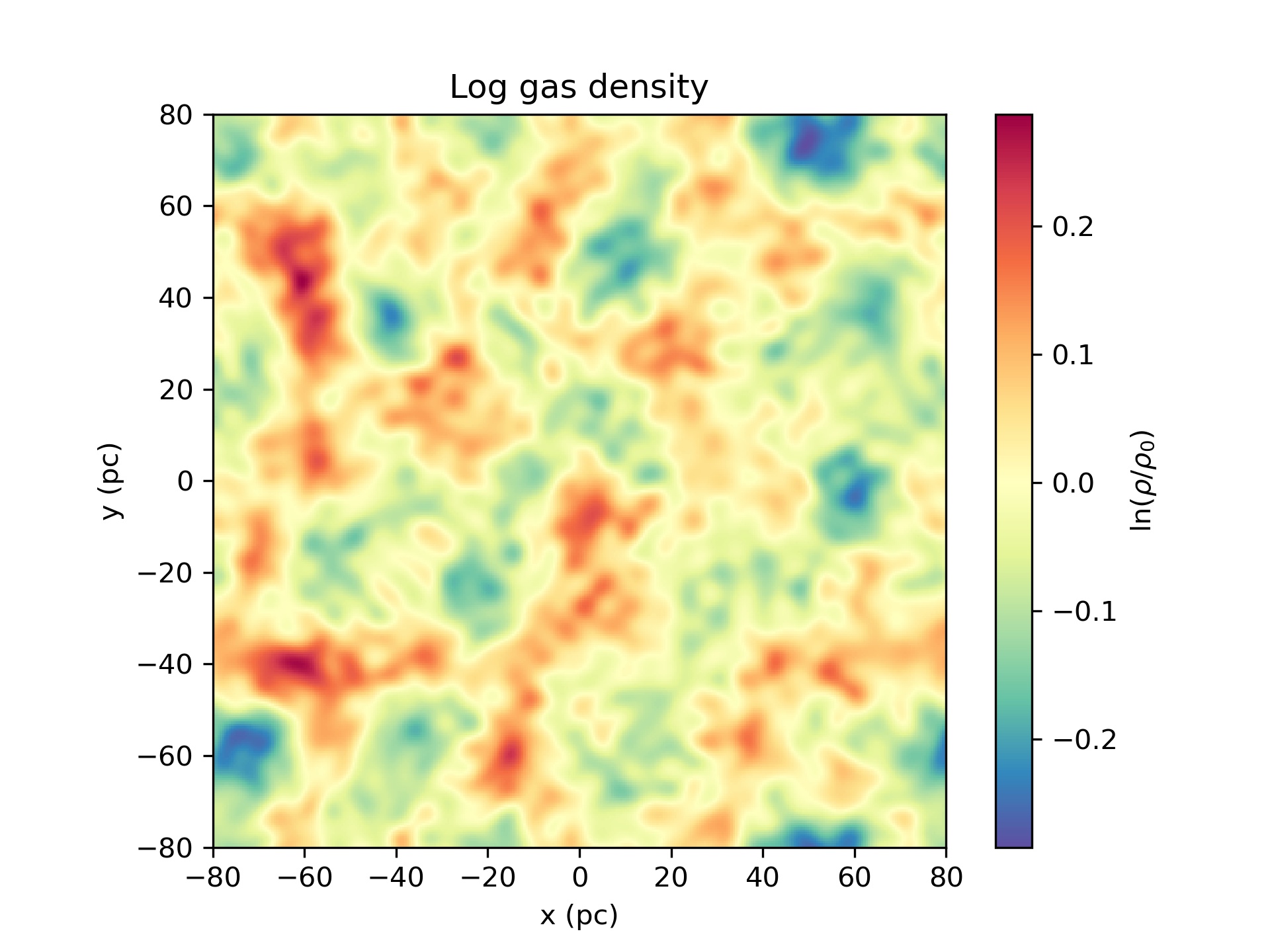}
      }
  \resizebox{\hsize}{!}{
      \includegraphics{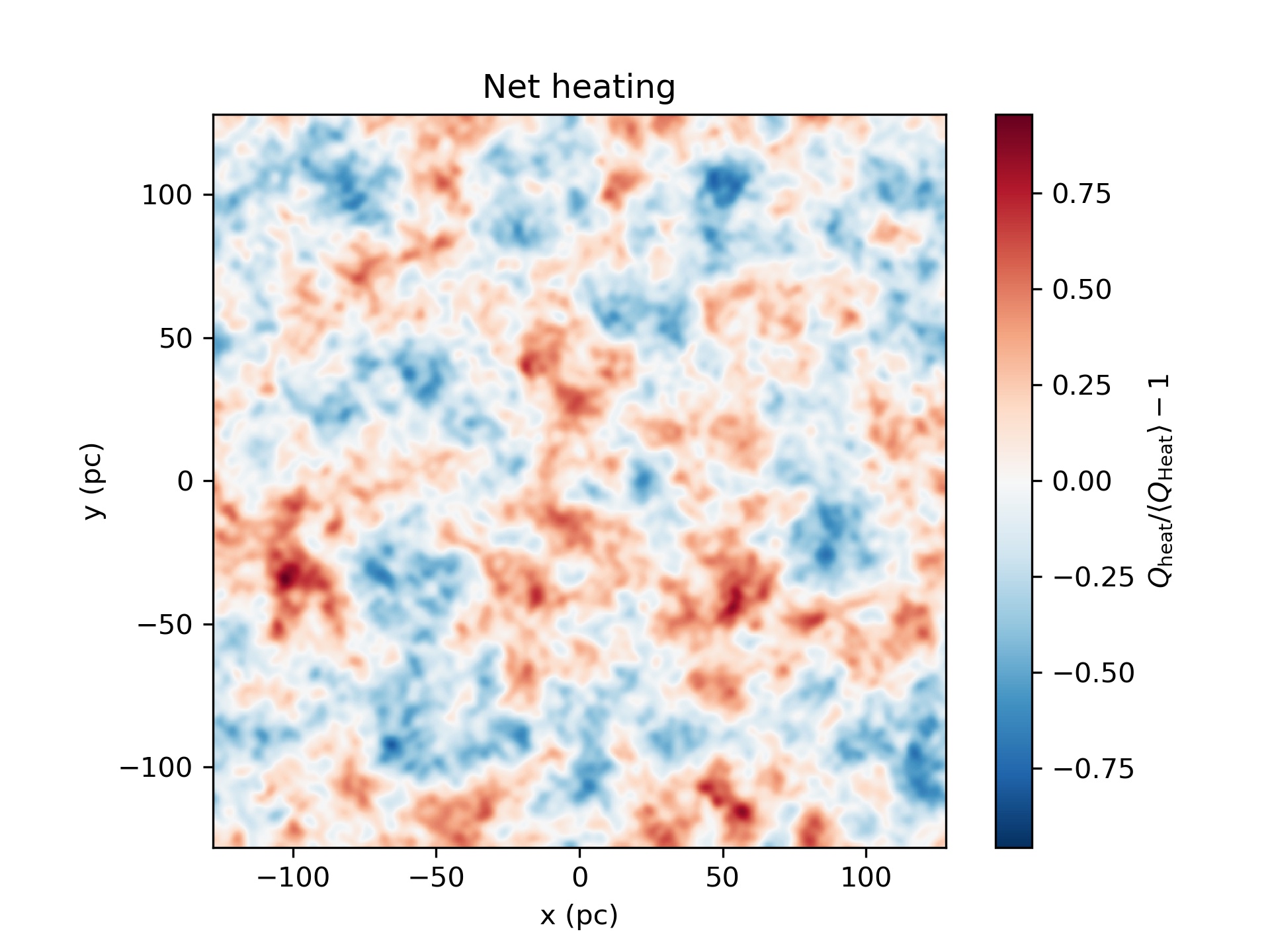}
      \includegraphics{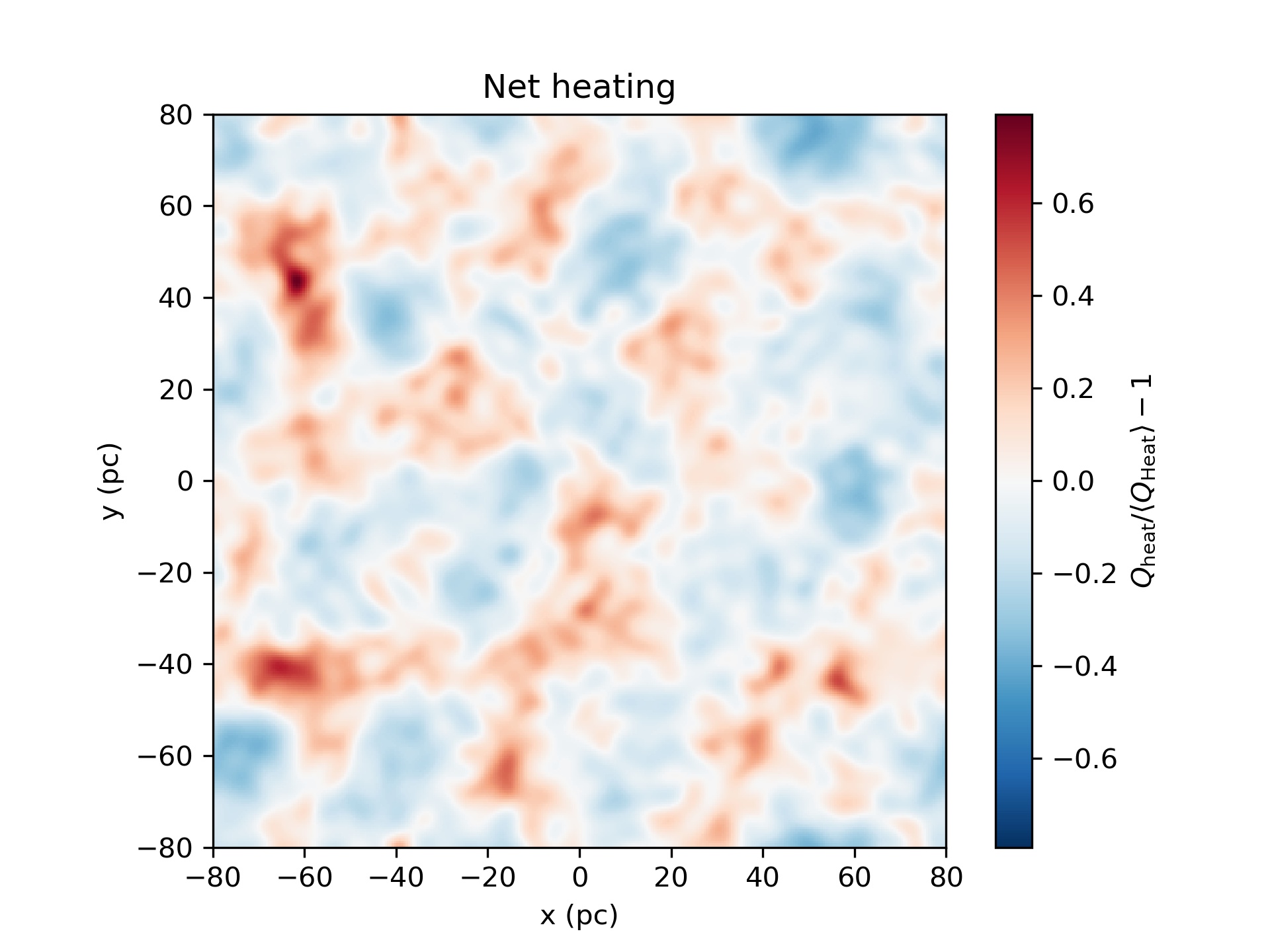}
      }
  \resizebox{\hsize}{!}{
      \includegraphics{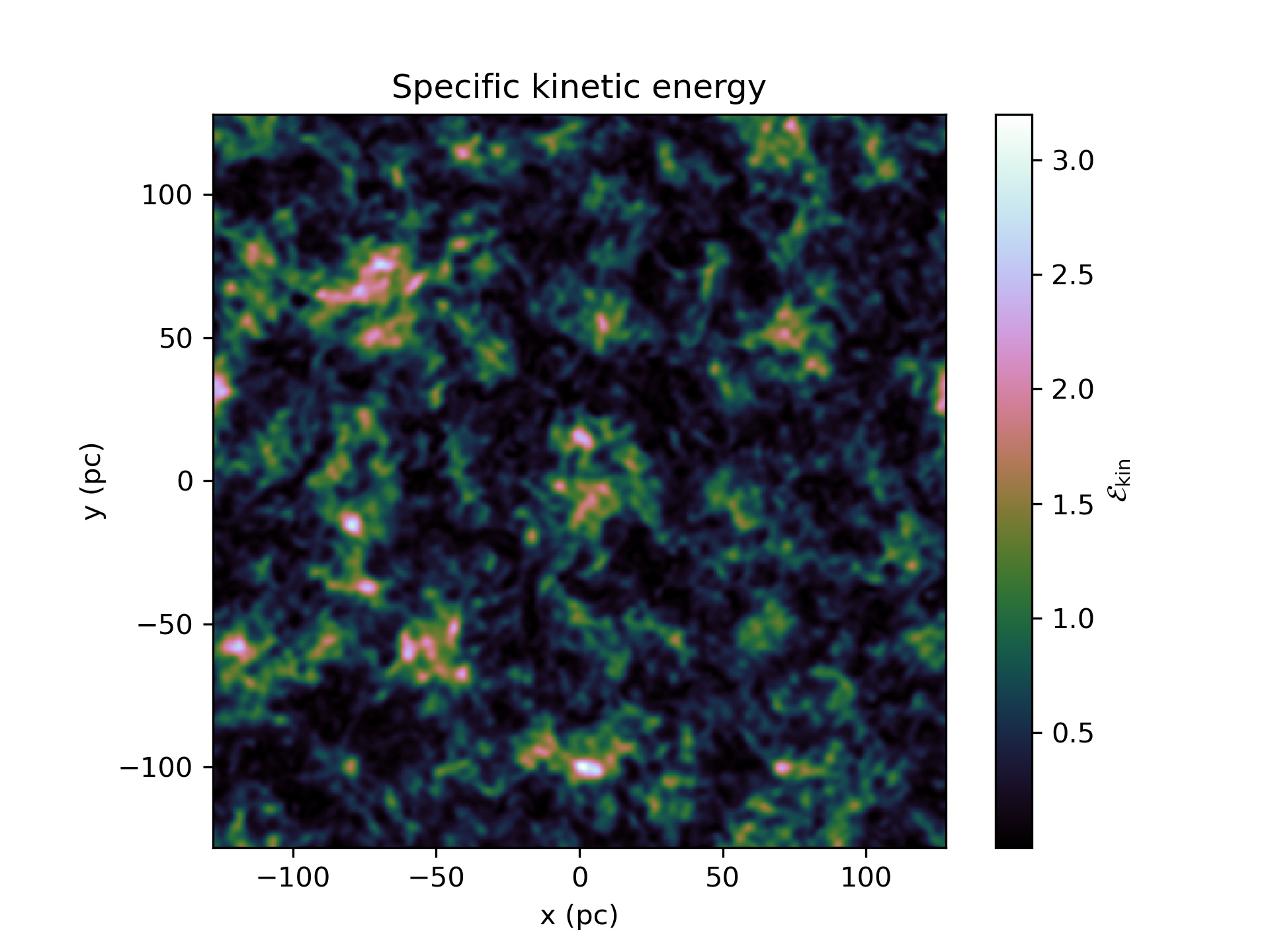}
      \includegraphics{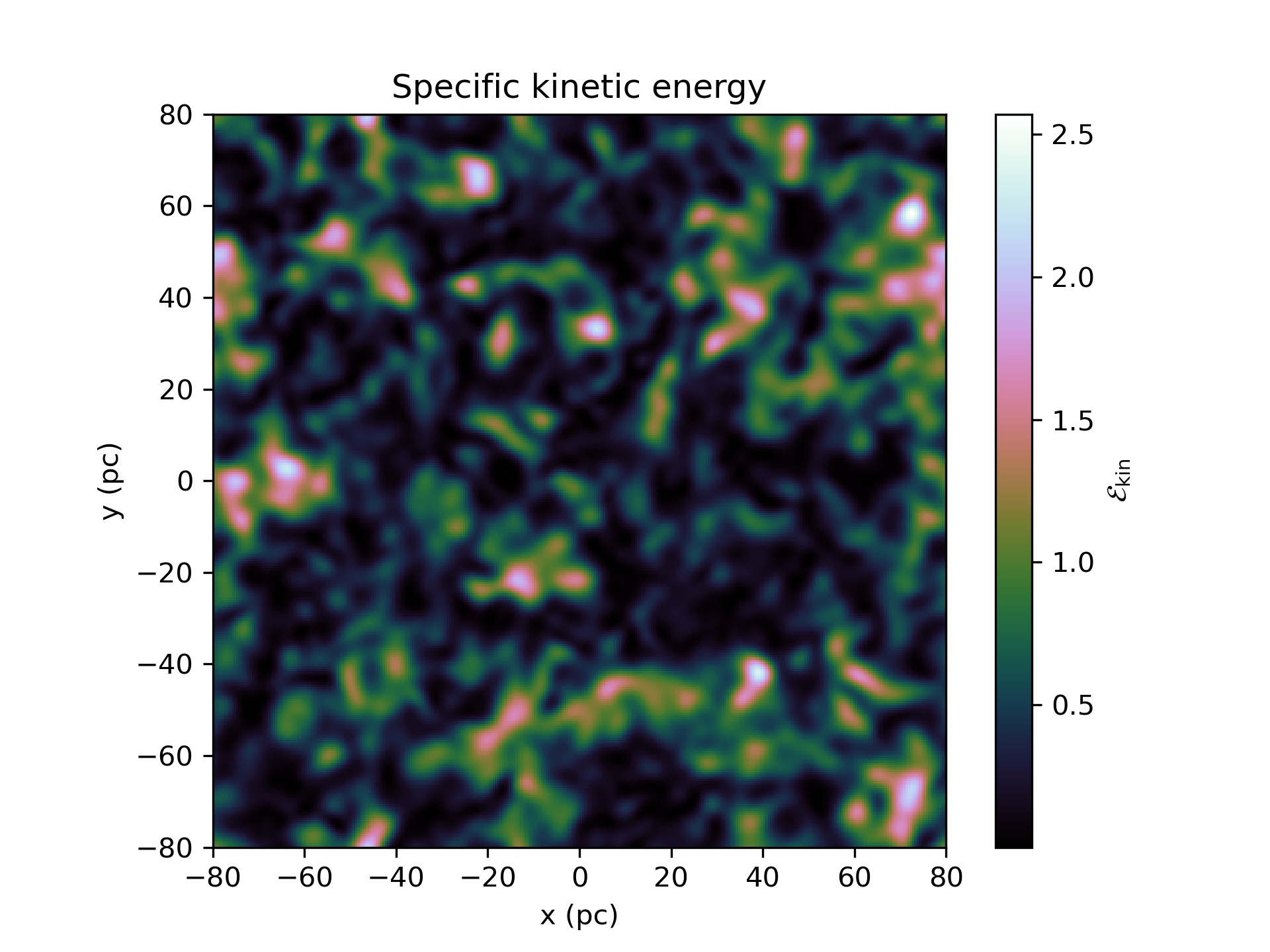}}
  \caption{Gas-density structure (upper panels), normalised net heating rate (middle panels) and specific kinetic energy ($\mathcal{E}_{\rm kin} = {1\over 2}|\,\mathbfit{u}|^2$) distribution (lower panels) without SN blast wave after 1~Myr of decay from the initial state with uniform density and a \citet{Kolmogorov41} velocity spectrum. \textit{Left:} $n_{\rm gas,0} = 0.1$~g~cm$^{-3}$. \textit{Right:}  $n_{\rm gas,0} = 1.0$~g~cm$^{-3}$.\label{fig:fake}}
  \end{figure*}

To quantify the variance in the ambient ISM density distribution that is caused by turbulence, we have performed simulations corresponding to C and D but without the SN blast wave (Fig.~\ref{fig:fake}). Obviously, the velocity variations produce fluctuations in the gas. The logarithm of the normalised gas density varies up to ${\sim}10$\% for $n_{\rm gas,0} = 0.1$~g~cm$^{-3}$ and ${\sim}30$\% for $n_{\rm gas,0} = 1.0$~g~cm$^{-3}$. The normalised heating rate for $n_{\rm gas,0} = 1.0$~g~cm$^{-3}$ is essentially proportional to $\ln\rho$, thus indicating a polytropic behaviour of the gas. The case with $n_{\rm gas,0} = 0.1$~g~cm$^{-3}$, on the other hand, shows a heating pattern which is much less correlated with $\ln\rho$. This is the reason why simulation C has its peculiar pattern of instabilities, since a non-polytropic gas can be expected to be prone to small-scale instabilities.  The range of the specific kinetic energy, $\mathcal{E}_{\rm kin} = {1\over 2}|\,\mathbfit{u}|^2$, of the flow is, on the other hand, ${\sim}25$\% greater for $n_{\rm gas,0} = 0.1$~g~cm$^{-3}$ compared to $n_{\rm gas,0} = 1.0$~g~cm$^{-3}$, which is crucial for the dust processing.

\section{Dust destruction in the first 100 kyr}
\label{apx:zoomindustdestruction}
\begin{figure}
 \includegraphics[trim=2.5cm 1.5cm 2.2cm 2.3cm, clip=true,page=5,height = 6cm]{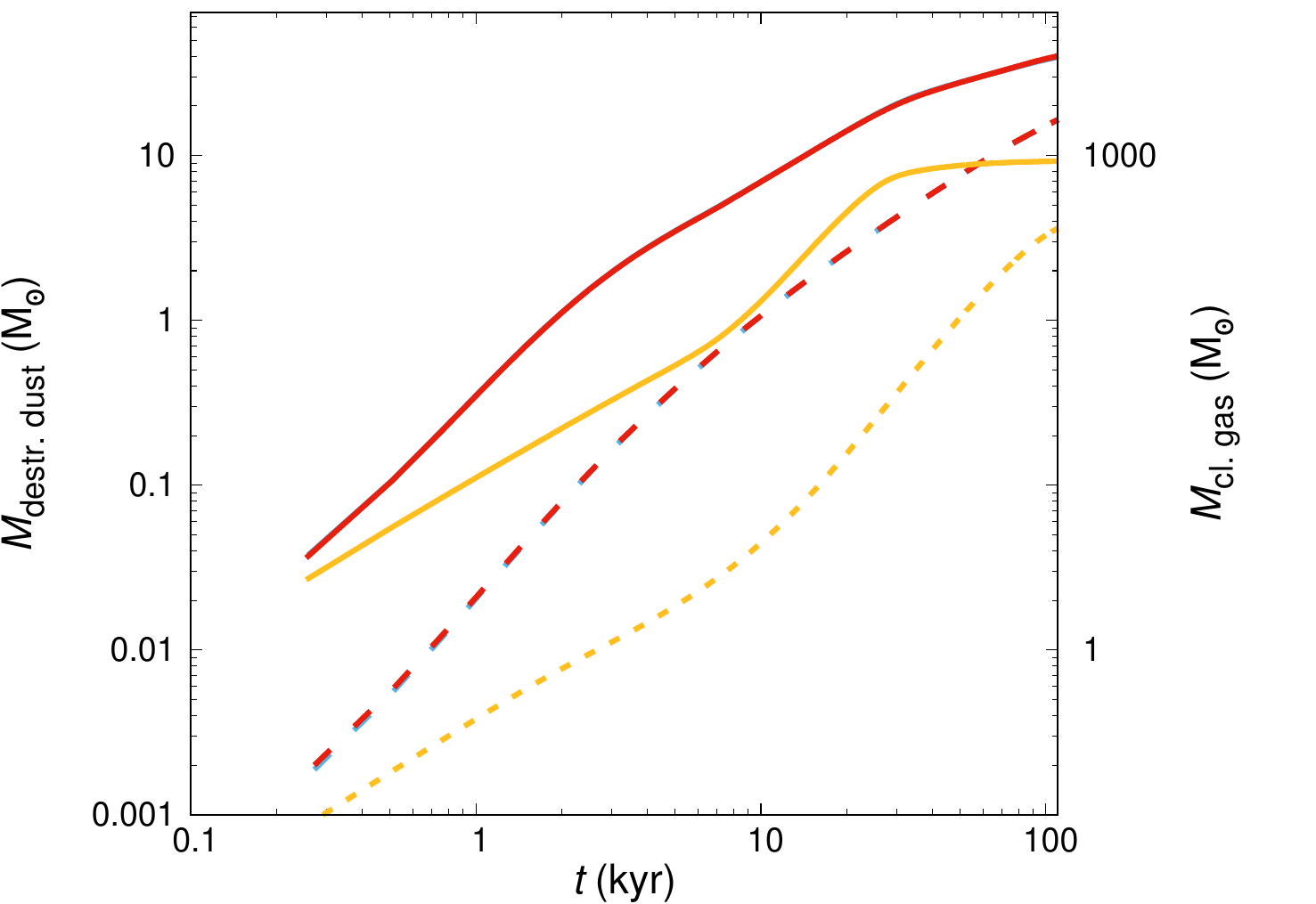}\\
  \caption{The first $\unit[100]{kyr}$ of the destroyed dust masses and cleared gas masses on a logarithmic scale (line colours and types are the same as in Fig.~\ref{fig_sputtering}).}
  \label{fig_sputtering_100kyr}  
  \end{figure}

In order to highlight the dust destruction 
during the early years, we show in Fig.~\ref{fig_sputtering_100kyr} the destruction curves (Fig.~\ref{fig_sputtering}) on a logarithmic scale  zoomed-in to the first $\unit[100]{kyr}$. We can see that for all simulation set-ups less than $\unit[0.3]{M_\odot}$ of dust are destroyed within the first $\unit[1]{kyr}$ and less than $\unit[40]{M_\odot}$ of dust within the first $\unit[100]{kyr}$.

\bsp	
\label{lastpage}
\end{document}